\documentclass[a4paper,11pt]{article}
\pdfoutput=1 
\usepackage{jcappub} 
\usepackage{afterpage,array,booktabs,longtable,multirow} 
\usepackage{amsmath,amssymb,amsxtra,amsthm} 
\usepackage[pass]{geometry}[2010/09/12]
\savegeometry{default-geometry}
\loadgeometry{default-geometry}
\usepackage{graphicx,subfig} 
\graphicspath{{./plots/}}

\usepackage{pdflscape}
\usepackage{expl3,xstring,xparse} 
\usepackage[usenames,dvipsnames]{xcolor}
\usepackage[capitalise]{cleveref}
\crefname{equation}{Equation}{Equations}
\crefname{figure}{Figure}{Figures}
\crefname{table}{Table}{Tables}

\NewDocumentCommand\cf{}{cf.}%
\NewDocumentCommand\etal{}{et al.}%
\NewDocumentCommand\etc{}{etc.}%
\NewDocumentCommand\eg{}{e.g.}%
\NewDocumentCommand\ie{}{i.e.}%

\catcode`\@=11 
\def\@capital#1{\uppercase{#1}}  
\DeclareRobustCommand{\acap}[1]{
        \ifnum\ifhmode\spacefactor\else2000\fi>1000\unskip\@capital#1%
        \else\unskip#1\fi%
}
\catcode`\@=12 

\newcommand{\abbrev}[2]{\newcommand{#1}{\acap{#2}}}
\abbrev{\DGP}{Dvali-Gabadadze-Porrati}
\abbrev{\fcd}{first-crossing distribution}
\abbrev{\HS}{Hu \& Sawicki}
\abbrev{\pdf}{probability density function}
\abbrev{\PS}{Press-Schechter}

\newcommand{\abs}[1]{\left| {#1} \right|}
\NewDocumentCommand\assb{}{a_{\textrm{SSB}}}
\NewDocumentCommand\denv{o}{\delta_{\textrm{env}\IfValueT{#1}{,#1}}}
\DeclareMathOperator\myd{d}
\NewDocumentCommand\dx{d<>o}{
    \myd{} \IfValueT{#1}{^{#1}}\IfValueT{#2}{_{#2}}
}
\NewDocumentCommand\Feff{}{F_{\mathrm{eff}}}
\ExplSyntaxOn
\NewDocumentCommand\fr{o}
{
  \IfValueTF{#1} 
  {\abs{f\sb{\R 0}} 
    \tl_if_empty:nF{#1}{= 10^{-#1}} 
  }
  {f(\R)} 
}
\ExplSyntaxOff
\NewDocumentCommand\given{}{\ensuremath{\, \middle| \,}}
\NewDocumentCommand\lcdm{}{\text{$\Lambda$CDM}}
\NewDocumentCommand\nueff{}{\nu_{\mathrm{eff}}}
\newcommand{\notimplies}{\mathrel{{\ooalign{\hidewidth$\not\phantom{=}$\hidewidth\cr$\implies$}}}}
\NewDocumentCommand\R{}{\mathcal{R}} 
\NewDocumentCommand\Senv{}{S_{\textrm{env}}}

\NewDocumentCommand\vectdiffop{d()d<>om}{
    \IfValueT{#1} { {}^{ {(#1)}} \!} 
    {#4}
    \IfValueT{#2} {^{#2}}  
    \IfValueT{#3} {\IfValueTF{#2}{_{\phantom{#2} #3}} {_{#3}}} 
}

\NewDocumentCommand{\scalar}{d()m}{\vectdiffop(#1){#2}}

\NewDocumentCommand\four{d()d<>om}{
    \IfValueTF{#2}
        {\vectdiffop(#1)<#2>[#3]{#4}}
        {\IfValueTF{#3}{\vectdiffop(#1)[#3]{#4}}
            {
            \IfSubStr*{ABCDEFGHIJKLMNOPQRSTUVWXYZabcdefghijklmnopqrstuvwxyz}{#4}
                {\vectdiffop(#1){\mathbf{#4}}} 
                {\vectdiffop(#1){\pmb{#4}}}      
            }
        }
}

\title{Universality of the halo mass function in screened gravity theories}
\author[a,1]{F. von Braun-Bates,\note{Corresponding author.}}
\author[a,b]{and J. Devriendt}
\affiliation[a]{Astrophysics, University of Oxford, Denys Wilkinson Building, Keble Road, Oxford OX1 3RH, UK}
\affiliation[b]{Universit\'e de Lyon, Universit\'e Lyon 1, ENS de Lyon, CNRS, Centre de Recherche Astrophysique de Lyon, UMR 5574, F-69230 Saint-Genis-Laval France}

\emailAdd{francesca.vonbraun-bates@physics.ox.ac.uk}
\emailAdd{julien.devriendt@physics.ox.ac.uk}

\abstract{
We investigate the efficacy of the halo mass function (HMF) as a probe of $f(\R)$, symmetron and Dvali-Gabadadze-Porrati (DGP) gravity. In this regard, we develop an excursion-set method to generalise a range of popular HMF fitting functions from General Relativity (GR) to screened Modified Gravity (MG), considering the HMF dependence on the critical density parameter $\delta_{c}$.  In particular, we propose a variety of new methods to account for the environmental dependence of chameleon screening and compare their accuracy to existing ones by submitting them to $N$-body simulation acid tests.  Using the nested sampling routine \texttt{MultiNest}, we then examine two propositions: can the MG $N$-body results be accurately described by a \lcdm{} GR HMF, and do MG HMFs display 
universality.

More specifically, the values of the free parameters defining a given HMF can always be determined by fitting $N$-body simulation data, but the question arises as to 
the dependency of these values on cosmological parameters and halo finding algorithm in the GR case, and indeed on the gravitational force itself when considering MG theories. We find that haloes extracted from any one of the MG $N$-body simulations can plausibly be fit by a \lcdm{} GR HMF, albeit with unusual best fit parameter values. In other words, despite the MG $N$-body data failing to support a clear deviation in the functional form of the HMF, it indicates that one needs to carefully establish \lcdm{}-expected credible regions for the values of the best fit parameters before attributing any difference in these values to MG. 

Alternatively, one can ask whether the MG $N$-body data is compatible with the existence of a truly universal (as in completely independent of the gravity model) HMF. Although we find that some HMFs display more universality than others --- in the sense that they exhibit better overlap of their best fit parameters posterior probability distribution functions (PDFs) for \emph{different} screening models ---, the general trend, regardless of the HMF considered, is for DGP HMF parameter PDFs to overlap with those of \lcdm{}, symmetron HMF parameter PDFs to overlap with those of $\fr[6]$, and $\fr[5]$ HMF parameter PDFs to remain somewhat distinct from all the others. These results strongly suggest that a truly universal HMF will remain elusive. }

\begin{document}
\maketitle
\flushbottom

\section{Introduction} 
\label{sec:introduction}

The era of precision cosmology will enable us to distinguish between competing gravitational models.  While the concordance cosmology of \lcdm{} has performed remarkably well over the past two decades \cite{2016PDU....12...56B}, there exist theoretical problems\footnote{Examples include fine-tuning of $\Lambda$, the lack of direct detection of CDM, possible fine-tuning of inflation \cite{Clifton:2011jh}.} and tensions between observations measured at early and late times.\footnote{Compare measurements with Planck 2015 \cite{2016A&A...594A..13P} to the values of $\sigma_8$ from weak lensing and cluster number counts, $H_0$ from Lyman-$\alpha$ measurements of BAO, $f\sigma_8$ from redshift-space distortions \cite{2016PDU....12...56B}.}  Hence a plethora of MG theories (\eg{} \cite{Clifton:2011jh}), and the surge in interest to probe a range of MG models which may alleviate these issues. 

The HMF, i.e. the number density of haloes within a given mass range, is the simplest (one point) statistic one can use to study structure formation 
and evolution in the Universe. It also possesses the remarkable property that, given GR, it is (nearly-)universal, in the sense that its functional form does 
not depend on redshift or cosmology \cite{Zentner:2006vw,2017MNRAS.467.3454B}. These features make the HMF the ideal candidate to investigate 
MG theories. Can we apply GR HMFs (and specifically \lcdm{}) to these other theories of gravity? In other words is the 
HMF also universal across (\ie{} insensitive to) different gravity theories?  

The simplest theoretical model to understand the origin and evolution of the HMF is called the Press-Schechter ansatz \cite{1974ApJ...187..425P} which derived the fraction of haloes above a given mass for the first time.  Excursion set theory was then proposed in \cite{1991ApJ...379..440B} to set the Press-Schechter result 
on a more solid theoretical footing.  This approach relies upon two key quantities.  The \lq\lq{}collapse density\rq\rq{} $\delta_{c}$ is the solution to the boundary value problem of spherical collapse of an overdensity to form a halo.  The \lq\lq{}resolution\rq\rq{}\footnote{We retain the term used by Bond et al. in their excursion set paper \cite{1991ApJ...379..440B}: other authors use various different names for this quantity.} $S$ is the variance in the (linear) overdensity field $\delta$ for a given radius.  Then one can map $S$ from a radius to a halo mass $M$ via a window function.  The fundamental idea behind the excursion set formalism is the realisation that the fraction of haloes above a certain mass can be obtained by solving a diffusion problem in the phase space $(\delta, S)$ with an absorbing barrier given by the collapse density $\delta_{c}$.  This model has already been applied to a number of MG scenarios \cite{2012MNRAS.425..730L,2012MNRAS.421.1431L,2013JCAP...11..056B,2014PhRvD..89b3523T,2011PhRvD..83f3511P}.

A pre-requisite for using the HMF as a probe of MG is a firm understanding of how it is typically derived in GR.  Superficially this may be obvious: the analytical HMF of Press-Schechter has long been supplanted by a range of fitting functions (summarised in \eg{} \cite{2013A&C.....3...23M}) whose free parameters are calibrated using $N$-body simulations.  However, this calibration only performed for certain values of the cosmological parameters and a certain range in mass and redshift.  It is the fitting function  which is deemed universal ---in the sense that it can be applied to a range of redshifts and cosmologies \cite{2008ApJ...688..709T}--- rather than the HMF itself.  There has been disagreement over whether \cite{2014MNRAS.439.3156J,Zentner:2006vw,2007ApJ...671.1160L} or not \cite{2007MNRAS.374....2R,2006ApJ...646..881W,2003MNRAS.346..565R} this universality holds --- and these papers only test a handful of the competing fitting functions.  Despite these shortcomings, it is widely believed that once various obscuring factors are addressed, the deviations from non-universality in the calibrated HMFs are of a few percent \cite{2008ApJ...688..709T}.  In particular, if one uses the correct measurement of halo over-density \cite{Despali:2015yla} and includes $z$-dependence in both the over-density and the resolution, it is thought that the best-fit parameter values provide an HMF which applies over several decades of mass and up to $z = 30$,\cite{2007MNRAS.374....2R}, at least for a fixed cosmology \cite{2011MNRAS.410.1911C}.  

Unfortunately, due to the additional computation time and complexity required to run $N$-body simulations in MG, one cannot take the GR approach described in the previous paragraph.  Instead, one typically resorts to using either the Press-Schechter or the Sheth-Tormen functions (although \cite{2017JCAP...03..012V} uses the Peacock function).  However, there appears to be little consensus on how the fifth-force degree of freedom affects the suitability of the fitting functions.  Some authors use the Sheth-Tormen fit on the grounds that this is considered satisfactory in the GR simulations (symmetron \cite{2014PhRvD..89b3523T}, $\fr$ \cite{Lombriser:2013wta}, DGP \cite{2010PhRvD..81f3005S}).  A counter-example is the Galileon HMF: the authors of \cite{2013JCAP...11..056B} reason that the assumptions used in the Sheth-Tormen model are unlikely to hold, given the Galileon fifth-force modification to GR, and so use the Press-Schechter fit. 
These examples illustrate the additional theoretical and practical complications when deriving an appropriate HMF in MG.


This paper thus seeks to address these issues.  Given the profuse and rapidly evolving landscape of MG models \footnote{The recent neutron star merger detection GW170817 (and its electromagnetic counterpart GRB 170817A) seemingly ruled out complex Horndeski gravity models (but not simpler models such as $f(\R)$ because these latter do not change the speed at which gravitational waves propagate) \cite{Baker2017}}, we choose to restrict ourselves to \HS{}-$f(\R)$, Symmetron and DGP gravity.  This selection is intended to represent a range of MG families and screening mechanisms, while ensuring that each theory is sufficiently related to take a unified approach. First we outline the various screened models of MG used in \cref{ch:screened_mg_theories}.  More detail is provided on: $f(\R)$ in \cref{sub:fr_gravity}, Symmetron in \cref{sub:symmetron_gravity} and DGP in \cref{sub:dgp_gravity}.  Then \cref{sec:the_halo_mass_function} describes the theoretical approach to the HMF.  \cref{sub:spherical_collapse_in_mg} describes how to extend the spherical collapse model from GR to MG; \cref{sec:excursion_set_theory} outlines the excursion set theorem, which is still used to derive the HMF in MG; \cref{sub:fitting_functions} summarises the fitting functions used in this paper and why they are universal.  Some technical details are passed over in this section and described in \cref{app:technical_aspects_of_spherical_collapse_in_mg}.  Next we describe the numerical work. \cref{sub:n_body_simulations_haloes} outlines the simulation parameters and algorithms used for the various $N$-body simulations and the extraction of the HMF from the simulations; \cref{sub:bayesian_inference} outlines the theory used to calculate the best-fit values for the free parameters in the HMFs.  (We calibrate the \lq\lq{}nested sampling\rq\rq{} algorithm and describe the data processing effects in \cref{app:calibration_of_multinest}).  The results section discusses the questions raised at the beginning of the introduction: \cref{sub:assuming_concordance_cosmology} shows how MG manifests itself in the free parameter values even when assuming a \lcdm{} HMF; \cref{sub:recalibrating_best_fit_parameters} recalibrates the fitting functions using the same gravity model in both the $N$-body simulations and the HMFs.  Finally, we summarise our method and key results in \cref{sub:summary} and discuss avenues for further work in \cref{sub:further_work}.  

The conventions used throughout this paper are: 
\begin{itemize}
\item Units: $c = 1$, Einstein constant $\kappa = 8\pi G/c^{2}$, Planck mass $M_{Pl} = \kappa^{-1/2}$ 
\item  Gauge choice is conformal Newtonian with metric signature $(-+++)$ 
 \begin{equation*}
  \dx s^{2} = a^{2}(\tau) \left[ -(1+2\Psi) \dx \tau^2 + (1 - 2\Phi) \delta_{ij} \dx \four<i>{x} \dx \four<j>{x} \right] 
 \end{equation*}
\item The full spacetime metric has Greek indices ranging from $0$ to $3$ while flat spatial hypersurfaces have Roman indices ranging from $1$ to $3$. 
\end{itemize}
This accounts for a variety of sign changes in some formulae compared to the equations in the citations.


\section{Screened gravity theories} 
\label{ch:screened_mg_theories}

The MG theories studied in this paper share a common trait to ensure that the successes of GR are unaffected by any modifications.  The \emph{screening mechanism} is a technique whereby fifth-force modifications to GR are \lq\lq{}screened away\rq\rq{} in regions where the theory must mimic GR in order to be experimentally viable.  A range of such techniques exists, so we limit ourselves to three popularly-held categories of screening---\emph{chameleon} \cite{Khoury:2003rn}, \emph{symmetron} \cite{2010PhRvL.104w1301H,2011PhRvD..84j3521H} and \emph{Vainshtein} \cite{Vainshtein:1972sx}---by selecting one theory from each family, namely: \HS{} $\fr$ (\cite{0705.1158}; chameleon), Hinterbichler-Khoury symmetron \cite{2010PhRvL.104w1301H} and \DGP{} (\cite{2000PhLB..485..208D}; Vainshtein).  Our specific MG models and parameter values are detailed in \cref{app:screened_gravity_theories}.

The background evolution of these models is always close to \lcdm{} for viable values of the free parameters.  Accordingly, we only employ the MG modification in computing the spherical collapse of the haloes within a \lcdm{} background.

By applying the quasi-static approximation in the weak-field limit, we can express the effect of the scalar field in the halo model without evolving the full scalar field equations of motion.  Does this produces a reasonable approximation to the perturbed field equations and the equation of motion of the scalar field?  In model-specific contexts, \cite{PhysRevD.80.043001} asserts that the QSA is appropriate on sub-horizon scales for DGP, \cite{PhysRevD.89.084023} finds $< 1\%$ changes in the local power spectrum for Symmetron models and \cite{Noller:2013wca} conclude that errors in the matter overdensity are $< 5\%$ for Hu-Sawicki $\fr$ (albeit including super-horizon scales where one expects the QSA to break down).  In general, the slow-rolling nature of the scalar field which produces \lcdm{}-like behaviour also causes the QSA matter overdensity to be accurate at the percent level on sub-horizon scales \cite{Noller:2013wca}.  

The resulting equations relate (after copious algebra: see \cite{2011PhLB..706..123D} for details) the metric perturbations $\Phi$, $\Psi$ to the perturbed field $\chi$ and gauge-invariant matter density $\delta$.  More algebraic manipulation (for details see \cite{2011PhLB..706..123D}) allows us to recast this as a Poisson equation.  Following \cite{Lombriser:2013wta,2012MNRAS.421.1431L} one can use the excursion set paradigm to transform this into an equation for the evolution of the critical overdensity for halo collapse $\delta_{c}$, which we call the \lq\lq{}barrier density\rq\rq{} in this paper.  Thus we obtain our effective multiplication of $G_{\mathrm{eff}} = (1 + F_{\mathrm{eff}}) G_{N}$ in the excursion set ODE which ultimately determines the value of $\delta_{c}$ in the theoretical HMF.  The exact expressions for $F_{\mathrm{eff}}$ are lengthy: we present them in \cref{app:screened_gravity_theories} for each MG model considered here.

Let us characterise the various types of screening used in this paper.  Screening mechanisms add new terms to the GR action, which are designed to suppress the non-GR modifications under certain conditions.  By construction, this suppression of the fifth-force modification happens on non-linear regimes.  The conditions required for screening divide the mechanism into three groups: chameleon, symmetron and Vainshtein. The different screening methods correspond to different behaviours of the Lagrangian, which we can classify by expanding about some value $\phi_0$ by an infinitesimal value $\left( \delta \phi \right)$ \cite{2015CQGra..32x3001B}: 
\begin{equation} \label{eq:Lagrangian_screening}
  \mathcal{L} = \frac{1}{2} \R M^{2}_{\mathrm{Pl}} 
  + \four[\mu]{\partial} \left( \delta\phi \right)  \four[\nu]{\partial} \left( \delta\phi \right) \four<\mu\nu>{Z} \left( \phi_0 \right)
  + \left( \delta\phi \right) m^{2} \left( \phi_0 \right) 
  + \frac{\rho_{m}}{M_{\mathrm{Pl}}} \beta \left( \phi_0 \right)
\end{equation}
In low density regions the scalar field takes a value $\phi_0 = \phi_{\mathrm{low}}$, for which:@
\begin{equation} \label{eq:gamma_defn}
    \gamma \equiv \abs{ \frac{\vec{F}_{\phi}}{\vec{F}_{N}} } 
  \propto \beta^{2} \left( \phi_{\mathrm{low}} \right) \sim 0
\end{equation}
and we see that the contribution of the fifth force $\lvert \vec{F}_{\phi} \rvert$ is non-negligible compared to the Newtonian value $\lvert \vec{F}_{N} \rvert$.  Compare this to the high-density value $\phi_0 = \phi_{\mathrm{high}}$, for which:
\begin{subequations} \label{eq:gamma}
\begin{align}  
  \gamma &\ll 1 &&\text{by definition}
  \intertext{which can only be produced by:}
  \beta \left( \phi_{\mathrm{high}} \right) &\ll \beta \left( \phi_{\mathrm{low}} \right)  &&\text{matter coupling suppressed} &&\text{(symmetron)} \\
  m \left( \phi_{\mathrm{high}} \right) &\gg m \left( \phi_{\mathrm{low}} \right)  &&\text{large local mass} &&\text{(chameleon)} \\
  \four<\mu\nu>{Z} \left( \phi_{\mathrm{high}} \right) &\gg \four<\mu\nu>{Z}\left( \phi_{\mathrm{low}} \right)  &&\text{weakened matter source} &&\text{(Vainshtein)}
\end{align}
\end{subequations}
This permits us to classify a screened MG theory not by abstract considerations (\textit{i.e.} which conditions are relaxed in Lovelock's Theorem) but rather by the practicalities of the mechanism by which it evades local tests of gravity. 

\cref{app:screened_gravity_theories} provides more detail on the fifth-force modification caused by various MG theories used in this paper.  We have already demonstrated how to use the effective fifth-force contribution to $G$ to find the density required for collapse of a spherical top-hat in \cite{2017JCAP...03..012V}.  We now turn our attention to the halo mass functions in the next section.


\section{Generalising the halo mass function from GR to MG} 
\label{sec:the_halo_mass_function}

The halo mass function $n(M)$ is defined to be the number density of dark matter haloes in a given mass interval at a certain redshift.  This is closely related to the first-crossing distribution $f(S)$ from excursion-set theory.  This is the probability for a given random walk in the excursion-set phase space $\left( \delta , S \right)$ to be absorbed at resolution $S$ when the over-density $\delta$ reaches the collapse density $\delta_c$.  In turn, $f(S)$ can be expressed in terms of a \lq\lq{}universal\rq\rq{} fitting function $F(\nu)$, which is invariant under changes in redshift and cosmological parameters.  The aim of this next section is to \lq\lq{}unpack\rq\rq{} the details of this process, including precisely defining the key quantities $\delta, S, \delta_{c}, \nu$ and the functions $f(S)$ and $F(\nu)$.  This enables us to extend the existing formalism to calculate the HMF from GR to the context of screened MG.  

In \cref{sub:spherical_collapse_in_mg} we show how the critical density $\delta_{c}$ forms a drifting-and-diffusing barrier which absorbs the trajectories in the excursion-set formalism.  We show how to determine this \lq\lq{}barrier density\rq\rq{} via the solution to an ODE which can be tailored to each theory of gravity using an effective Newton's constant.  This equation encapsulates the modifications to non-linear collapse from MG.  

In \cref{sec:excursion_set_theory} we build upon this result by deriving the first-crossing distribution $f(S)$ from the barrier density.  We generalise the theoretical approach provided by excursion set theory from GR to MG.  

In \cref{sub:fitting_functions} we apply this method to a variety of empirical fitting functions.  These are derived from and calibrated using $N$-body simulations, and we discuss the additional complexities which this presents in MG.  In particular we focus on the various ways to account for environment dependence---a problem peculiar to chameleon MG theories---and in \cref{sub:including_denv_barrier_in_mg} we analyse the accuracy of both existing methods and the novel methods which we propose in this paper.

\subsection{Spherical collapse in MG} 
\label{sub:spherical_collapse_in_mg}

\begin{figure}
\captionsetup{skip=-1pt}
  \begin{center}
    \subfloat[{\lcdm{}, DGP1: $r_{c} = 1.2 c/H_{0}$, DGP2: $r_{c} = 5.6 c/H_{0}$, SymA: $a_{\mathrm{SSB}} = 0.5$, SymB: $a_{\mathrm{SSB}} = 0.33$}]{\label{subfig:deltac_lnM_all}\includegraphics[keepaspectratio,width=0.95\textwidth]{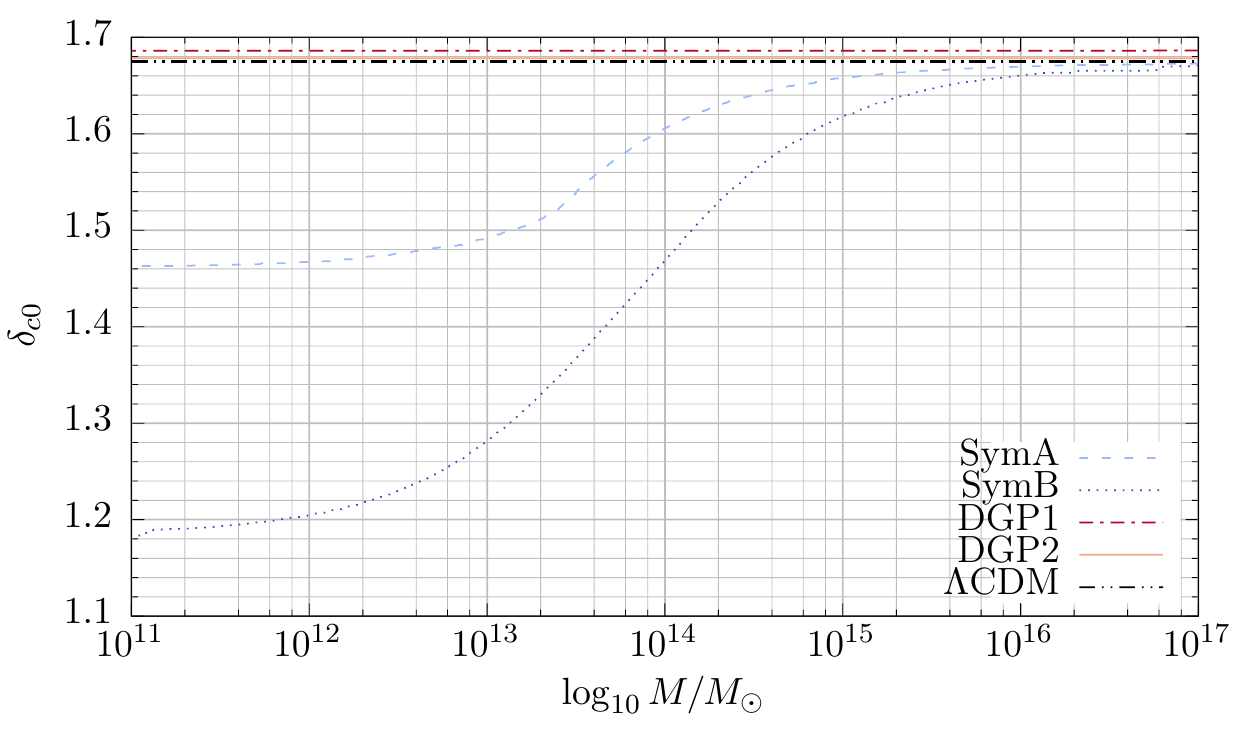}} \\
    \subfloat[{F6: $\fr[6]$}]{\label{subfig:deltac_lnM_fr6}\includegraphics[keepaspectratio,width=0.49\textwidth]{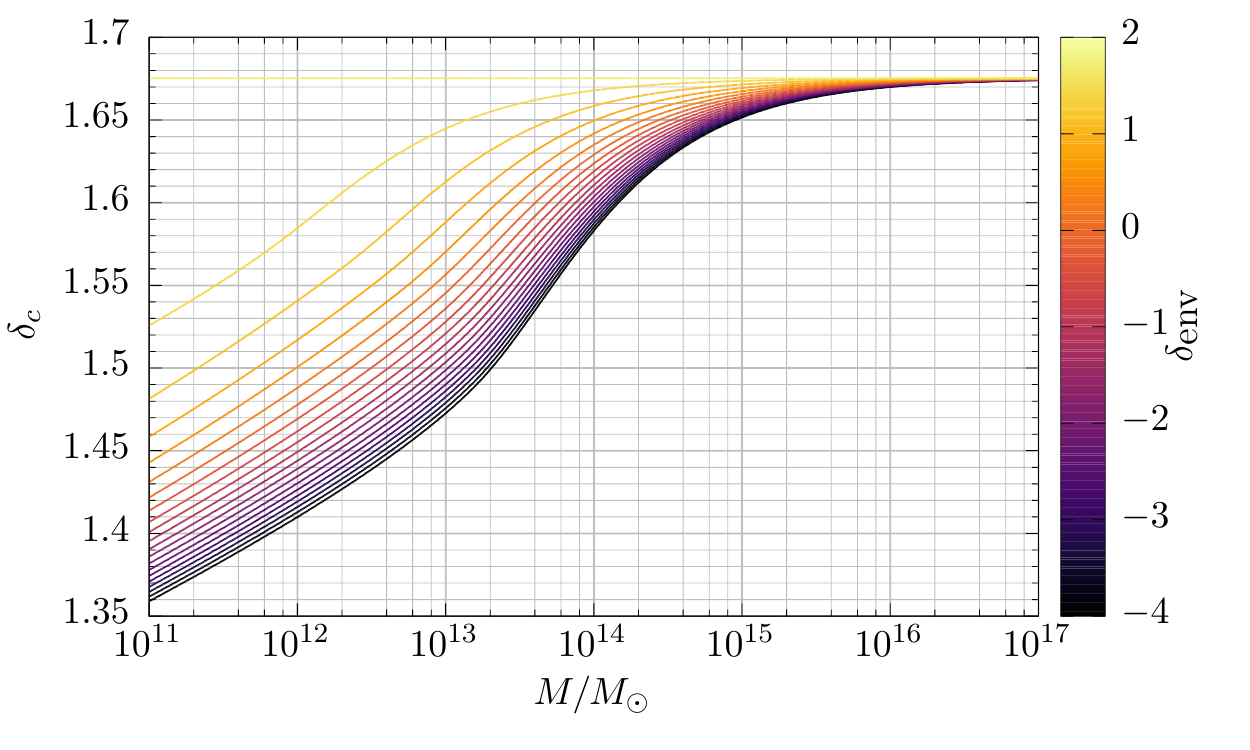}}
    \subfloat[{F6: $\fr[6]$}]{\label{subfig:deltac_denv_fr6}\includegraphics[keepaspectratio,width=0.49\textwidth]{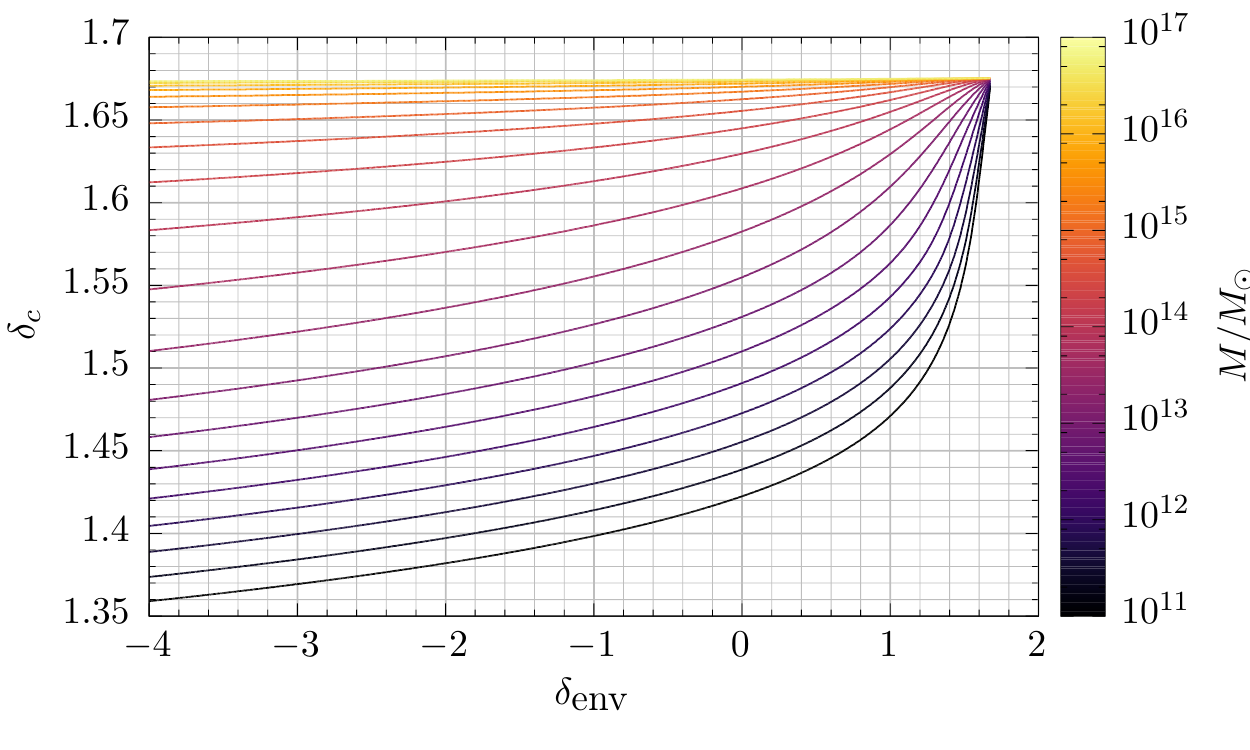}} \\
    \subfloat[{F5: $\fr[5]$}]{\label{subfig:deltac_lnM_fr5}\includegraphics[keepaspectratio,width=0.49\textwidth]{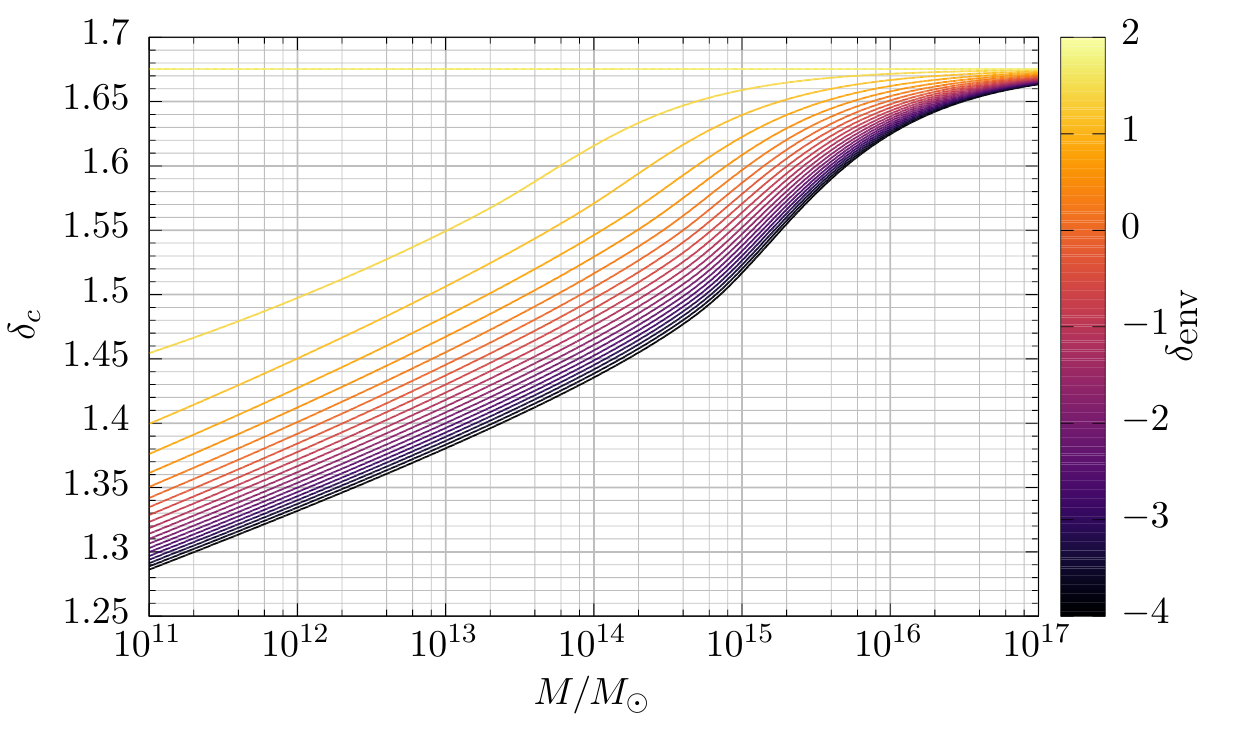}}
    \subfloat[{F5: $\fr[5]$}]{\label{subfig:deltac_denv_fr5}\includegraphics[keepaspectratio,width=0.49\textwidth]{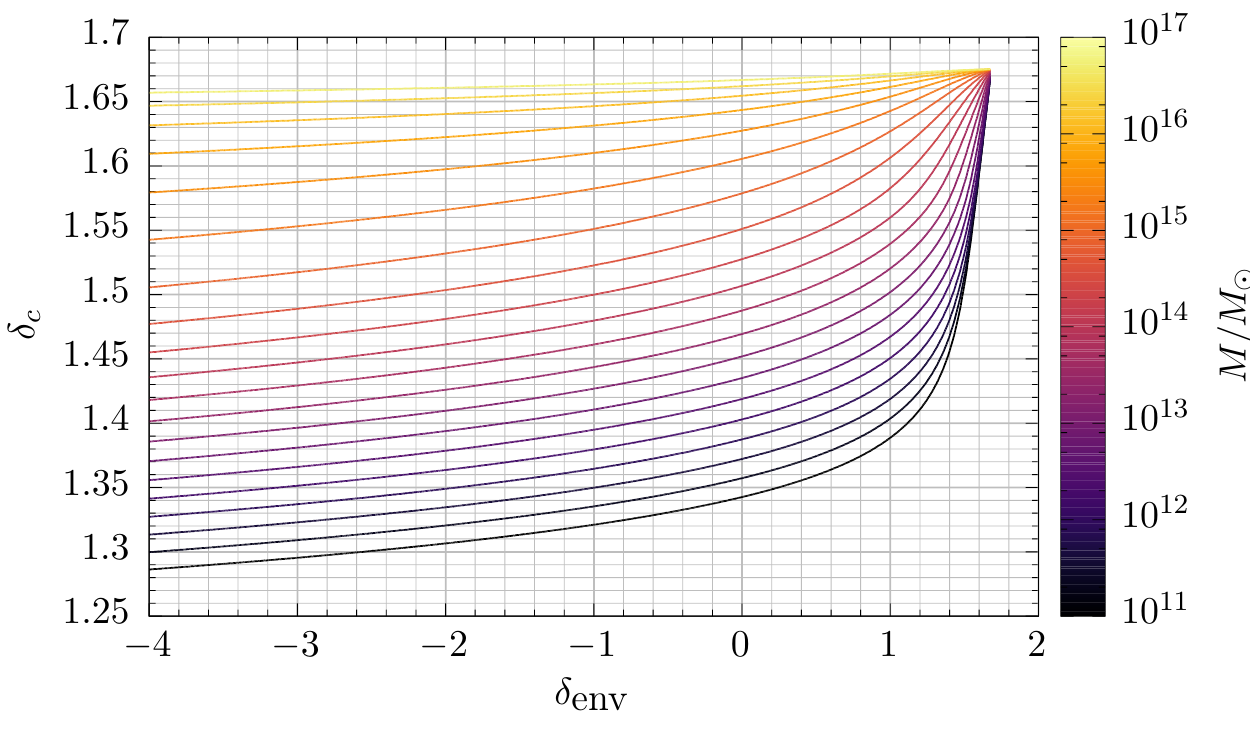}} \\
  \end{center}
  \caption{The barrier density for each of the MG models in this paper.  In \cref{subfig:deltac_lnM_all} the barrier is constant for \lcdm{} and only a function of mass for DGP and Symmetron models.  In \cref{subfig:deltac_denv_fr5,subfig:deltac_denv_fr6,subfig:deltac_lnM_fr5,subfig:deltac_lnM_fr6} we show cross-sections of the $\fr$ barrier \lq\lq{}surface\rq\rq{} at constant environment density (\cref{subfig:deltac_lnM_fr5,subfig:deltac_lnM_fr6}) and mass (\cref{subfig:deltac_denv_fr5,subfig:deltac_denv_fr6}).
  }
  \label{fig:deltac_all}
\end{figure}

\begin{figure}
  \begin{center}
    \includegraphics[keepaspectratio,width=0.75\textwidth]{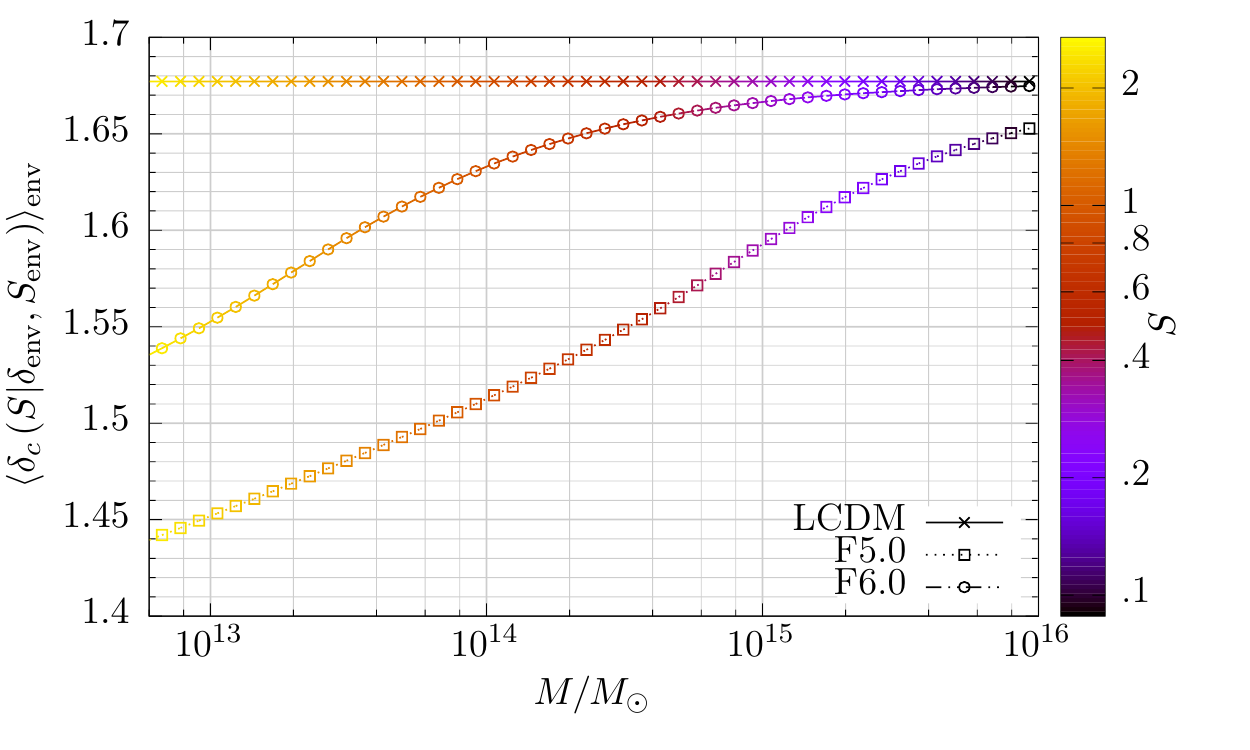}
  \end{center}
  \caption{The collapse density for \lcdm{}, $\fr[5]$ and $\fr[6]$ averaged over the Eulerian environment distribution \cref{eq:pdf_denv_E} converts the drifting-and-diffusing barrier density to a drifting $\delta_{c}(S)$. }
  \label{fig:plot_density_average}
\end{figure}

In this section we quantify how screening affects the critical density $\delta_c$ required for halo collapse in the excursion set formalism.  We present a formula for $\delta_c$ which is sufficiently general for all of our screened MG theories. The main difference in modified gravity is that $\delta_c$ depends both on mass and environment density as well as depending on redshift as in $\Lambda$CDM.  Throughout this paper we use the terms \lq\lq{}critical density\rq\rq{}, \lq\lq{}barrier density\rq\rq{} and \lq\lq{}drifting-and-diffusing barrier\rq\rq{} synonymously, to indicate $\delta_{c}(S, \Senv, \denv)$.  This is they key result of this section: the solution to \cref{eq:sph_collapse_all} (coupled to \cref{eq:sph_collapse_lcdm} if necessary), subject to the initial conditions \cref{eq:sph_collapse_linear,eq:mass_conservation} and the boundary value problem \cref{eq:delta_c_defn}.

The density field $\delta$ is the linearly-extrapolated over-density at the present epoch, convolved with a window function.\footnote{Originally \cite{1991ApJ...379..440B} denoted this by $F$ to avoid confusion with $\delta_{c}$ and to emphasise that it is not purely $\rho / \bar{\rho}$, the usual cosmological definition of over-density compared to the background density.  However, present convention dictates that $\delta$ be used instead.}  Assuming that the density field exhibits Gaussian fluctuations on all scales, we scale the smoothed density field in units of the variance $\delta = \nu\sigma(R)$ for any positive constant $\nu$ and $\sigma = \sqrt{S}$ defines the resolution $S$ over the scale $R$:
\begin{equation} \label{eq:resolution}
    S \equiv \sigma^{2} 
    = \frac{1}{\left( 2 \pi{} \right)^{3}} \int \dx<3>{k} \left< \abs{\delta \left( \vec{k}; R \right)} \right>
    = \frac{1}{2 \pi{}} \int_{0}^{\infty} \dx{\ln k} \; P(k,z) W(k;R)^{2} k^{3}
\end{equation}
where $P(k,z)$ is the power spectrum which completely describes the original Gaussian density field $\delta$ at redshift $z$.  

Now we require a condition for the redshift evolution of the density field: we choose a linear evolution $\delta(k,z;R) = D(z) \delta(k;R)$ where $D(z)$ is the growing mode normalised to unity at the present epoch.  Since all of our MG models have background evolution close to \lcdm{}, we utilise the usual growing mode \textsc{ode}:
\begin{subequations} \label{eq:d_linear}
\begin{align}  
    0 &= \frac{\dx<2>{D}}{\dx \ln a^{2}} 
    + \left(2 - \frac{3}{2} \Omega_{m}(\ln a) \right) \frac{\dx{D}}{\dx \ln a} 
    - \left( \frac{3}{2}  \Omega_{m}(\ln a) \right) D(\ln a) \\
    \Omega_{m}(\ln a) &= \frac {\Omega_{m0} \exp{(-3 \ln a)}} {\Omega_{m0} \exp{(-3 \ln a)} + \Omega_{\Lambda 0}} \\
    \Omega_{\Lambda}(\ln a) &= \frac {\Omega_{\Lambda 0}} {\Omega_{m0} \exp{(-3 \ln a)} + \Omega_{\Lambda 0}} 
\end{align}
\end{subequations}
Let us view the present-day $\delta$ as a fixed density field, with a critical over-density for collapse $\delta_{c0}$: where $\delta > \delta_{c0} / D(z)$, we expect these over-densities to have already collapsed at redshift $z$; conversely where  $\delta < \delta_{c0}$, these parts of the density field have yet to collapse.  Our aim is to calculate $\delta_{c0}$ for both GR and our screened MG theories.

The collapse of the environment surrounding the halo is equivalent to collapse in general relativity.  Let us assume that the initial over-density is a spherical top-hat in Eulerian space.  We can utilise the resulting axisymmetry to simplify the gravitational collapse equation to:
\begin{equation} \label{eq:sph_collapse_lcdm}
    0 = \frac{d^2{y}}{d \ln a^{2}}
    + \left(2 - \frac{3}{2} \Omega_{m}(a) \right) \frac{d{y}}{d \ln a}
    + \frac{1}{2}  \Omega_{m}(a) \left( \frac{1}{y^{3}} - 1 \right) y
\end{equation}
where $y(a)$ is the ratio of the physical radius of the halo $R_{\rm TH}(a)$ to the physical radius of the filter $a(t)R$ \cite{2012MNRAS.421.1431L}.   Now we have an expression for the halo density in general relativity and the environment density in modified gravity.

Since all of our MG theories obey a modified Poisson equation in the weak-field limit, we can use the standard equations for collapse of a spherical top-hat over-density.  We need only replace $G_{N}$ by $G_{N}(1 + \Feff)$ in the Poisson equation.  Following the same steps as for the environment collapse, we obtain:
\begin{equation} \label{eq:sph_collapse_all}
      0 = \frac{\dx<2> y}{\dx \ln a^{2}}
    + \left(2 - \frac{3}{2} \Omega_{m}(a) \right) \frac{\dx y}{\dx \ln a}
    + \frac{1}{2} \left( 1 + \Feff \right)  \Omega_{m}(a) \left( \frac{1}{y^{3}} - 1 \right) y 
\end{equation}
The enhancement factors  $\Feff$ are derived in \cref{app:screened_gravity_theories}.  Specifically, they are \cref{eq:feff_fr} for $\fr$, \cref{eq:symm_feff} for Symmetron and \cref{eq:dgp_feff} for DGP.  For Symmetron gravity $\Feff$ depends upon the halo over-density in \cref{eq:symm_feff}; while in \DGP{} models $\Feff$ depends upon the halo mass and radius in \cref{eq:dgp_feff}.  Therefore for Symmetron and Vainshtein screening we need only solve \cref{eq:sph_collapse_all}.  In contrast, the chameleon-screened $\fr$ theory with $\Feff$ in \cref{eq:feff_fr} depends both upon the halo density and the environment density: in this case \cref{eq:sph_collapse_all} (for the halo collapse) must be coupled to \cref{eq:sph_collapse_lcdm} (for the collapse of the surrounding environment).  This is the general form for modified gravity collapse of a spherical top-hat.

Finally we set the initial conditions.  Since $\Feff  \ll 1$ at early times, we can use the same initial conditions in MG and GR.  Mass conservation determines $y_{i}$ via equating the physical radii of the physical halo and the top-hat window function at the initial time $a_{i}$:
\begin{equation} \label{eq:mass_conservation}
   M = \frac{4}{3} \pi \rho_{m0} a_{i}^{3} R_{\mathrm{TH}}^{3} = \frac{4}{3} \pi \rho_{m0} \left( 1 + \delta_{i} \right)  r_{i}^{3}
   \implies y(a_{i}) = 1 - \frac{\delta(a_i)}{3}
\end{equation}
At early times we expect the ODE \cref{eq:sph_collapse_all,eq:sph_collapse_lcdm} to be well-approximated by its linearised equivalent.  Without loss of generality, we can set the initial redshift of the ODE (we used $z = 500$) to before the modification to GR, hence $F_{\mathrm {eff}} = 0$.  Substituting the result for $y_{i}$ gives the corresponding first derivative:
\begin{equation} \label{eq:sph_collapse_linear}
  \frac{\dx y(a_{i})}{\dx \ln a} = - \frac{\delta(a_{i})}{3}
\end{equation}
An important corollary of \cref{eq:mass_conservation} is that $\delta = y^{-3} - 1$, which contributes the nonlinearity in the final term of \cref{eq:sph_collapse_all,eq:sph_collapse_lcdm}.

Despite the fact that our (coupled) ODEs are in terms of $y(a)$, our aim is to calculate $\delta_{c}(z_{c})$.  We define this via the boundary-value problem \cite{Kopp:2013lea}:
\begin{equation} \label{eq:delta_c_defn}
  \delta_{c} \left( z_{c} \right) = \left. \delta_{i} \left( z_{i} \right) \frac{D \left( z_{c} \right)}{D \left( z_{i} \right)} \given y \left( z_{c} \right) = 0 \right.
\end{equation}
This condition specifies that the value of $\delta_{i}$ which causes $y_{c}$ to vanish at the collapse redshift $z_{c}$ is the initial over-density which can be linearly-extrapolated to find $\delta_{c}(z_{c})$.  We now have all the requirements to compute the present-day collapse density (in fact, the barrier density required for collapse at any redshift).

The collapse densities using our screening parameters are shown in \cref{fig:deltac_all}.  The DGP models produce a flat result with similar values to \lcdm{}.  This suggests that these results may be indistinguishable from a \lcdm{} model with non-standard cosmological parameters.  In contrast, the two Symmetron models produce a collapse density which is a function of mass. While it asymptotes  to \lcdm{} density at high masses, it is substantially lower for small masses and appears to tend to a constant value on sub-cluster scales. As expected, the MG model parameters are correlated with the collapse density divergence from \lcdm{}: DGP models recover the \lcdm{} collapse density as the cross-over radius tends to infinity, whereas symmetron models do so as the symmetry-breaking scale factor tends to unity. 

\begin{figure}
  \begin{center}
  \subfloat[{$\fr[5]$}]{\label{subfig:surf_deltac_linear_F5}\includegraphics[keepaspectratio,width=0.9\textwidth]{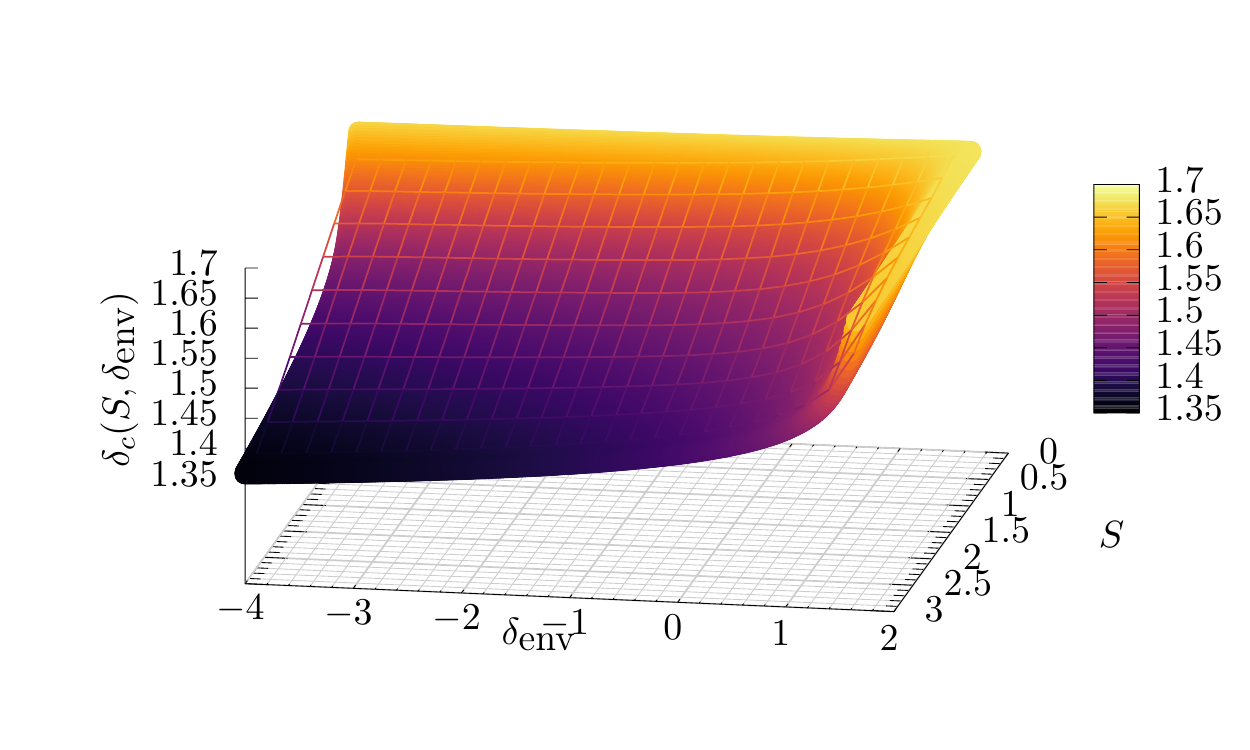}} \\[-1ex]
    \subfloat[{$\fr[6]$}]{\label{subfig:surf_deltac_linear_F6}\includegraphics[keepaspectratio,width=0.9\textwidth]{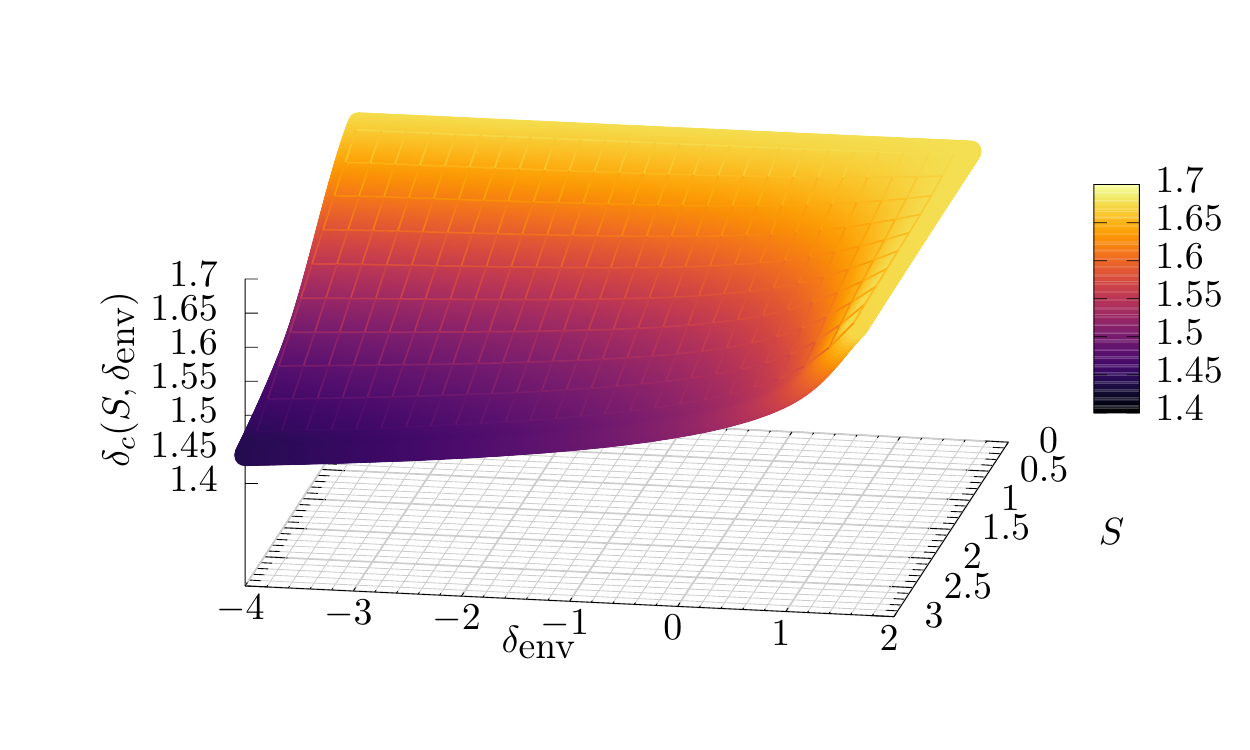}} 
  \end{center}
\caption{The full barrier (surface) and linear approximation (grid) for $\fr$ models with $\fr[5]$ and $\fr[6]$. Both the surface and the grid are coloured according to the value of $\delta_{c}$, so any areas in which the grid is visible indicates a discrepancy between the linear estimate and the true value.  This is particularly pronounced for $\fr[5]$ compared to $\fr[6]$. }
 \label{fig:surf_deltac_linear}
\end{figure}

The $\fr$ collapse density exhibits a more complex behaviour.  We shows its dependence on mass in \cref{subfig:deltac_lnM_fr6,subfig:deltac_lnM_fr5} and environment in \cref{subfig:deltac_denv_fr6,subfig:deltac_denv_fr5}.  The collapse density is no longer flat: $\delta_{c}(M, \denv)$ is a monotonically-increasing function of $M$, and the peak-background split $\delta_{c} - \denv$ is a monotonically-decreasing function of $\denv$.  Compared to $\Lambda$CDM, we expect haloes to form at higher masses and to have more in low-density regions.  \cref{subfig:deltac_denv_fr6,subfig:deltac_denv_fr5} demonstrate that the collapse density is always bounded from below by the environment density and from above by the \lcdm{} result.  We have marginalised over the environment distribution according to \cref{eq:pdf_denv_L,eq:pdf_denv_E}. The resulting environment-averaged collapse density is shown in \cref{fig:plot_density_average}.  It is also possible to linearise the collapse density in both $M$ (or $S$) and $\denv$, as shown in \cref{fig:surf_deltac_linear}.  We will use both of these approximations in \cref{sec:the_halo_mass_function}.  This completes our analysis of spherical collapse in MG.

\subsection{Excursion set theory} 
\label{sec:excursion_set_theory}

In this section we relate the halo mass function $n(M)$ to the \fcd{} $f(S)$ via excursion-set theory, including \lq\lq{}nuisance parameters\rq\rq{} other than the halo mass $M$.  We outline the excursion set formalism in general relativity and then discuss the modifications induced by modified gravity.  

The \textit{Ansatz} of the excursion set formalism is the relation of the collapse of the over-density to a halo of mass $M$ in real space to the absorption of trajectories in the density-resolution space by a barrier density at resolution $S$. As shown in \cite{1991ApJ...379..440B,1993MNRAS.262..627L}, the fraction of random walks in the plane $(\delta,S)$ that are absorbed by the collapse over-density $\delta_c$ at resolutions earlier than $S$ is equivalent to the cumulative fraction $F(>M)$ of mass contained in haloes above mass $M$.

The ingredients of the excursion set formalism are:
\begin{enumerate}
  \item The window function $W(kR)$ used to smooth the density field.  We utilise a Gaussian filter and a top-hat filter in both real and Fourier space.
  \item The over-density field $\delta$ corresponding to the fractional, linearly-evolved over-density smoothed over a scale $R$ defined by the aforementioned window function.
  \item The resolution $S$, which is related to the linear matter power spectrum $P(k)$ and the halo mass $M$.
  \item The collapse density $\delta_{c}$ which acts as a drifting-and-diffusing barrier in the excursion-set parameter space.
\end{enumerate}

Following \cite{1991ApJ...379..440B}, we obtain a diffusion equation for the probability density function $\Pi(S,\delta)$ that a Markovian trajectory which moves randomly in the linearly-extrapolated density field $\delta$ and moving linearly forwards in $S$ from the origin will first exceed the barrier density $\delta_{c}(S)$ at resolution $S$.  In GR the barrier is flat, so we obtain an analytical solution for $\Pi$ provided that we use a top-hat filter in Fourier space for $S$.  This is given by the Press-Schechter mass function \cite{1974ApJ...187..425P}:
\begin{equation} \label{eq:ps_erfc_integral}
    F(>M) 
    = 2 \int_{\delta_{c}}^{\infty} \dx{} \delta \; \Pi (\delta(S)) 
    = \int_{0}^{S} \dx{} S^{\prime} f (S^{\prime})
    = \textrm{erfc} \left(\frac{\nu_h}{\sqrt{2}} \right)
\end{equation}
where $\nu_h\equiv\delta_c/\sqrt{S}$.  (A more detailed derivation of this result is in \cite{1991ApJ...379..440B}.)

However, our aim is to express this solely in terms of $S$---the resulting function $f(S)$ is known as the \emph{\fcd{}}---and as the name suggests is the probability of first up-crossing the barrier density at $S$.  

By assuming that trajectories in $(\delta , S)$ are uncorrelated, the diffusion equation admits the solution \cite{2011PhRvD..83f3511P}:
\begin{subequations} \label{eq:volterra_all}
\begin{align}  
    f(S|\Senv , \denv) 
    &= g(S) + \int_{\Senv}^{S} \! \dx x \; k(S,x) \, f \left( x \given \Senv , \denv \right) \label{eq:volterra_fcd} \\
    k(S,x) &= \left[ \frac{\delta_{c}(S) - \delta_{c}(x)}{S - x} - 2\frac{\dx \delta_{c} (S)}{\dx S} \right]
    \frac{1}{\sqrt{2\pi (S-x)}} \exp\left\{ -\frac{(\delta_{c}(S) - \delta_{c}(x))^{2}}{2 (S-x)} \right\} \\
    g(S) & = \left[ \frac{\delta_{c}(S) - \denv }{S - \Senv } - 2\frac{\dx \delta_{c} (S)}{\dx S} \right] 
    \frac{1}{\sqrt{2\pi (S - \Senv)}} \exp \left\{ -\frac{(\delta_{c}(S) - \denv)^{2}}{2(S - \Senv)} \right\}
\end{align}
\end{subequations}
This is valid for theories with a non-flat barrier density $\delta_{c}(S,\Senv,\denv)$ which depends upon the starting point of the random walk $(\Senv,\denv)$ and the variance $S$ at which the random walk crosses the barrier.

It is straightforward to show that this has an analytical solution for a linear barrier density \cite{Zhang:2005ar,2011PhRvD..83f3511P}, including the constant case $\delta_{c0}^{\Lambda} \approx 1.676$ which is the \lcdm{} solution:
\begin{equation} \label{eq:ps_conditional}
  f(S|\Senv , \denv) = \frac{\delta_{c}^{\Lambda} - \denv}{\sqrt{2 \pi (S - \Senv)^3}} \exp \left[ - \frac{1}{2} \frac{ \left( \delta_{c}^{\Lambda} - \denv \right)^{2} }{S - \Senv} \right]
\end{equation}
We shall refer to this as the (environment-)conditional HMF from now on.  This generalises directly from GR to MG: we merely replace the \lcdm{} value $\delta_{c}^{\Lambda}$ with the appropriate density from \cref{fig:deltac_all}.  This holds for flat barriers, \lq\lq{}drifting\rq\rq{} barriers $\delta_{c}(S)$ and the full \lq\lq{}drifting-and-diffusing\rq\rq{} barrier $\delta_{c}(S,\Senv,\denv)$.

Finally we must marginalise over the environment.  We require a probability distribution for $\denv$, described in \cref{sec:choice_of_environment_density_function}.  Then the unconditional mass functions is simply the conditional one marginalised over the nuisance parameter \cite{2012MNRAS.421.1431L,2012MNRAS.425..730L}:
\begin{equation} \label{eq:fcd_ave_integral}
    f \left( S \right)
    = \langle f \left( S \given \Senv , \denv \right) \rangle_{\mathrm{env}} = \int_{-\infty}^{\delta_{\Lambda}} \! \dx{\denv} \;
    p \left( \denv \given \Senv \right) \; f \left( S \given S_{\mathrm{env}} , \delta_{\mathrm{env}} \right)
\end{equation}
In \lcdm{} we obtain an analytic solution which is precisely the \PS{} function:
\begin{equation} \label{eq:ps_unconditional}
    f \left( S \right)
    = \frac{\delta_{c}^{\Lambda} }{\sqrt{2 \pi S^3}} \exp \left[ - \frac{1}{2} \frac{ \left( \delta_{c}^{\Lambda} \right)^{2} }{S} \right]
\end{equation}
The unconditional HMF \cref{eq:ps_unconditional} has the same functional form as the conditional one \cref{eq:ps_conditional}, albeit with the substitutions $\delta_{c} \rightarrow (\delta_{c} - \denv)$ and $S \rightarrow S - \Senv$.  In \cref{sec:the_halo_mass_function} we shall assume that this applies to other fitting functions beyond \PS{}.

So far we have neglected the additional complications that the resolution $S$ is sensitive to the MG parameters (\eg{} $\fr$) via the power spectrum and that $S(R)$ should be based upon a scale $R$ which is Eulerian rather than Lagrangian.  

We have chosen to use the linear $P(k)$ computed in the $\Lambda$CDM model in \cref{eq:resolution} when computing the $f(\R)$ gravity predictions.  This is in agreement with \cite{2016JCAP...12..024C,2017JCAP...03..012V}.  \cref{sec:justification_for_using_the_lcdm_power_spectrum} summarises our reasoning.

Throughout this paper we set $\Senv = S(10\; \mathrm{Mpc}/h) \approx 0.1$, \ie{} an environment radius of $\approx 14 \, \mathrm{Mpc}$.  This is the Eulerian size of the environment: see \cref{sub:the_scale_of_the_environment} for details of the Eulerian-Lagrangian distinction.  Using a Gaussian window function limits the maximum halo mass to $\lesssim 10^{16} M_{\odot}$.  We confirmed that this is above the maximum halo mass found in our $N$-body simulations, so that we do not exclude any data by imposing the excursion-set condition $S < \Senv$.  


\subsection{Generalising HMF fitting functions} 
\label{sub:fitting_functions}

This section summarises the various halo mass functions used in this paper.  In GR a variety of empirical fits to {\it N}-body simulations have been proposed in a \lq\lq{}universal\rq\rq{} form $f(\nu)$.  The aim of these fits is to find a fitting function over a broad range of masses which is independent of redshift and cosmology, and the extent to which each fitting function exhibits this universality is a controversial one even in GR.  Nonetheless, we show how to generalise the HMFs proposed in GR to three types of screened MG.  Furthermore, we show a variety of methods to include environment dependence in the HMF, as required by chameleon-screened MG.  A summary of the functions used in this paper is in \cref{tab:hmf_functions}.

Two of the fitting functions used here (\textit{viz.} Press-Schechter and Sheth-Mo-Tormen) can be derived analytically.  The Press-Schechter function results from assuming spherical symmetry and a flat (scale-independent) barrier density. That Press-Schechter can be derived analytically in screened MG (\eg{} for a chameleon model \cite{2012MNRAS.421.1431L}) via a simple change of variables is the very reason that we can extend the other mass functions in this paper to MG as well.  The Sheth-Mo-Tormen function in \cite{2001MNRAS.323....1S,Sheth:2001dp} was originally a fit to data with the substitution $\delta_{c} \rightarrow \sqrt{a} \delta_{c}$ made to produce a better fit than \PS{}.  Later \cite{Maggiore:2009rx} showed that this is equivalent to using excursion set theory with the moving barriers which result from ellipsoidal collapse.  To our knowledge, no-one has shown that the assumptions used to construct the Sheth-Tormen function in GR also hold in MG.  Nonetheless, the analyticity in GR makes these appealing functions for use in MG.

The remaining fitting functions are purely empirical fits, derived from \lcdm{} N-body simulations.  
Having extracted the discrete approximation to the HMF $n(M)$, this can be converted to a discrete \fcd{} and a continuous \lq\lq{}best-fit\rq\rq{} approximation found---either in terms of $\sigma$ or in terms of $\nu$---which holds for a given redshift (range), mass range and cosmology (or family of cosmologies).  The resulting HMF also depends upon the halo finder used to extract the halo masses, as well as any subsequent calibration or corrections.  These factors all contribute to the final parameter values adopted by a particular function.  This is particularly notable for the Sheth-Mo-Tormen fit, for which several authors\footnote{This is why Jenkins appears twice in \cref{tab:hmf_functions}: one instance is their calibration for the SMT function (as with Courtin) and the other is their new fitting function.} \cite{2001MNRAS.323....1S,2001MNRAS.321..372J,2011MNRAS.410.1911C,2006ApJ...646..881W} have proposed their own \lq\lq{}improved\rq\rq{} values for the best-fit parameters of this function.  This emphasises that while the same functional form can provide a good fit to data in different background cosmologies, the same $N$-body data and the same fitting function will produce different best-fit parameters and varying degrees of invariance depending upon the halo extraction tehcniques and the theoretical assumptions of the authors.  For this reason, we have separated the HMFs into families which share the same functional form.  We summarise the HMFs $f(\nu)$ in \cref{tab:hmf_functions}. 

There are four steps involved in generalising the \lcdm{} HMFs to MG:
\begin{enumerate}
\item \label{item:hmf_deltac} Calculating the appropriate barrier density;
\item \label{item:hmf_pk} Selecting the appropriate linearised power spectrum;
\item \label{item:hmf_sigma_to_nu} Converting from $\sigma$ (or $S$) to $\nu$ as the dependent variable;
\item \label{item:hmf_params} Rescaling the free parameters accordingly.
\end{enumerate}

We discussed \cref{item:hmf_deltac} in \cref{sub:spherical_collapse_in_mg}.  The barrier densities for the models in this paper are in \cref{fig:deltac_all}.

In \cref{item:hmf_pk} we must decide whether to keep the \lcdm{} power spectrum or to adjust it according to the MG theory.  We keep the linear $P(k)$ from \lcdm{}, in agreement with \cite{2017JCAP...03..012V,2016JCAP...12..024C}.  We discuss the reasons for this in \cref{sec:justification_for_using_the_lcdm_power_spectrum}.

\cref{item:hmf_sigma_to_nu} is concerned with making explicit the dependence on the collapse density.  In GR, there are two opinions on the \lq\lq{}correct\rq\rq{} independent variable for the HMF.   Those arguing for $\nu$ (\textit{inter alia} \cite{Sheth:2001dp,2001MNRAS.323....1S,Peacock2007,Zentner:2006vw}) assert that according to excursion set theory the (albeit weak) cosmological and redshift dependence of both $\delta_{c}$ and $\sigma$ cancel when the \fcd{} is expressed in terms of $\nu \equiv \delta_{c} / \sigma$ in GR cosmologies.  Those arguing for $\sigma$ (or $S$, or $\sigma^{-1}$, or $\ln \sigma$) (\textit{inter alia} \cite{2001MNRAS.321..372J,2007MNRAS.374....2R,2003MNRAS.346..565R,2008ApJ...688..709T,2010MNRAS.403.1353C,2006ApJ...646..881W}) assert that $\delta_{c}$ is a sufficiently weak function of $\Omega_{m0}$ and $z$ that this dependence can be ignored; in fact \cite{2001MNRAS.321..372J} go so far as to say that \lq\lq{}taking $\delta_{c} = 1.686$ in all cosmologies leads to excellent agreement with our numerical data if halos are defined at fixed over-density\rq\rq{}.  In GR cosmologies there is an  invertible mapping between these two options because $\delta_{c}$ is assumed to be a constant.  

This is clearly not the case in our extended gravity theories.  Again, there is more than one opinion on this topic.  Do we change only the resolution $S(M)$ \cite{2011PhRvD..84h4033L} or only the barrier density $\delta_{c}$ \cite{2017JCAP...03..012V} or both \cite{Lombriser:2013wta}, or do we need to evolve the full scalar field equations \cite{Kopp:2013lea}?  The only choice which fulfills the three objectives:
\begin{enumerate}
  \item A consistent model applicable in all our MG theories
  \item Incorporating the mass- and $\denv$-dependence for $\delta_{c}$ in $\fr$
  \item Our choice of the \lcdm{} $P(k)$ to calculate $S(M)$, so that we have the same map between mass and resolution for all haloes
\end{enumerate}
is to use the universal parameter $\nu$ and change only the barrier density from $\delta_{c}^{\Lambda}$ to the full barrier density appropriate for a given MG theory. 

\cref{item:hmf_params} is the consequence of \cref{item:hmf_sigma_to_nu}.  All the free parameters in the HMFs\footnote{The \lq\lq{}Reed 2007\rq\rq{} fit is the 2003-like fit from \cite{2007MNRAS.374....2R}, rather than the one with $n_{\mathrm{eff}}(\sigma)$ dependence, which creates ambiguity about which terms to convert to $\nu$.  In the fit we use the $\sigma$-dependence is clearly only caused by treating $\delta_{c}$ as a constant.} need to be converted from $\ln \left( \sigma^{-1} \right) $ to $\nu$, by absorbing factors of $\delta_{c}$. This requires paying particular attention to whether the SCDM or \lcdm{} collapse density is used in the original papers, as this has implicitly been absorbed into the best-fit parameters.\footnote{If we were concerned with redshift evolution, we would have to decide whether to absorb $\delta_{c0}$ or $\delta_{c}(z)$.  Some authors fix $\delta_{c}$ and allow the free parameters to vary with redshift; others do the opposite.}

This is sufficient for MG theories which only have a drifting barrier $\delta_{c}(S)$, but not for the drifting-and-diffusing barrier $\delta_{c}(S,\denv,\Senv)$.  We treat this additional generalisation in the next few paragraphs.

The drifting-and-diffusing barrier can be accounted for using a wide variety of methods:
\begin{enumerate}
\item \label{item:scaling_Volterra} Scaling using the Volterra integral solution in \cref{eq:volterra_fcd}
\item \label{item:cosmic_web} Averaging over the cosmic web (described in \cite{2017JCAP...03..012V})
\item \label{item:deltac_ave} Calculating a $\denv$-averaged collapse density, \ie{} converting to a drifting barrier
\item \label{item:barrier_approx} A flat or linear-barrier approximation to the full excursion-set problem
\end{enumerate}
We address each of these in turn.

Technique~\ref{item:scaling_Volterra} utilises the fact that we already know how to account for the extra barrier complexity in the excursion set approach: whereas the flat barrier produces the \PS{} distribution, the drifting-and-diffusing barrier leads to the solution \cref{eq:volterra_fcd}.  So we may calculate the unconditional HMF, i.e. assuming $(\denv = 0, \, \Senv = 0)$, then accounting for the effects of the drifting and diffusing barrier using excursion set theory.  This results in a rescaling of the unconditional HMF:
\begin{equation} \label{eq:scaling_Volterra}
f(S)_{\mathrm{MG}} = f(S | \Senv = 0, \denv = 0 )_{\mathrm{MG}} \quad \frac{\text{Volterra solution}}{\text{Press-Schechter}}
\end{equation}
This is efficient, as we only need to solve the Volterra equation once for each drifting-and-diffusing barrier density, rather than re-calculating for each HMF as well.

Technique~\ref{item:cosmic_web} uses the $\nueff$ prescription of \cite{2017JCAP...03..012V}.  This is substituted directly into the HMF:
\begin{equation}
    \nu_{\mathrm{eff}} = \mathrm{max} \left\{ 0, \frac{ \nu_{h} \left( S, \denv \right) - \epsilon^{2} \left( S, \Senv \right) \nu_{\text{env}}
    \left( \Senv, \denv \right) }{ \sqrt{ 1 - \epsilon^{2} \left( S, \Senv \right) }} \right\}
\end{equation}
Since we are not interested in restricting ourselves to a particular environment, we can simplify the relevant equations in \cite{2017JCAP...03..012V} to:
\begin{equation} \label{eq:cosmic_web}
    f \left( S \right) = 
    \int_{-\infty}^{\infty} \dx \nu_{\text{env}} \;  f \left( \nu_{\mathrm{eff}} \right)
    \int_{0}^{\infty} \dx \rho \int_{-\rho}^{\rho} \dx \theta \; p(\rho, \theta, \nu_{\text{env}})
\end{equation}
We have rearranged the order of integration to highlight that the conditional mass function $f \left( \nu_{\mathrm{eff}} \right)$ is independent of $\rho$ and $\theta$.

Technique~\ref{item:deltac_ave} approximates the drifting-and-diffusing barrier by a drifting-only barrier.  Effectively we integrate over the environment density at the stage of calculating the collapse density:
\begin{equation} \label{eq:deltac_ave}
\nueff = \frac{\langle \delta_{c} \rangle_{\mathrm{env}} }{S}
\quad \text{where} \quad
\langle \delta_{c} \rangle_{\mathrm{env}} = 
\int_{-\infty}^{\delta_{\Lambda}} \! \dx{\denv} \; 
    p \left( \denv \given \Senv \right) \; \delta_{c} \left( S \given S_{\mathrm{env}} , \delta_{\mathrm{env}} \right)
\end{equation}
This effective-$\nu$ is substituted directly into the unconditional HMF.  In contrast to the full barrier solution---where the random walk must up-cross $\delta_c$ at $S$ having started at $(\denv,\Senv)$---here there is no accounting for the environment-dependent absorption of the Markovian trajectories in $(\delta , S)$ caused by the drifting-and-diffusing barrier.

Technique~\ref{item:barrier_approx} approximates the full solution to the Volterra equation in \cref{eq:volterra_fcd} by a linear barrier.  This is motivated by the fact that the Volterra equation with a linear barrier $\delta_{c}(S) = \omega - \beta S $ is calculable analytically \cite{2011PhRvD..83f3511P}:
\begin{equation} \label{eq:flat_barrier}
  f \left( S, \delta_{c}(S) \given \Senv, \denv \right) = \frac{ \delta_{c} - \denv }{ \sqrt{2 \pi \left( S - \Senv \right)^{3}} } \exp \left[ -\frac{1}{2} \frac{\left( \delta_{c} - \denv - \beta \left( S - \Senv \right) \right)^{2} }{S - \Senv} \right]
\end{equation}
so we can substitute the effective arguments into the fitting function, in the same way as for the cosmic web.  However, \cref{fig:surf_deltac_linear} shows that the $\fr$ barriers are not linear in $S$ (except when $\denv \rightarrow \delta_{c}^{\Lambda}$).  The linear approximation is always an overestimate and is particularly poor at approximating the sharp rise at $\denv \approx \delta_{c}^{\Lambda} $.  Therefore we discard this approach due to its poor approximation.

A \textit{caveat} for all of these methods is that there is no way to eliminate dependence on $\Senv$.  The excursion set condition $\Senv < S$ prevents us from marginalising directly.  The environment distributions (whether cosmic web or the PDFs for the environment over-density) in the limit $\Senv \rightarrow 0$ do not have a finite limit.  Instead we take a sufficiently large environment radius (\eg{} $R_{\text{env}} = 10,\, 20 \text{Mpc}/h$) that $S > \Senv$ is always obeyed (but not too large, otherwise this is no longer a sufficiently local description of the halo surrounds \cite{2012MNRAS.425..730L}).

The main result of \cref{sub:including_denv_barrier_in_mg} is to compare these methods for generating the conditional HMFs for $\fr[5]$ and $\fr[6]$.  For the other MG models, the barriers do not have an environment-dependent component, so their calculation is straightforward.

The process described in this section updates the $\lcdm$-calibrated fits to a format which is compatible both with GR and MG.  Our fitting functions are shown in \cref{tab:hmf_functions}.  The independent parameter is $\nu$ and the other variables are free parameters.  It is these free parameters which we vary, in order to optimise the fit between the fitting functions and the HMF derived from $N$-body simulations.  In \cref{sub:assuming_concordance_cosmology} we assume \lcdm{} values for $\nu$ in our fitting functions, whereas in \cref{sub:recalibrating_best_fit_parameters} we use the full MG values of $\nu$.

\setlength{\aboverulesep}{0.3ex} 
\setlength{\belowrulesep}{0.3ex} 
\addtocounter{table}{-1} 

\begin{figure}
\begin{longtable}{l *{4}{m{1.4cm}} l}
\toprule 
	\multirow{2}{*}{ Cosmology }
	& 
	\multicolumn{4}{c}{ Parameters } & \multirow{2}{*}{ HMF paper(s)  } \\ 
\cmidrule{2-5}
	& $\Omega_{m0}$ & $\Omega_{\Lambda 0}$ & $\sigma_{8}$ & $h$ & 
	\\
\midrule
\endhead 
\bottomrule \\[-.1in]
\endfoot 
	\multirow{16}{*}{\lcdm{}} 
	& 0.269 & 0.731 & 0.80 & 0.704 & This paper \\
\cmidrule{2-6}
	& 0.29 & 0.71 & 0.90 & 0.72 & WMAP1 (Courtin) \\
	& 0.30 & 0.70 & 0.90 & 0.70 & WMAP1 (Sheth-Tormen, Jenkins, Tinker, Warren) \\ 
\cmidrule{2-6}
	& 0.24 & 0.76 & 0.74 & 0.73 & WMAP3 (Courtin, Reed) \\
	& 0.24 & 0.76 & 0.75 & 0.73 & WMAP3 (Tinker) \\
	& 0.24 & 0.76 & 0.8 & 0.73 & WMAP3 (Tinker) \\
\cmidrule{2-6}
	& 0.26 & 0.74 & 0.79 & 0.72 & WMAP5 (Courtin) \\
	& 0.27 & 0.73 & 0.80 & 0.70 & WMAP5 (Watson) \\
	& 0.28 & 0.72 & 0.80 & 0.70 & WMAP5 (Watson) \\
\cmidrule{2-6}
	& 0.27 & 0.73 & 0.90 & 0.70 & \multirow{5}{*}{Tinker}  \\
	& 0.27 & 0.73 & 0.79 & 0.70 &  \\
	& 0.26 & 0.74 & 0.75 & 0.71 &  \\
	& 0.23 & 0.77 & 0:75 & 0:73 &  \\  
	& 0.20 & 0.80 & 0:90 & 0:70 &  \\  
\cmidrule{2-6}
	& 0.25 & 0.75 & 0.80 & 0.70 & Crocce \\
\cmidrule{2-6}
	& 0.25 & 0.75 & 0.80 & 0.73 & Peacock \\
\cmidrule{2-6}
	& 0.25 & 0.75 & 0.90 & 0.73 & Angulo, Reed \\
\midrule
	& 1.0 & 0.0 & 0.60 & 0.50 & Sheth-Tormen \\
\cmidrule{2-6}
	SCDM 
	& 1.0 & 0.0 & 0.51 & 0.50 & Jenkins \\
\cmidrule{2-6}
	& 1.0 & 0.0 & 0.79 & 0.72 & Courtin  \\
\midrule
	OCDM
	& 0.30 & 0.0 & 0.85 & 0.70 & Sheth-Tormen, Jenkins \\
\midrule
\multirow{2}{*}{$\tau$CDM} & 1.0 & 0.0 & 0.51 & 0.5 & Jenkins \\ 
\cmidrule{2-6}
	& 1.0 & 0.0 & 0.60 & 0.50 & Sheth-Tormen, Jenkins \\
\midrule
L-\lcdm{} & 0.10 & 0.90 & 0.79 & 0.72 & Courtin  \\ 
\midrule
LRP-CDM & 0.26 & 0.74 & 0.79 & 0.72 & Courtin  \\ 
\bottomrule \\[-.1in] 
\end{longtable}
\captionof{table}{\label{tab:cosmologies} Cosmological parameters used to derive each of the HMF fitting functions in GR. }
\end{figure}


\begin{landscape}
\begin{longtable}{m{8.25cm} m{4cm} l l}
\toprule 
	Fitting function $f(\nu)$ 
	& 
	Mass calibration
	&
	Cosmology
	&
	Reference
	\\
\midrule
\endhead 
\bottomrule \\[-.1in]
\caption{\label{tab:hmf_functions} Details of the HMF fitting functions used in this paper.  The various \lq\lq{}$n$CDM\rq\rq{} cold dark matter cosmologies are described in \cref{tab:cosmologies}.  For some fits, we have rewritten the function in terms of $\nu$ by substituting for $\sigma$ and absorbing factors of $\delta_{c}$ into the original free parameters. }
\endfoot 
	$\displaystyle	\sqrt{\frac{2}{\pi}} \nu \exp\left[-\frac{ \nu^{2} }{2} \right]$
	&
	--
	&
	EdS
	&
	Press-Schechter \cite{1974ApJ...187..425P} \\ 
\midrule
	\multirow{3}{6.25cm}{$\displaystyle
		A\sqrt{\frac{2a}{\pi}}\left[1+\left( a\nu^{2} \right)^{-p} \right] \nu \exp\left[-\frac{a\nu^{2}}{2}  \right]
	$} 
	&
	$ \nu^{2} \in [0.5, 10] $
	&
	\lcdm{} SCDM OCDM
	&
	Sheth-Tormen \cite{Sheth:2001dp} \\
\cmidrule{2-4}
	&
	$ \ln \sigma \in [ -0.7 , 0.8 ] $
	&
	\lcdm{}, OCDM, SCDM, $\tau$CDM
	&
	Jenkins \cite{2001MNRAS.321..372J}  \\
\cmidrule{2-4}
	&
	$ \ln \sigma \in [ -0.7 , 0.8 ] $
	&
	\lcdm{} 
	SCDM
	L-\lcdm{}
	LRP-CDM
	&
	Courtin \cite{2011MNRAS.410.1911C}  \\
\midrule
	\multirow{2}{6.25cm}{$\displaystyle
		A \left[ \left( \frac{\nu}{\delta_{c}^{\Lambda}} \right)^{a} + b \right] \exp\left[ - c\nu^{2} \right]
	$} 
	&
	$ M \in [ 10^{10} , 10^{15} ] M_{\odot} $
	&
	\lcdm{}
	&
	Warren \cite{2006ApJ...646..881W} \\
\cmidrule{2-4}
	&
	$ M \in [ 10^{10.5} , 10^{15.5} ] M_{\odot} $
	&
	\lcdm{}
	&
	Crocce \cite{2010MNRAS.403.1353C} \\
\midrule
	$ \displaystyle 
		A \exp \left[ - \abs{\ln \nu + b}^{a} \right]
	$ 
	&
	$ \ln \sigma \in [ -1.05 , 1.2 ] $
	&
	\lcdm{} OCDM SCDM $\tau$CDM
	&
	Jenkins \cite{2001MNRAS.321..372J} \\
\midrule
	$ \displaystyle 
		\frac{ \nu^{2} \exp(-c \nu^{2})}{ \left( 1 + a \nu^{b} \right)^{2}} \! 
		\left[ a b \nu^{b-1} \!+ c (1 + a \nu^{b}) \right]
	$  
	&
	$ M \in [ 10^{10} , 10^{15} ] M_{\odot} $
	&
	\lcdm{} 
	&
	Peacock \cite{Peacock2007} \\
\midrule
$\displaystyle
	A\sqrt{\frac{2a}{\pi}} \nu \exp\left[-\frac{ca \nu^{2} }{2} \right] \newline
	\left[ 1+ \left( a\nu^{2} \right)^{-p} + Q \exp \left[ - \frac{1}{2} \left( \frac{ \ln \nu - q }{0.6} \right)^{2} \right] \right] 
$ &
	$ \ln \sigma \in [ -0.9 , 1.7 ] $
	&
	\lcdm{}
& Reed 2007 \cite{2007MNRAS.374....2R} \\    
\midrule
\multirow{3}{6cm}{$ \displaystyle \newline\newline
	A \left[ \left( b \nu \right)^{a} + 1 \right] \exp \left(- c \nu^2 \right)
$} & 
	$\sigma \in [ 10^{-0.4} , 10^{0.6} ] $
	&
	\lcdm{} 
	&
	Tinker 2008 \cite{2008ApJ...688..709T} \\
\cmidrule{2-4}
	&
      $ M \in [ 10^{8} , 10^{16} ] M_{\odot} $
	&
	\lcdm{} 
	&
	Angulo \cite{1203.3216} \\
\cmidrule{2-4}
	&
      $ \ln \sigma \in [ - 1.31 , 0.55 ] $
	&
	\lcdm{} 
	&
	Watson \cite{2013MNRAS.433.1230W} \\
\end{longtable}
\end{landscape}

\clearpage

\section{Data processing} 
\label{sec:algorithms}
 
This section summarises the N-body simulations and the process of extracting the halo catalogues which form our data.  This includes our choice of halo finder, corrections and mass cuts to the binned halo counts and quantification of the uncertainties in the resulting discrete HMF.
 
\subsection{$N$-body simulations} 
\label{sub:n_body_simulations_haloes}
 
The {\it N}-body simulations were run using the \texttt{ISIS} and \texttt{ECOSMOG} code \cite{2014A&A...562A..78L, 2012JCAP...01..051L}, which is a modified gravity modification of the high-resolution {\it N}-body code \texttt{RAMSES} \cite{2002A&A...385..337T}. For details about the implementation and for a comparison of these codes see the modified gravity {\it N}-body code comparison project \cite{2015MNRAS.454.4208W}.
 
We ran two sets of simulations with different mass resolution.  Simulation set 1 has $N = 512^3$ particles of mass $8.75 \times 10^{9} M_{\odot}/h$ in a box of $B = 250~\text{Mpc}/h$. The background cosmology is a flat $\Lambda$CDM model with $\Omega_m = 0.269$, $\Omega_\Lambda = 0.732$, $h = 0.704$, $n_s = 0.966$ and $\sigma_8 = 0.8$ . These simulations were presented in \cite{2015MNRAS.454.4208W}. Simulation set 2 has $N = 256^3$ particles of mass $3.531 \times 10^{10} M_{\odot}/h$ in a box of $B = 200~\text{Mpc}/h$. The background cosmology is a flat $\Lambda$CDM model with $\Omega_m = 0.267$, $\Omega_\Lambda = 0.733$, $h = 0.719 $, $n_s = 1.0$ and $\sigma_8 = 0.8$. These simulations were presented in \cite{2012JCAP...10..002B}.
 
Dark matter {\it N}-body simulations are performed by evolving two equations. The first one is the Poisson equation which gives us the gravitational potential $\Phi$ in terms of the particle positions (which determines the density field $\rho_m$)
\begin{align}\label{eq:poisson}
\nabla_x^2\Phi = 4\pi G(\rho_m - \overline{\rho}_m) a^{2}
\end{align}
and the second is the geodesic equation
\begin{align}\label{eq:geodesic}
\ddot{{\bf x}} + 2H\dot{{\bf x}} = - \frac{\nabla_x\Phi}{a^{2}}
\end{align}
which determines the evolution of the particles. For the modified gravity simulations we consider here the only change is that we have a fifth force $-\nabla\varphi$ that contributes to the right hand side of \cref{eq:geodesic} and we have to solve a field equation similar to \cref{eq:poisson}, but highly non-linear, to get the fifth-force potential $\varphi$. More details about the implementation for the models considered in this paper are contained in \cite{2014A&A...562A..78L,2015MNRAS.454.4208W,2012JCAP...10..002B}.
 
We utilised two different halo finders of differing complexity:
\begin{enumerate}
  \item The friend-of-friend halo-finder \texttt{MatchMaker}\footnote{MatchMaker can be found at https://github.com/damonge/MatchMaker} with linking-length $b=0.2$.
  \item The 6D phase-space friend-of-friend halo-finder \texttt{RockStar} \cite{2013ApJ...762..109B}.
\end{enumerate}
Both halo finders use the Friends-of-Friends (FoF) algorithm developed in \cite{1985ApJ...292..371D}, albeit with different distance measures.  Particles are formed into connected graphs by drawing an edge between vertex particles if the distance between them is less than some fraction $b$ of the mean inter-particle distance.  Each connected graph is defined to be a halo if it is not a subgraph of a larger halo.  

The \texttt{MatchMaker} finder is a parallel 3d-FoF finder.  We used the canonical linking length $b = 0.2$.  The distance between particles is the usual 3d Euclidean distance.

The \texttt{RockStar} finder uses the normal 3d FoF (albeit with $b = 0.28$) to identify groups of particles, within which it uses FoF in the 6d phase space to identify subgroups. After conversion from subgroups to subhaloes, any unbinding of particles from haloes is performed using the halo potentials.  (This algorithm is summarised in Figure 1 of \cite{2013ApJ...762..109B}.)

Subsequently halo properties are extracted.  The halo property with which we are concerned in this paper is the halo mass of the parent haloes only (\ie{} we ignore subhaloes because we are only interested in the largest mass ranges).  This is defined to be $M_{200}$, the total mass of all particles within the over-density satisfying $\rho \geq 200 \rho_{\mathrm{crit}}$, where $\rho_{\mathrm{crit}}$ is the background critical density (not the matter density) \cite{2013ApJ...762..109B}.

A comparative analysis of halo finder performance in \lcdm{} can be found in \cite{2011MNRAS.415.2293K}.
 
\subsection{Simulation corrections} 
\label{sub:simulations_corrections}
 
Having obtained our halo catalogues we now approximate the continuous HMF using a histogram.
 
We do not make any corrections to the data.  Some authors propose adjusting the values of $\sigma (M)$ in the simulation data. The aim is to correct for the finite box size, which precludes modes with $k \leq 2\pi / L_{\mathrm{box}}$ from contributing to the over-density fluctuations in the halo.  This effect can be approximated by the extended Press-Schechter approach (amongst other methods: see \cite{2007ApJ...671.1160L} for details).  We avoid corrections for the mass variance due to our large box size, for which corrections are negligible.
 
It is necessary to remove simulation artifacts from the low-mass end.  We truncate the mass function at a lower bound of $100$ particles, where one particle is the mass resolution of the $N$-body simulations.  (This is independent of the minimum number of particles required in the halo identification process.)  Compared to the cuts of \cite{2007ApJ...671.1160L} this is a conservative cut: faced by a relatively small box size, we wish to retain as much of the HMF as possible.  However, the cut is sufficient to remove the low-mass \lq\lq{}tail\rq\rq{} where the mass function---which should be monotonically-decreasing with mass---actually increases with mass.  Such a phenomenon arises from the finite (mass) resolution of the simulations.  At the lowest masses, there is insufficient resolution to identify all of the bound objects with few particles, so the number density is increasingly suppressed at masses below a characteristic turnover mass.  This limitation cannot be alleviated without sub-sampling the simulation box at finer resolution.  Slightly higher, at haloes with tens of particles, the uncertainty on the mass values is a significant fraction of the total halo mass.  Consequently, the loss or addition of one particle can move the halo between bins.  There are two possible ways of accounting for this: either incorporating a mass uncertainty in the likelihood function, or minimising the effect by judicious bin optimisation.  We opt for the latter.  This low-mass effect is well-known and we do not discuss it further.
 
At the high mass end our cutoff is artifically imposed by the finite box size.  The finite box size curtails the number of large mass haloes found in a finite sub-volume of the horizon (this underestimation is quantified for \lcdm{} in \cite{2013A&C.....3...23M}).  In addition, our excursion set technique prevents us from calculating the HMF for masses of $S < \Senv$ (for the excursion-set method) or $\nueff < 0$ (for the cosmic web).  Using a Gaussian window function with a radius of $10 \, \mathrm{Mpc}/h$ this corresponds to a mass of $M \approx 10^{16} M_{\odot}$. We have confirmed that this does not remove any haloes from our data.
 
The remaining factor in our simulated HMF is the bin width of the histogram.  For simplicity we adopted constant bin widths, as is common across the literature (although we discuss this and other options in more depth in \cref{sub:number_of_bins}).  We verified that the results we obtain are independent of the bin width in \cref{sub:number_of_bins}.  Consequently we adopted a value of $N = 30$ bins as an average value across all our data sets.
 
\subsection{Uncertainty in the data} 
\label{sub:data_errors}
 
Finally we quantify the uncertainty in the HMF. 
 
The uncertainty in the bin occupation is assumed to be Poissonian.  The usual uncertainty on bin occupations is taken to be the well-known result\footnote{Eq. (11) in \cite{2011MNRAS.410.1911C} is not true, but is related to the standard deviation as a \emph{fraction} of bin occupation: $\sigma/N = 1 / \sqrt{N} \notimplies \sigma = 1/N$.} that the Poisson standard deviation on a bin containing $N$ haloes is $\sigma = \sqrt{N}$.  Various HMF papers (\eg{} \cite{2007ApJ...671.1160L}) use an \lq\lq{}improved\rq\rq{} Poisson error defined to be:
\begin{equation} \label{eq:poisson_err}
  \sigma_{\pm} = \sqrt{N + \frac{1}{4}} \pm \frac{1}{2}
\end{equation}
This asymmetric error asymptotes to the usual one for large $N$ but is better-behaved for small $N$, particularly for empty bins.  While this does not affect the Poisson-based likelihood, it does enter the Gaussian-based likelihood.  The reason is straightforward: the number of counts per bin is known precisely, so there is no uncertainty.  The Poisson \lq\lq{}noise\rq\rq{} expresses the uncertainty in the mean of the underlying Poisson \pdf{}, which is equal to the variance term which does enter into the Gaussian \pdf{}.  Since the error does not enter the likelihood function, we make no cuts when using constant-width bins (explained below).  For variable-width bins, we tolerate an error of $10\%$ or less, which is in line with the choice from other papers (\eg{} \cite{2011MNRAS.410.1911C}).
We must pay attention to combining the asymmetric errors via the method of \cite{2004physics...6120B}.  The combined variance is then:
\begin{equation} \label{eq:asymmetric_error}
  V \left( x \given \hat{x} \right) = V_{0} + V_{1}(x - \hat{x})
   \quad \text{where} \quad V_{0} = \sigma_{+}\sigma_{-}
   \quad \text{and} \quad
   V_{1} = \sigma_{+} + \sigma_{-}
\end{equation}
This method tightens the uncertainty on the low-occupation bins, weighting the Gaussian likelihood more favourably towards the high-mass end than with symmetric error bars.
 
 
\section{Bayesian inference} 
\label{sub:bayesian_inference}

Our aim is to find the posterior probability distribution $p \left( H \given D,I \right)$ for a given hypothesis $H$ when we take into account our prior information $I$ and the data $D$.  We have already examined $D$, the counts per mass bin from our simulation data in \cref{sub:simulations_corrections}.  In this section we define our priors, the appropriate likelihood function and characterise different measures of the high-probability regions in our probability density functions for the HMF free parameters.  The effects of the bin width, choice of likelihood function and nested sampling settings are discussed in \cref{app:calibration_of_multinest}.
 
The priors $I$ contain our assumptions about the prior distribution for the parameters of the hypothesis.  We set uniform priors, \ie{} we have no reason to favour any regions of parameter space over another.  The lower limit for the priors is zero, because we must avoid an HMF which is negative (if the scaling parameter is $A < 0$), complex (fractional powers of $a\nu$ where $a < 0$), or non--monotonically-decreasing (since $\nu$ is an increasing function of mass, $\nu^{p}$ must have $p > 0$). The upper limit is arbitrary.  Given that most published values for the parameters lie in $[0,2]$ (the exception being Jenkins $a = 3.8$), we used an upper limit of $10$ to ensure that the credible regions were well-contained within the prior region.

Now we require an expression for the likelihood function $p \left( D \given I,H \right)$.  We apply the Poisson likelihood:
\begin{equation} \label{eq:likelihood_Poisson}
  \ln \mathcal{L} = - \sum_{i} \left( \mu_{i} - n_{i} + n_{i} \ln \frac{n_{i}}{\mu_{i}}  \right)
\end{equation}
where $\mu_{i}(\mathbf{q})$ is the number of counts given by the parameter set $\mathbf{q}$ and $n_{i}$ is that given by the data in the $i$-th bin.  The last term is zero when $n_{i}$ is zero: otherwise $n$ is always a positive integer, whereas $\mu$ is a positive real number.
 
The likelihood $\mathcal{L}(H)$ is the probability that the hypothesis $H$ produces the data $D$.  However, we are interested in the probability that the data are consistent with the hypothesis.  Via Bayes\rq{} theorem, we obtain: 
\begin{equation}
\label{eq:posterior}
  p \left( H \given D,I \right) = \frac{ p \left( H \given I \right)  \mathcal{L} \left( D \given H \right)}{ p\left( D \given I \right) }
\end{equation}
This is the complete posterior PDF for our hypothesis $H$ in light of the data $D$ given prior information $I$.
 
Having chosen a particular HMF, we are interested in the universality of the HMF across different MG theories.  One way in which to measure this is to see how the best-fit parameter values change depending on the MG model.  \texttt{MultiNest} gives three different options to define \lq\lq{}best-fit\rq\rq{}: the maximum-likelihood (ML), maximum-a-posteriori (MAP) and posterior mean (PM) values.
 
The ML value for a given parameter $\theta$ is defined to be that which maximises $p \left( D \given \theta,I \right)$, \ie{} the most probable value for the model to give the observed data.  In contrast the MAP value maximises the probability $p \left( \theta \given D,I \right)$.  Since we have chosen uniform priors, the MAP value is equal to the ML value via \cref{eq:posterior}.  Since the ML value is the fastest to converge to the true value in MultiNest \cite{2009MNRAS.398.1601F}, we used this value for our best-fit parameters.  We are also interested in the PM value
\begin{equation} \label{eq:posterior_mean}
  \langle \theta \rangle = \int \dx \bar\theta \; p \left( \bar\theta \given D, I \right) \bar\theta
\end{equation}
because the $1\sigma$ credible regions given by MultiNest are only provided for the posterior mean. 
 
If the full PDF is not well-described by a single value (\eg{} the three values above are very different, or the posterior is very flat \etc{}), then we are better off examining the credible region $R$ of credibility $C$, which is the set:
\begin{equation} \label{eq:credible_region}
  R = \left\{ \four{\theta} : \int_{p \left( \bar{\four{\theta}} \given D, I \right) > c } d \bar{\four{\theta}} \;
  p \left( \bar{\four{\theta}} \given D, I \right) = C \right\}
\end{equation}
where $c$ is the level set forming the boundary $\partial R$, inside which the probability is greater than $c$ and outside which it is less than $c$.   We refer to the $1\sigma$ and $2\sigma$ credible regions for $C = 0.68$ and $C = 0.95$ respectively.  
 
\section{Results and discussion} 
\label{sec:results_and_discussion}

This section addresses the questions posed in the introduction.  \cref{sub:including_denv_barrier_in_mg} present our new results for incorporating MG into \lcdm{}-calibrated HMFs.  In \cref{sub:assuming_concordance_cosmology} we discuss whether the presence of screened gravity can be mistaken for a change in best-fit parameter values for the \lcdm{} HMF.  In \cref{sub:recalibrating_best_fit_parameters} we apply the procedure used in \lcdm{} to calibrate the MG HMF using the full excursion set approach.  We compare to existing values from the literature and assess the deviation.  In particular we address the universality of the halo mass function, \ie{} the invariance of the best fit free parameters to changes in the underlying gravity model.  \cref{app:calibration_of_multinest} confirms that our parameter estimation technique is reliable for the problem at hand and describes the settings we used in the nested sampling algorithm.
 
\subsection{Accounting for the drifting-and-diffusing barrier in MG} 
\label{sub:including_denv_barrier_in_mg}

\cref{sub:including_denv_barrier_in_mg} presents our new results for incorporating MG into \lcdm{}-calibrated HMFs.  

In \cref{sec:the_halo_mass_function} we suggested a variety of different methods to implement the full effects of environment dependence in MG for a fitting function.  We compare the data from our $\fr$ simulations with $\fr[5]$ and $\fr[6]$ to the theoretical HMF using each method.  

\begin{figure}
  \begin{center}
    \subfloat[\lcdm{}]{\label{subfig:plots_danddbarrier_plot_fcd_lenm_30_LCDM_Reed-07}\includegraphics[keepaspectratio,height=0.3\textheight]{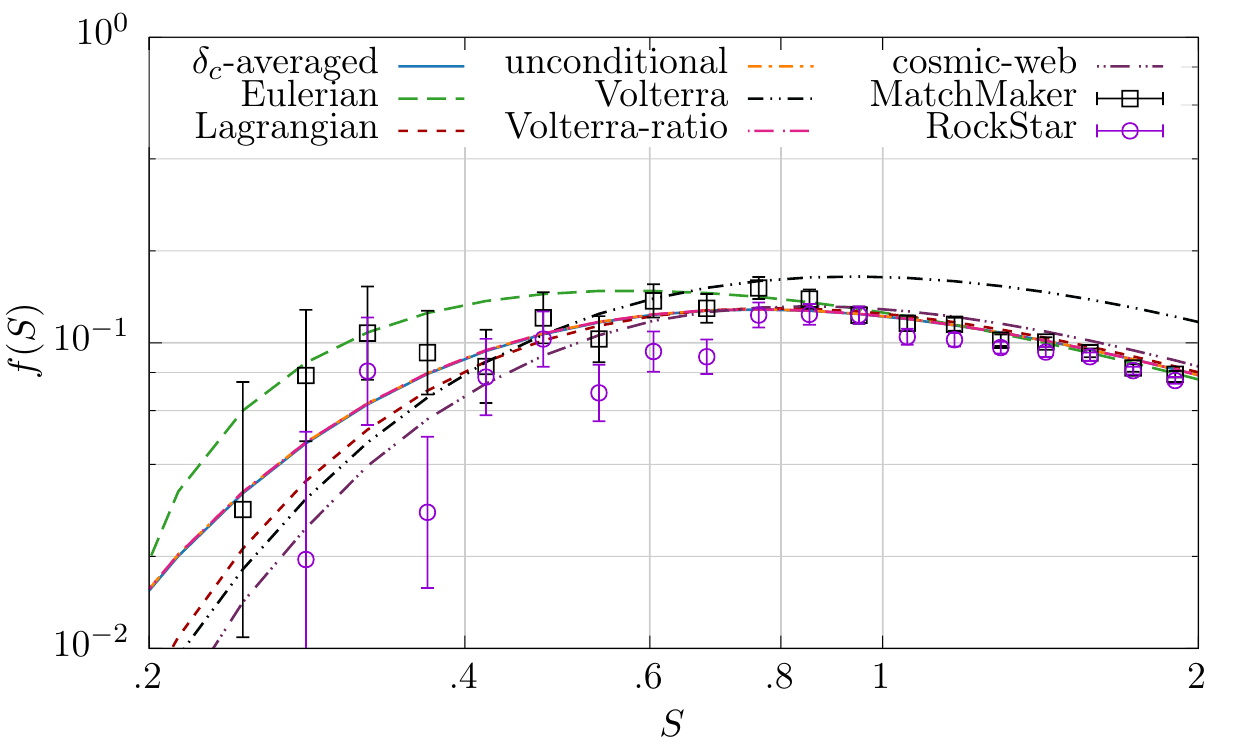}} \\[-1ex]
    \subfloat[{$\fr[5]$}]{\label{subfig:plots_danddbarrier_plot_fcd_lenm_30_F5_Reed-07}\includegraphics[keepaspectratio,height=0.3\textheight]{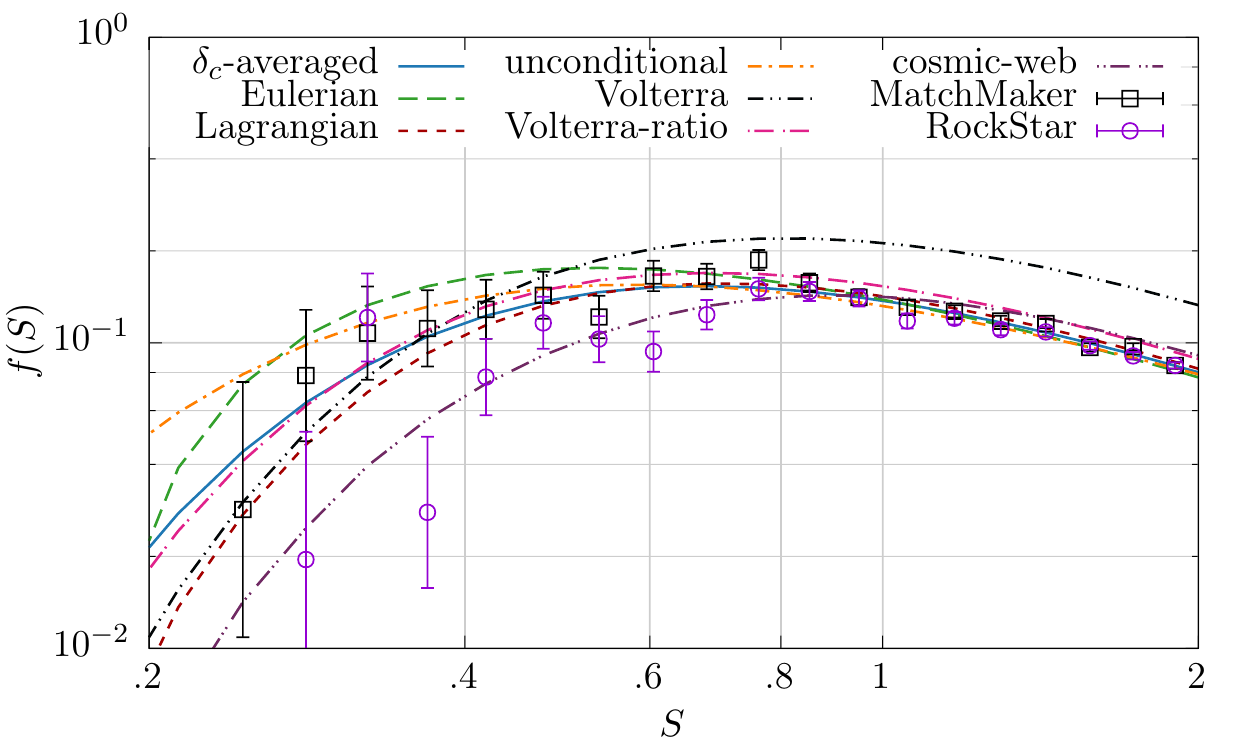}} \\[-1ex]
    \subfloat[{$\fr[6]$}]{\label{subfig:plots_danddbarrier_plot_fcd_lenm_30_F6_Reed-07}\includegraphics[keepaspectratio,height=0.3\textheight]{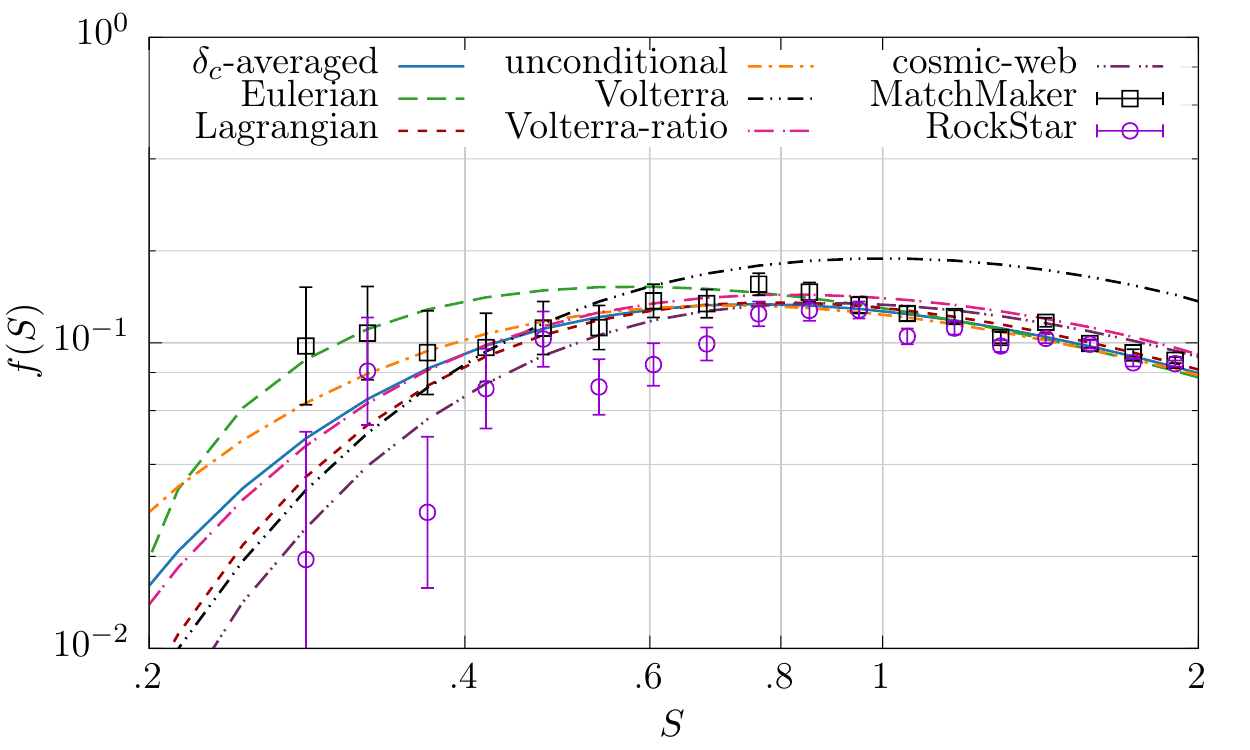}} 
  \end{center}
  \caption{The first-crossing distribution for a drifting-and-diffusing barrier using the variety of methods explored in this paper are shown as lines.  We also plot the data from the $N$-body simulations using the two halo finders.}
  \label{fig:plot_fcd_lenm_30_Reed-07_cont1}
\end{figure}

In the order given by \cref{fig:plot_fcd_lenm_30_Reed-07_cont1} we have:
\begin{enumerate}
\item The unconditional HMF using $\langle \delta_{c}\left( S \given \denv, \Senv \right) \rangle_{\mathrm{env}}$, the environment-averaged collapse density \cref{eq:deltac_ave}
\item The conditional HMF marginalised over the Eulerian $\denv$-distribution \cref{eq:pdf_denv_E}
\item The conditional HMF marginalised over the Lagrangian $\denv$-distribution \cref{eq:pdf_denv_L}
\item The unconditional HMF, \ie{} assuming both $\denv$ and $\Senv$ vanish
\item The Volterra solution to \cref{eq:volterra_all}
\item The unconditional HMF scaled with the Volterra solution using \cref{eq:scaling_Volterra}
\item The cosmic web HMF marginalised over the tidal tensor distribution \cref{eq:cosmic_web}
\end{enumerate}
\cref{fig:plot_fcd_lenm_30_Reed-07_cont1} shows these methods for the Reed-07 fitting function\footnote{We might have used any of the fitting functions, because our aim is to compare the behaviour of each technique for extending the mass function to MG.  The Reed fit produced the closest fit to the data given the default parameters.} (lines) and the $N$-body HMFs calculated using the \texttt{MatchMaker} and \texttt{RockStar} halo finders (points).  These simulations and the workings of the two halo finders are described in \cref{sub:n_body_simulations_haloes}.  We examine the \lcdm{} results before discussing each method over the next few paragraphs.

The \lcdm{} behaviour in \cref{subfig:plots_danddbarrier_plot_fcd_lenm_30_LCDM_Reed-07} is consistent with the excursion-set framework of \cite{1991ApJ...379..440B}.  The environment-averaged collapse density is $\delta_{c}^{\Lambda}$ (because the excursion-set barrier is flat) and the Volterra solution reduces to \PS{}.  Therefore both of these methods produce the same result as the unconditional HMF.  The other four methods all differ.  While both the Lagrangian and cosmic web methods do equal the unconditional HMF using \PS{}, this is due specifically to the design of this function from excursion set theory.  It is the only integrand for which the solution to the integral equation \cref{eq:volterra_fcd} is merely a rescaling between the conditional and unconditional forms of $\nu$.  We have now ensured that the different methods behave as expected in GR, before applying them to $\fr$.

The first option avoids using excursion set theory altogether, by pre-emptively converting the drifting-and-diffusing barrier to an average density.  The drifting barrier $\delta_{c}(S)$ can be incorporated straightforwardly, just as in the non-chameleon MG models, into the unconditional HMF.  Surprisingly, this gives a very good fit, superior to the purely unconditional HMF.  Although the peak of \cref{eq:pdf_denv_L} is at zero, this result illustrates that we need to use the entire PDF, rather than only using the peak to approximate the average.  In this way we can account for the peak-background split, whereby it is easier for haloes to form as $\denv \rightarrow \delta_{c}$ in dense regions and more difficult in under-dense regions.  This method has the advantage that we can compute the barrier density once, rather than re-computing the conditional function at every stage of the MCMC process.  We have managed to produce a good fit by considering only the barrier density, rather than accounting for the complex excursion set behaviour of the full drifting-and-diffusing barrier.

The conditional HMFs marginalised over the Lagrangian (\cref{eq:pdf_denv_L}) and Eulerian (\cref{eq:pdf_denv_E}) theoretical distributions $p(\denv)$ have appeared in the literature before\footnote{\cite{Lombriser:2013wta} only plotted the relative enhancement $n_{\mathrm{MG}}(M) - n_{\mathrm{GR}}(M) / n_{\mathrm{GR}}(M)$, rather than the HMF proper $n(M)$, so we cannot readily compare our results to theirs.} \cite{Lombriser:2013wta}, where they were applied to the Sheth-Tormen HMF.  The authors suggested discarding the Lagrangian (density) distribution in favour of the Eulerian, on the basis that a density distribution which better reflects the physical formation of over-densities, would correspondingly produce a more accurate HMF.  This is supported by the findings of \cite{2012MNRAS.425..730L}.  While our results agree using the \texttt{MatchMaker} halo finder, the \texttt{RockStar} halo finder predicts a systematically lower distribution of haloes, better suited to the Lagrangian model.  

  The question of which PDF to use is somewhat moot considering that neither fit performs particularly well.  This is probably due to generating the conditional from the unconditional HMF.  The rescaling at the end of \lcdm{} excursion set theory which is used in \cite{2012MNRAS.421.1431L,Lombriser:2013wta} implicitly assumes that a linear function of $\delta_{c}(S)$ is a good approximation for the actual barrier demsity.  We have already seen in \cref{fig:surf_deltac_linear} that this is not the case.  Therefore we cannot use methods which work for a flat barrier density in \lcdm{} to good effect in $\fr$.

The solution to the integral equation is the MG-equivalent of \PS{}.  For this reason, we do not expect it to be a good fit to the data.  Indeed the $\fr[5]$ anf $\fr[6]$ plots in \cref{subfig:plots_danddbarrier_plot_fcd_lenm_30_F5_Reed-07,subfig:plots_danddbarrier_plot_fcd_lenm_30_F6_Reed-07} share the recognised flaws of the \PS{} fit in \lcdm{} in \cref{subfig:plots_danddbarrier_plot_fcd_lenm_30_LCDM_Reed-07}, namely that it underpredicts at the high-mass end (low-$S$) and overpredicts at the low-mass end (high-$S$).  Nonetheless, it reproduces the general behaviour of the \fcd{}, which is remarkable for such a simple model.

Moreover, we can improve the result from the unconditional HMF by scaling by the ratio of the Volterra solution to \PS{}.  The two poor results combine to form a decent approximation.  At the low-$S$ end, the Volterra solution forms too few haloes becuase the random walks in excursion set theory are not absorbed early enough by the barrier density, so too many trajectories survive to produce haloes at high-$S$.  In contrast, the unconditional HMF assumes $\denv = 0$ (and $\Senv = 0$) so at small values of $S$ the value of $\nu = \delta_{c}(S)/\sqrt{S}$ is large and vice-versa at high-$S$.  Since the fitting function $f(\nu)$ is montonically increasing with $\nu$, we have too many haloes at small $S$ and too few at low $S$.  These two behaviours counteract one another to reduce the overall discrepancy of the fit.  This is because we have deliberately designed a method to combine different strengths of the analytical and empirical approaches.  The unconditional fitting function is designed to produce a good approximation to the \lcdm{} data.  The Volterra solution captures the excursion set behaviour of the barrier density, which incorporates the main effect of $\fr$ compared to \lcdm{} from a theoretical viewpoint.  Thus, we can combine two simple mechanisms to produce a relatively good solution, despite their individual predictions being ineffectual.

The performance of the cosmic web method is discussed in more detail in our previous paper \cite{2017JCAP...03..012V}.  In general it underpredicts the HMF at the high-mass end, whereas it works well at the low-mass end.  There are two contributing factors, namely the $\nueff$ approximation substituted into the fitting function and the $p(\rho, \theta, \nu_{\text{env}})$ distribution related to the tidal tensor in the cosmic web.  The former is not particularly successful even in \lcdm{} (\cite{Alonso:2014zfa}) so we ought not expect any better performance in $\fr$ where the spherical collapse in more complicated due to the fifth force.  The latter is the equivalent of the Lagrangian distribution for $\denv$ applied to all three eigenvalues of the tidal tensor rather than its trace.  We have already seen that the Lagrangian-$\denv$ fit does not produce an accurate fit.  This technique is useful for calculating the HMF in individual structures of the cosmic web (\eg{} voids, sheets \etc{}), wherein we have no other analytical treatment, but not the overall HMF.

Finally we comment on the distinction between the cosmic web results (\cref{eq:cosmic_web}) and those of the Lagrangian PDF.  Given that the conditional HMF in the cosmic web doesn't depend upon $(\rho,\theta)$, and that the distributions for $p(\nu_{\text{env}})$ and $p(\denv)$ have the same dependence on $\denv$, one may expect that these two methods should produce the same results.  Clearly \cref{fig:plot_fcd_lenm_30_Reed-07_cont1} does not agree.  This is due to the effect of the window functions in calculating $S(M)$ and $\Senv(R_{\text{env}})$.  In cosmic web we use a Gaussian window function for the environment and a (real-space) top-hat for the halo, whereas in the excursion set we used a top-hat window function for both halo and environment.  This affects the value of $\Senv$ obtained from the same environment radius.  Additionally, the Lagrangian (and Eulerian) result(s) both assume that the halo-environment correlation is $\epsilon = \Senv / S$.  This is due to the assumption in excursion set theory that our trajectories in $(\delta, S)$ are uncorrelated at successive values of $S$, for which one requires a sharp-$k$ window function, whence comes this expression for $\epsilon$.  In the cosmic web method we calculate $\epsilon$ numerically for a Gaussian and top-hat window function in the environment and halo respectively.  Therefore the same values of $\delta_{c}(S | \denv, \Senv)$ map differently onto the argument of the conditional HMF.  Once this is established, it does not matter that the marginalisation over the environment is the same in both methods.

Given the performance of the various technique for extending the fitting functions to chameleon MG, and the performance factor involved in re-calculating a conditional HMF (and marginalising over it) at every stage of the MCMC procedure, we shall use the Volterra-ratio method to calculate the $\fr$ HMF in \cref{sub:recalibrating_best_fit_parameters}.  Nevertheless, we have found a broad spectrum of possible methods by means of which we can incorporate MG into fitting functions originally designed for \lcdm{} alone.  Moreover, we have found that some are more suited to certain applications (\eg{} the cosmic web approach) or halo finders (\eg{} the two density-marginalised methods) than others.  This demonstrates the additional complexity which environment dependence produces in chameleon screening compared to symmetron- and Vainshtein-screened theories.  We cannot neglect this and simply substitute the unconditional HMF if we wish to produce a useful empirical function to use in lieu of deriving one from $N$-body simulations.


\begin{figure}
  \begin{center}
    \subfloat[\lcdm{}: MatchMaker halofinder, 30 bins]{\label{chains_new_lcdm/Jenkins_DGP1_lenm_30_nlive_500_MM_}\includegraphics[keepaspectratio,height=0.45\textheight]{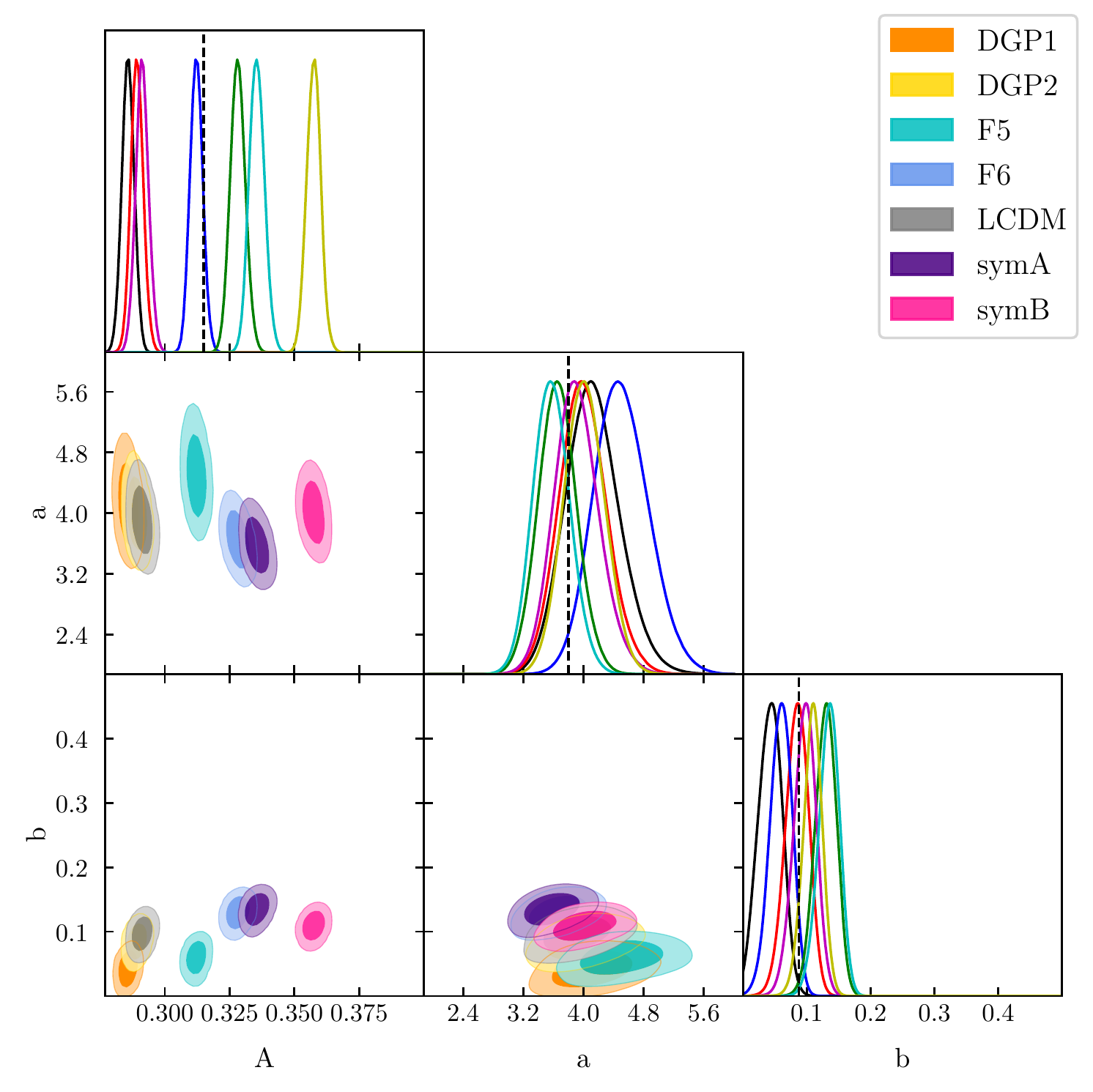}} \\
    \subfloat[\lcdm{}: RockStar halofinder, 30 bins]{\label{chains_new_lcdm/Jenkins_DGP1_lenm_30_nlive_500_RS_}\includegraphics[keepaspectratio,height=0.45\textheight]{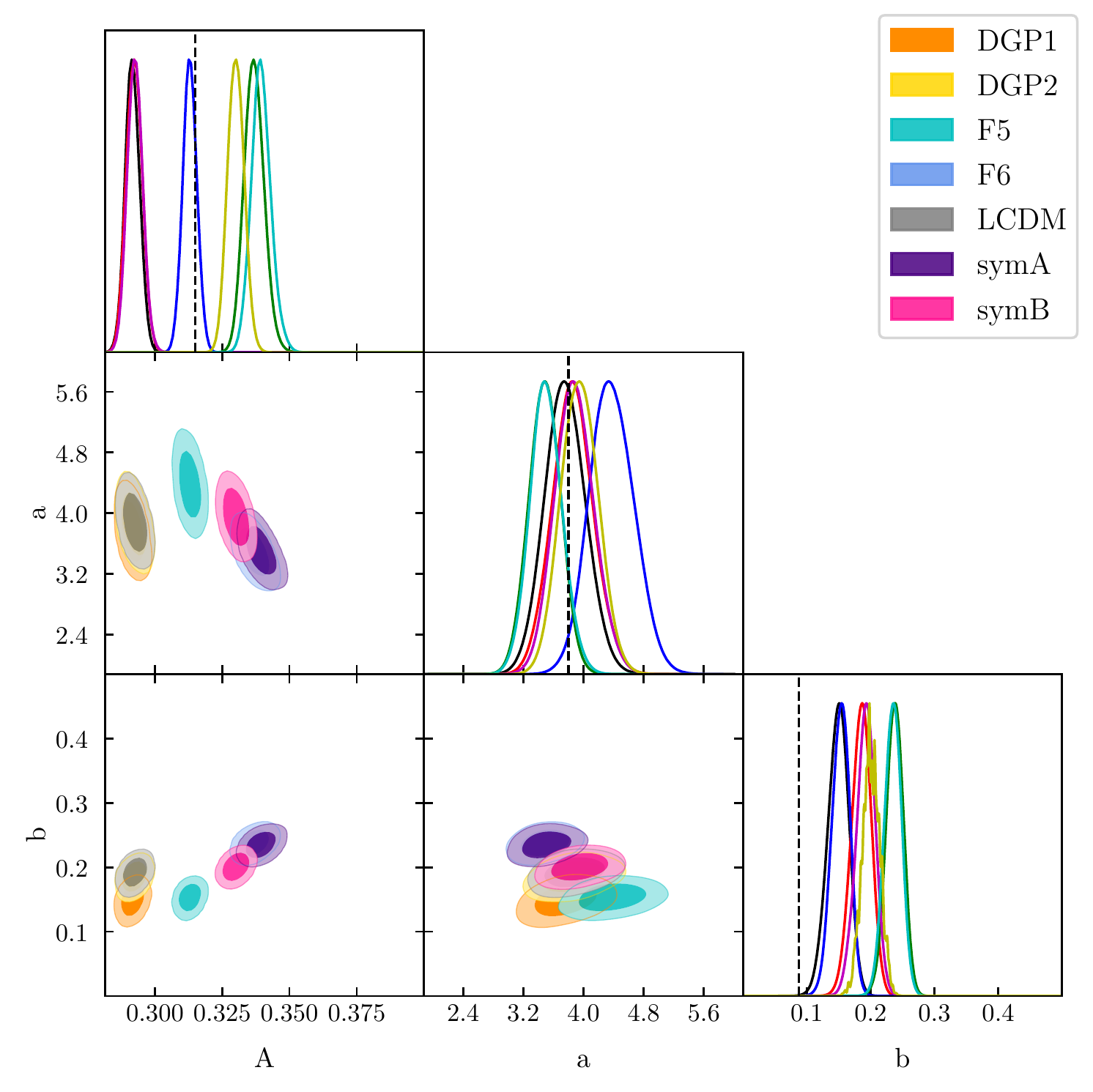}} 
  \end{center}
  \caption{Triangle plot showing the 1-$\sigma$ (dark) and 2-$\sigma$ (light) posterior credible regions for the Jenkins HMF assuming \lcdm{} fitting functions and using $N$-body data from GR and MG (coloured regions).  The black dashed lines show the values proposed by \cite{2001MNRAS.321..372J}.}
  \label{fig:.plots/chains_new_lcdm/Jenkins_DGP1_lenm_30_nlive_100}
\end{figure}

\begin{figure}
  \begin{center}
    \subfloat[\lcdm{}: MatchMaker halofinder, 30 bins]{\label{chains_new_lcdm/Peacock_DGP1_lenm_30_nlive_500_MM_}\includegraphics[keepaspectratio,height=0.45\textheight]{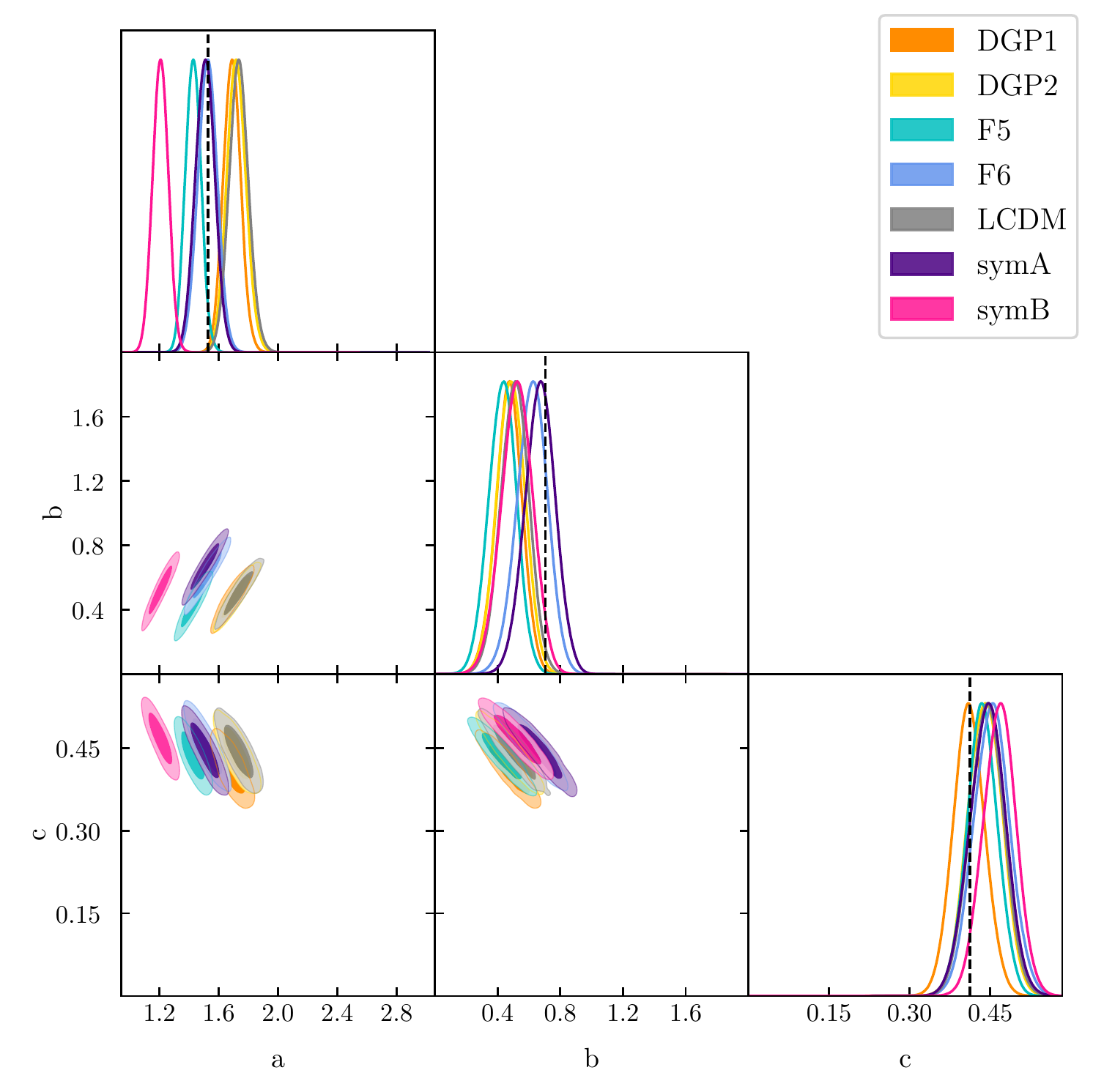}} \\
    \subfloat[\lcdm{}: RockStar halofinder, 30 bins]{\label{chains_new_lcdm/Peacock_DGP1_lenm_30_nlive_500_RS_}\includegraphics[keepaspectratio,height=0.45\textheight]{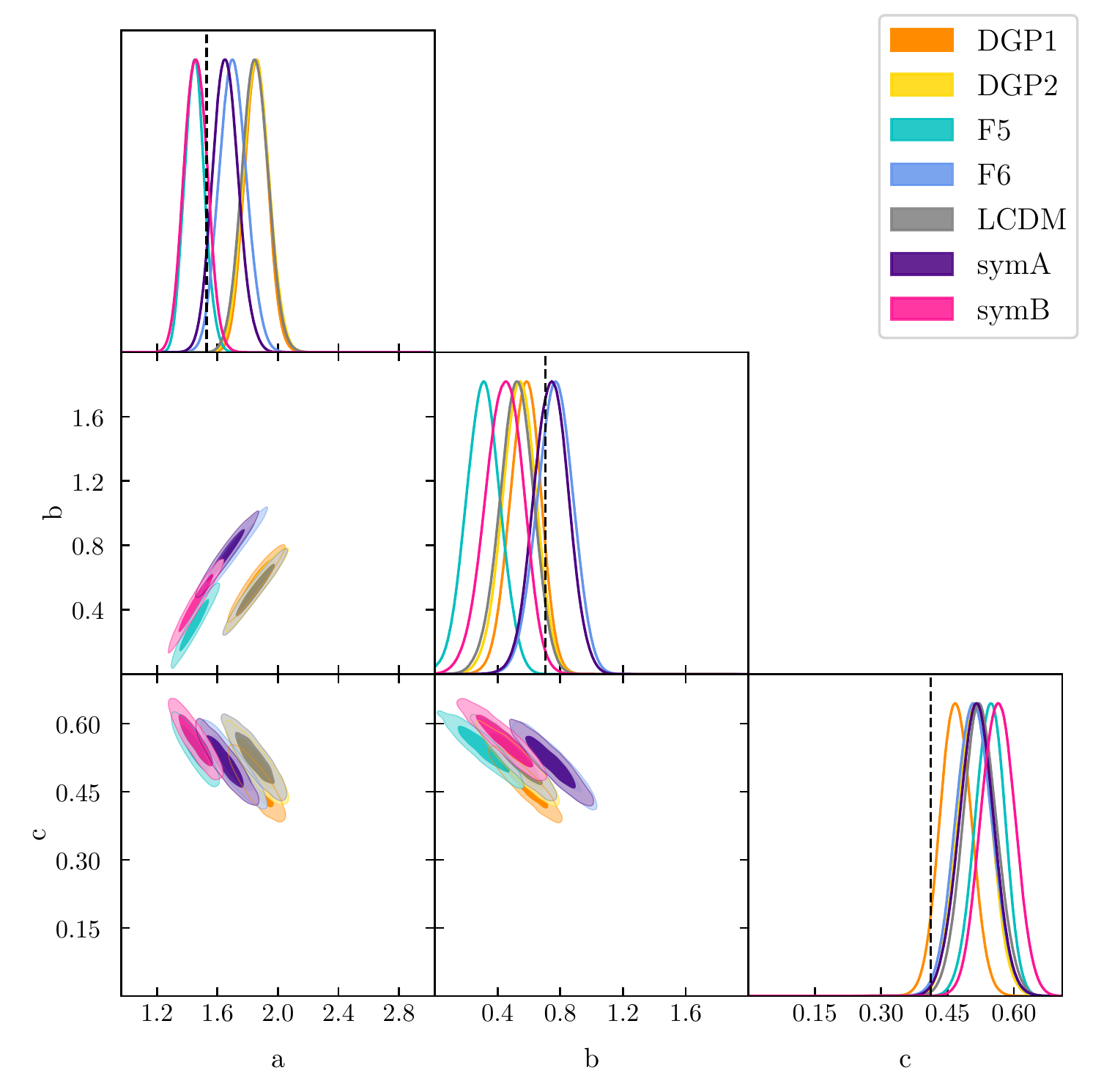}} 
  \end{center}
  \caption{Triangle plot showing the 1-$\sigma$ (dark) and 2-$\sigma$ (light) posterior credible regions for the Peacock HMF assuming \lcdm{} fitting functions and using $N$-body data from GR and MG (coloured regions).  The black dashed lines show the values proposed by \cite{Peacock2007}.}
  \label{fig:.plots/chains_new_lcdm/Peacock_DGP1_lenm_30_nlive_100}
\end{figure}

\begin{figure}
  \begin{center}
    \subfloat[\lcdm{}: MatchMaker halofinder, 30 bins]{\label{chains_new_lcdm/SMT-Courtin_DGP1_lenm_30_nlive_500_MM_}\includegraphics[keepaspectratio,height=0.45\textheight]{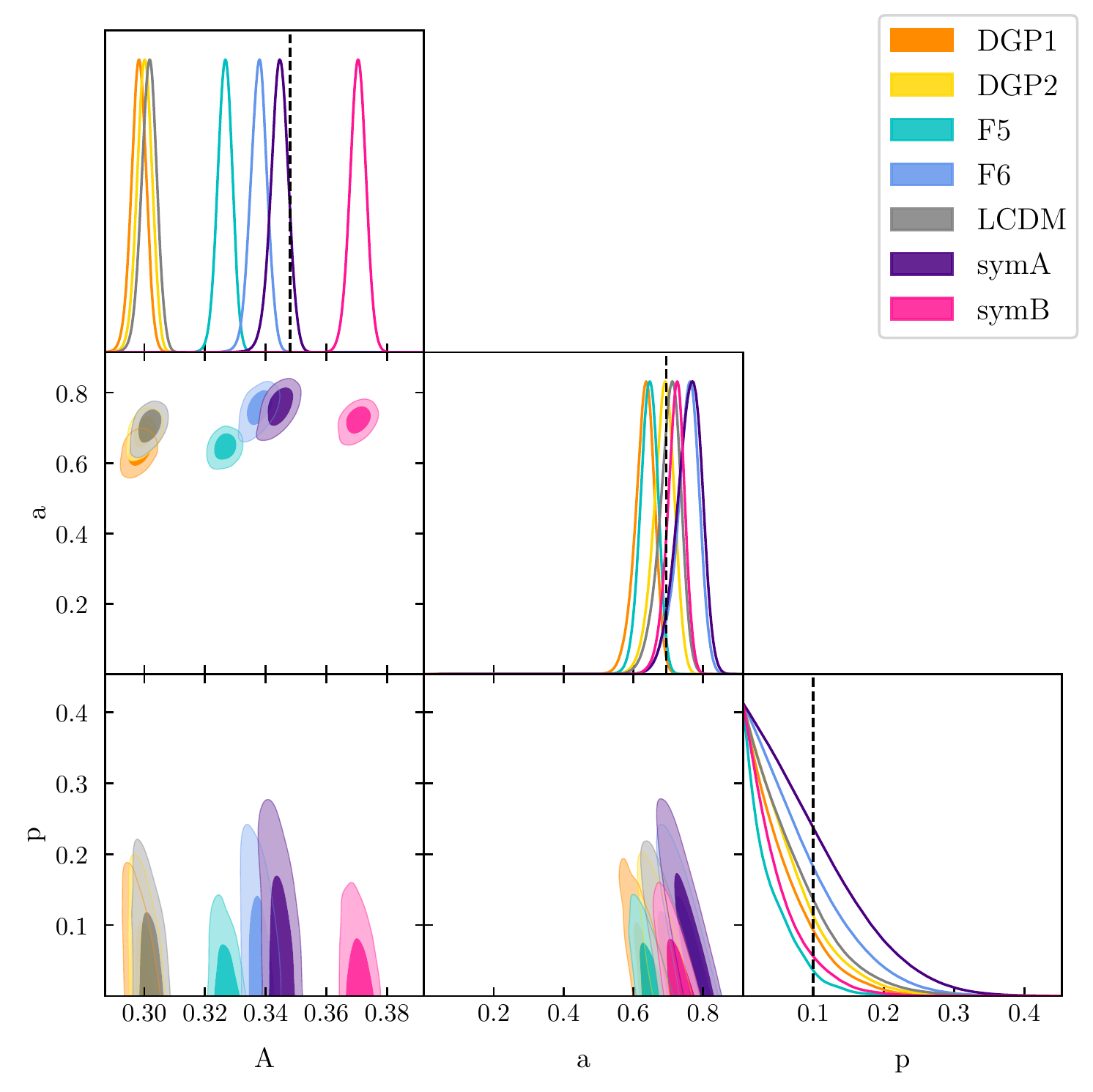}} \\ 
    \subfloat[\lcdm{}: RockStar halofinder, 30 bins]{\label{chains_new_lcdm/SMT-Courtin_DGP1_lenm_30_nlive_500_RS_}\includegraphics[keepaspectratio,height=0.45\textheight]{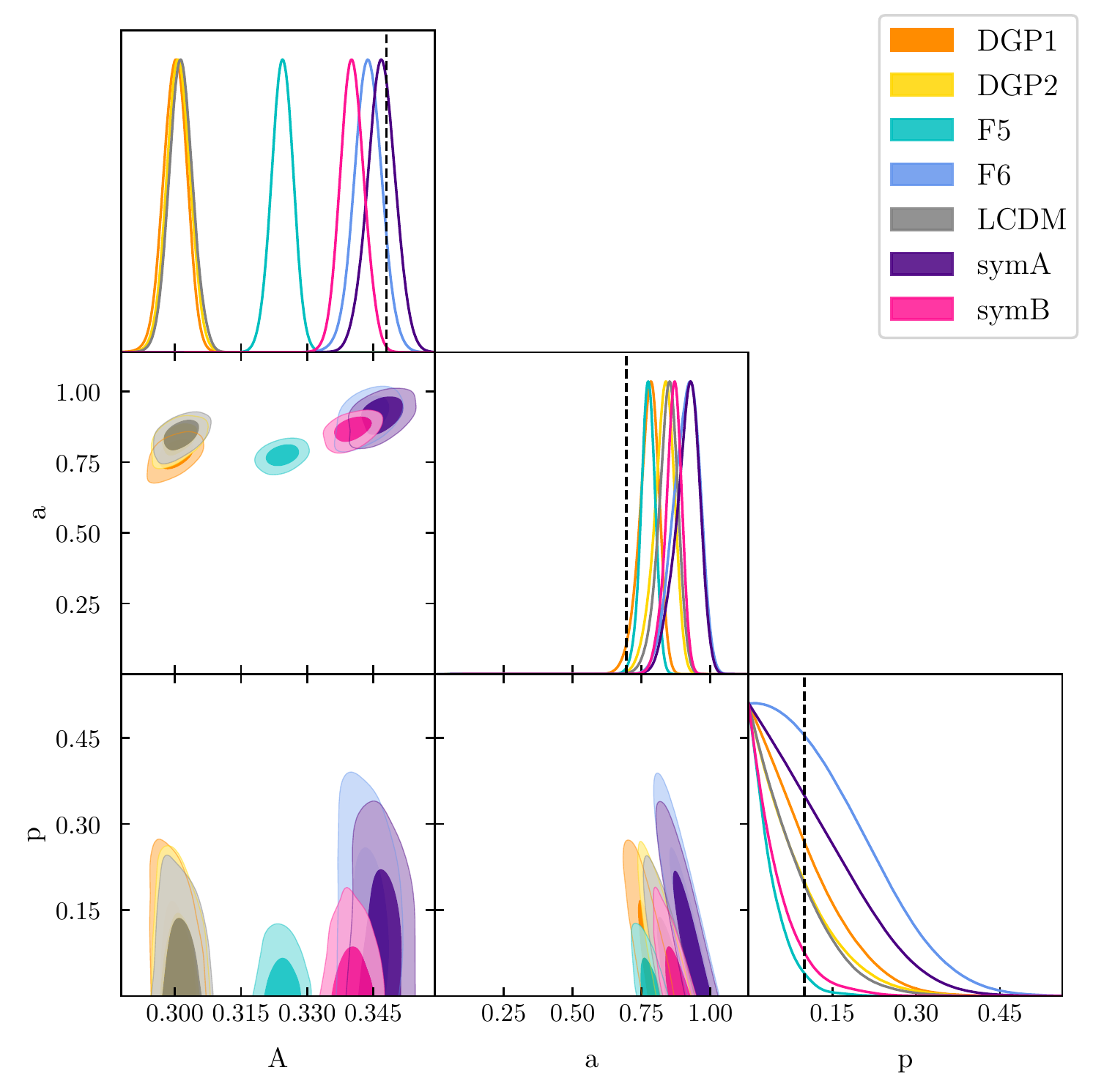}} 
  \end{center}
  \caption{Triangle plot showing the 1-$\sigma$ (dark) and 2-$\sigma$ (light) posterior credible regions for the SMT-Courtin HMF assuming \lcdm{} fitting functions and using $N$-body data from GR and MG (coloured regions).  The black dashed lines show the values proposed by \cite{2011MNRAS.410.1911C}.}
  \label{fig:.plots/chains_new_lcdm/SMT-Courtin_DGP1_lenm_30_nlive_100}
\end{figure}
  
\begin{figure}
  \begin{center}
    \subfloat[\lcdm{}: MatchMaker halofinder, 30 bins]{\label{chains_new_lcdm/Warren-Crocce_DGP1_lenm_30_nlive_500_MM_}\includegraphics[keepaspectratio,height=0.45\textheight]{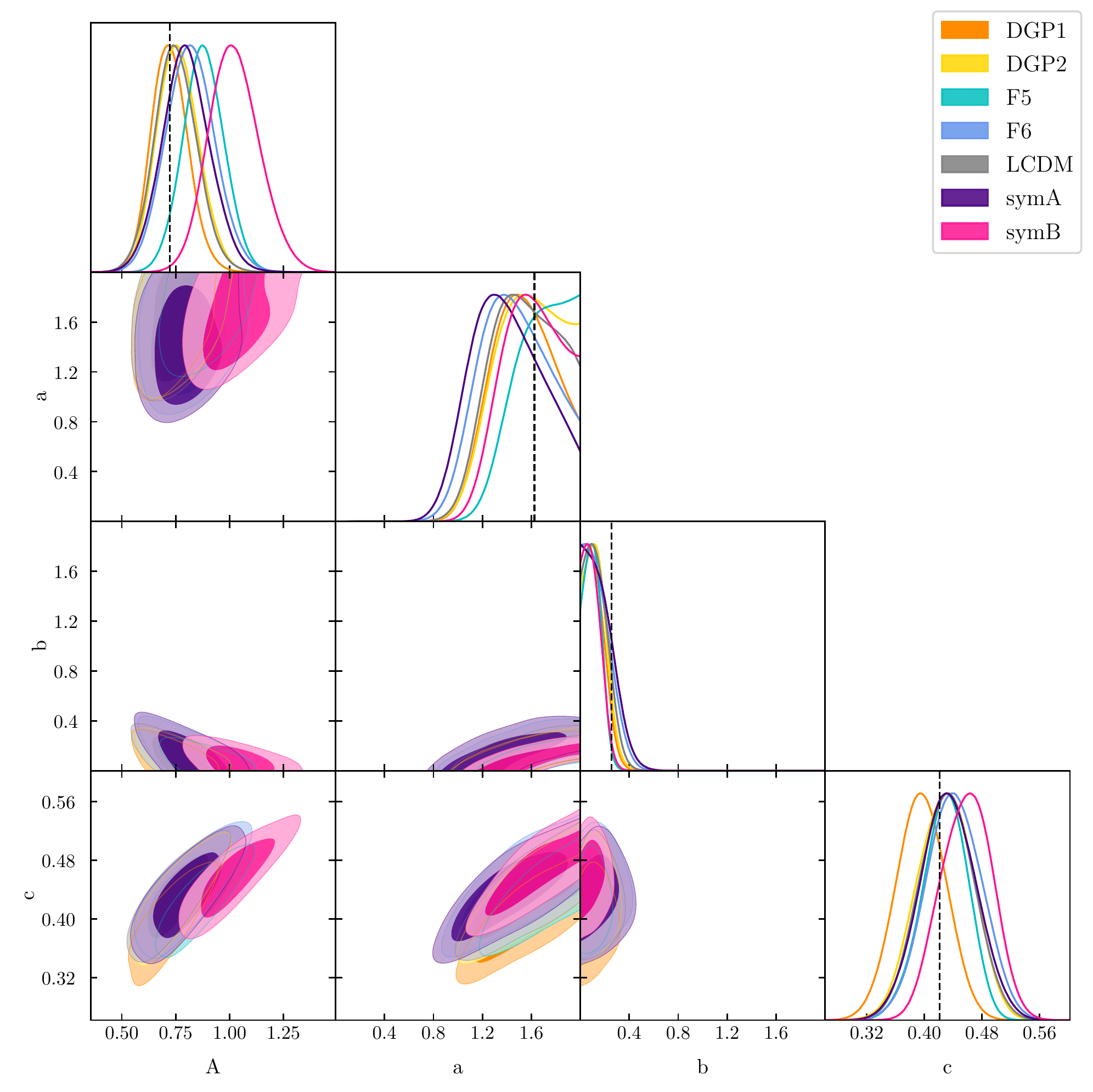}} \\
    \subfloat[\lcdm{}: RockStar halofinder, 30 bins]{\label{chains_new_lcdm/Warren-Crocce_DGP1_lenm_30_nlive_500_RS_}\includegraphics[keepaspectratio,height=0.45\textheight]{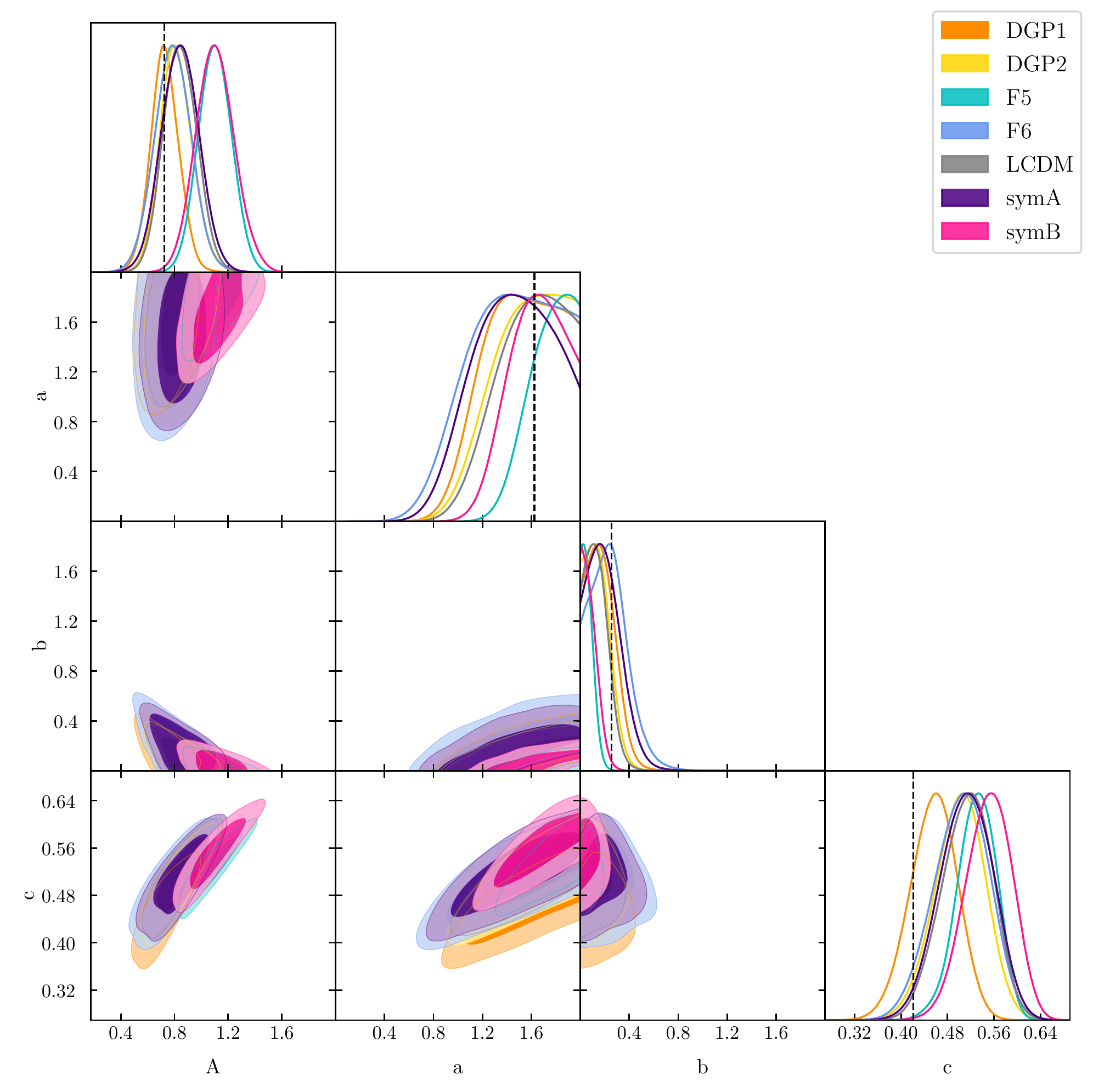}} 
  \end{center}
  \caption{Triangle plot showing the 1-$\sigma$ (dark) and 2-$\sigma$ (light) posterior credible regions for the Warren-Crocce HMF assuming \lcdm{} fitting functions and using $N$-body data from GR and MG (coloured regions). The black dashed lines show the values proposed by \cite{2006ApJ...646..881W}.}
  \label{fig:.plots/chains_new_lcdm/Warren-Crocce_DGP1_lenm_30_nlive_100}
\end{figure}


\subsection{Assuming concordance cosmology} 
\label{sub:assuming_concordance_cosmology}
  
The first problem is perhaps so obvious as to be invisible: if we do not set out to look for modified gravity, we may be unable to detect it.  In other words, having assumed a \lcdm{} cosmology \textit{a priori}, does modified gravity present itself \textit{a posteriori} as merely a different (\ie{}, non-standard) set of best-fit parameters in a \lcdm{} halo mass function?  In this subsection, we fit the N-body HMFs for each modified gravity model using the \lcdm{} mass functions.  We used $10^{2}$ and $10^{3}$ live points to ensure that our results had converged, \ie{} that our estimate of the posteriors PDFs was accurate.  We also examined the effects of both the \texttt{MatchMaker} and \texttt{RockStar} halo finders, which we expect to differ due to their varying criteria for finding bound objects, which translates into a different halo mass function.

Throughout this subsection, we will refer to \cref{fig:.plots/chains_new_lcdm/Warren-Crocce_DGP1_lenm_30_nlive_100,fig:.plots/chains_new_lcdm/SMT-Courtin_DGP1_lenm_30_nlive_100,fig:.plots/chains_new_lcdm/Peacock_DGP1_lenm_30_nlive_100,fig:.plots/chains_new_lcdm/Jenkins_DGP1_lenm_30_nlive_100}.  These figures show the 1,2-$\sigma$ credible regions for the free parameters in each of the fitting functions which passed our tests in \cref{app:calibration_of_multinest}.  The main diagonal shows the 1-$d$ posteriors marginalised over all other parameters, while the off-diagonal plots show correlations between pairs of parameters via the 2-$d$ credible regions at $1,2$-$\sigma$.  The black dashed lines show the published values for each HMF, \ie{} those proposed by their original authors.  (Where it is not visible, the value is outside the range of the plot provided by \texttt{GetDist}.)  The free parameters are shown in \cref{tab:hmf_functions} for each of the four families of HMF fitting function.
 
First we define appropriate criteria for a fitting function to produce a well-behaved fit to the data.  An ill-behaved fitting function, \eg{} with poorly constrained posteriors, suggests that this fitting function is sensitive to MG because our (incorrect) \lcdm{} hypothesis does not fit the data, no matter how far we vary the values of the free parameters.  A well-behaved fitting function is one for which the probability of the model given the data is well-constrained, \ie{} our hypothesis of an \lcdm{} cosmology \textit{with new parameter values} is likely \textit{despite} the fact that we know that our data emerge from a non-\lcdm{} simulation.   In this scenario, we would fail to detect the influence of MG if we attributed the shift in MAP-values to a factor such as differing cosmology or halo finders from the paper(s) in which the fitting function was first proposed.

We define a fit to be \lq\lq{}successful\rq\rq{} if the credible regions are continuous, smooth and peaked, tending to zero well within the prior parameter range.  We do not exclude a fit from being well-behaved if zero is within its $1\sigma$ credible region: instead we examine the proximity of the maximum-a-posteriori value, excluding the fit if the probability at zero is close to that of the MAP.  (A more quantitative option would be to exclude a fit if zero lies within the full-width-half-maximum of the Gaussian posterior, but we cannot guarantee that our posteriors are Gaussian.)  The upper limit is set by the priors, so our main criterion is the breadth of the posterior.  If it does not have a clear peak (or peaks), but a flat posterior, then there is too much uncertainty for the fit to be useful: it is not constrained by the data.
 
The majority of the mass functions are well-behaved for all MG models and both halo finders: Peacock (\cref{fig:.plots/chains_new_lcdm/Peacock_DGP1_lenm_30_nlive_100}), Jenkins (\cref{fig:.plots/chains_new_lcdm/Jenkins_DGP1_lenm_30_nlive_100}).  Initially, SMT-Courtin (\cref{fig:.plots/chains_new_lcdm/SMT-Courtin_DGP1_lenm_30_nlive_100}) appears to not be successful because both halo finders and all MG models have a peak as $p$ approaches zero.  (All the curves in the bottom right plot in the triangle in \cref{fig:.plots/chains_new_lcdm/SMT-Courtin_DGP1_lenm_30_nlive_100} increase as $p \rightarrow 0$; \cf{} the peaks on the 1-$d$ posteriors for $A$ and $p$.)  However, this lower bound is set by the physical requirement for the parameters to be positive, rather than an indication that we have not explored enough of the parameter space.  Moreover, considering that the prior volume of $p$ is in $[0,10]$, the motion of the maximum-likelihood from $0.1$ (Courtin \etal{}'s original value) to $0.0$ is only a $1\%$ shift relative to the size of the parameter space.  The Warren-Crocce fitting function exhibits slightly different behaviour in \cref{fig:.plots/chains_new_lcdm/Warren-Crocce_DGP1_lenm_30_nlive_100} for $b$.  Using the \texttt{MatchMaker} finder, only LCDM, DGP1 and F5 peak clearly away from $b \approx 0$; whereas using \texttt{RockStar} only SymB peaks at $b \approx 0$.  The same logic applies as with SMT-Courtin, except that we can be more reassured here because the majority of MG models produce a ML value at the published value (the black dashed line).  However, the posteriors are broader and more complicated in shape than the simple peaks for the other HMFs. This is probably due to the corrections made to the mass of each halo when Warren \etal{} derived the function: a halo of $N$ particles was corrected to $N(1 - N^{-0.6})$, producing a non-linear correction to the resulting $n(M)$. In this way we have found that all of our fitting functions can exhibit a degeneracy between a change in cosmology and a change in the underlying gravity theory.
 
We can now determine the sensitivity of these fitting functions to the halo data produced by the underlying theory of gravity.  We are interested in the overlap between credible regions as we change the MG $N$-body \lq\lq{}data\rq\rq{}.  Overlap indicates that these parameter pairs are insensitive to deviations from GR.  Conversely, should the values of (or degeneracies between) parameters change sufficiently between $N$-body simulation \lq\lq{}data\rq\rq{}, then we have an HMF in which deviations from \lcdm{} values may reliably indicate the underlying deviation from an \lcdm{} cosmology.  Over the next few paragraphs we examine each fitting function.  
  
In \cref{fig:.plots/chains_new_lcdm/Jenkins_DGP1_lenm_30_nlive_100}, the Jenkins HMF shows a high degree of overlap in parameters $a$ and $b$ (the bottom centre 2-$d$ posterior), so we must rely on $A$ to separate the various theories.  Both the posterior for $\left\{ A, a \right\}$ (centre-left) and for $\left\{ A, b \right\}$ (bottom-left), the MG theories clump into several groups.  Regardless of halo finder, DGP and LCDM are nearly indistinguishable, whereas F5 is distinct.  The overlap between F6 and the two symmetron models depends upon the halo finder: using \texttt{MatchMaker} (top triangle plot) F6 and SymA overlap and SymB is distinct; whereas using \texttt{RockStar} (bottom triangle plot) all three overlap.  Thus we can clearly identify whether a result is in one of the DGP-LCDM or Sym-FR \lq\lq{}families\rq\rq{} but not confidently be more specific.

The Peacock model shows similar behaviour in \cref{fig:.plots/chains_new_lcdm/Peacock_DGP1_lenm_30_nlive_100}.  Here the overlap is more severe.  (Consider the posteriors for $\left\{ a, c \right\}$ (bottom-left) and for $\left\{ b, c \right\}$ (bottom-centre).)   Thus only $A$ shows any spread in values, with bot DGP models overlapping \lcdm{} and the rest depending upon the halo finder.  Again for \texttt{RockStar} there is more overlap between the three \lq\lq{}families\rq\rq{}, whereas for \texttt{MatchMaker} SymB is distinct from the indistinguishable F6, F5 and SymA.
 
The situation is even more problematic for the SMT-Courtin fit in \cref{fig:.plots/chains_new_lcdm/SMT-Courtin_DGP1_lenm_30_nlive_100}.  In the 1-$d$ posteriors, not only does $a$ (centre) have significant overlap, but $p$ (bottom) peaks to the same value for all MG models.  Once again we find the DGP and LCDM models (themselves inseparable) not distinguishable from the $\fr$ and Symmetron ones.  This time, considering $\left\{ a, b \right\}$ (centre-left) \texttt{MatchMaker} (top triangle) separates SymB from the remaining three models, whereas \texttt{RockStar} (bottom triangle) separates F5 from the other three.  This demonstrates the impact of the halo finder on the question of universality, \ie{} invariance to MG.

The Warren-Crocce model has too much overlap to determine the underlying MG model.  \cref{fig:.plots/chains_new_lcdm/Warren-Crocce_DGP1_lenm_30_nlive_100} shows overlap for all of the 2-$d$ posteriors, to the extent that not all of the credible regions are visible.  The 1-$d$ posteriors (main diagonal) confirm that the parameters overlap, particularly $b$.  Although the MAP values do not overlap exactly for every MG model, they are sufficiently close that it would be difficult to distinguish between MG theories in this case.  We would not see the shift in parameter MAP-values in MG (coloured curves) compared to \lcdm{} (grey curves) which would signal a problem with our \lcdm{} hypothesis, using MG halo data.
 
Therefore we find that is it possible for MG to be mis-interpreted as a \lcdm{} result for all of the fitting functions which survived \cref{app:calibration_of_multinest}, bar Warren-Crocce.  However, we cannot distinguish between the underlying mechanisms for the deviation from \lcdm{}.  The same trends occur: the separation of LCDM and the two DGP models into one group and the $\fr$ and Symmetron ones into one or more others.  This latter group has behaviour which is halo-finder--dependent.  Once may extrapolate that this is due to the flat barrier of DGP, compared to the mass-dependent barriers of the other MG models (\cf{} \cref{fig:deltac_all}).  In practice, we cannot readily use this to test for varying theories of MG because a given point in parameter space can be occupied by the credible regions of multiple MG theories.  Nonetheless, it is remarkable that MG theories can be well-approximated by an HMF assuming an \lcdm{} cosmology.  The extra fifth-force interactions governing the halo collapse in the $N$-body data can be ignored for the purposes of excursion set theory.  Only a change in the best-fit parameters is required for the simple \lcdm{} excursion set model to match MG data.
 

\begin{figure}
  \begin{center}
    \subfloat[MG: MatchMaker halofinder, 30 bins]{\label{chains_new_mg/Jenkins_DGP1_lenm_30_nlive_500_MM_}
    \includegraphics[keepaspectratio,height=0.45\textheight]{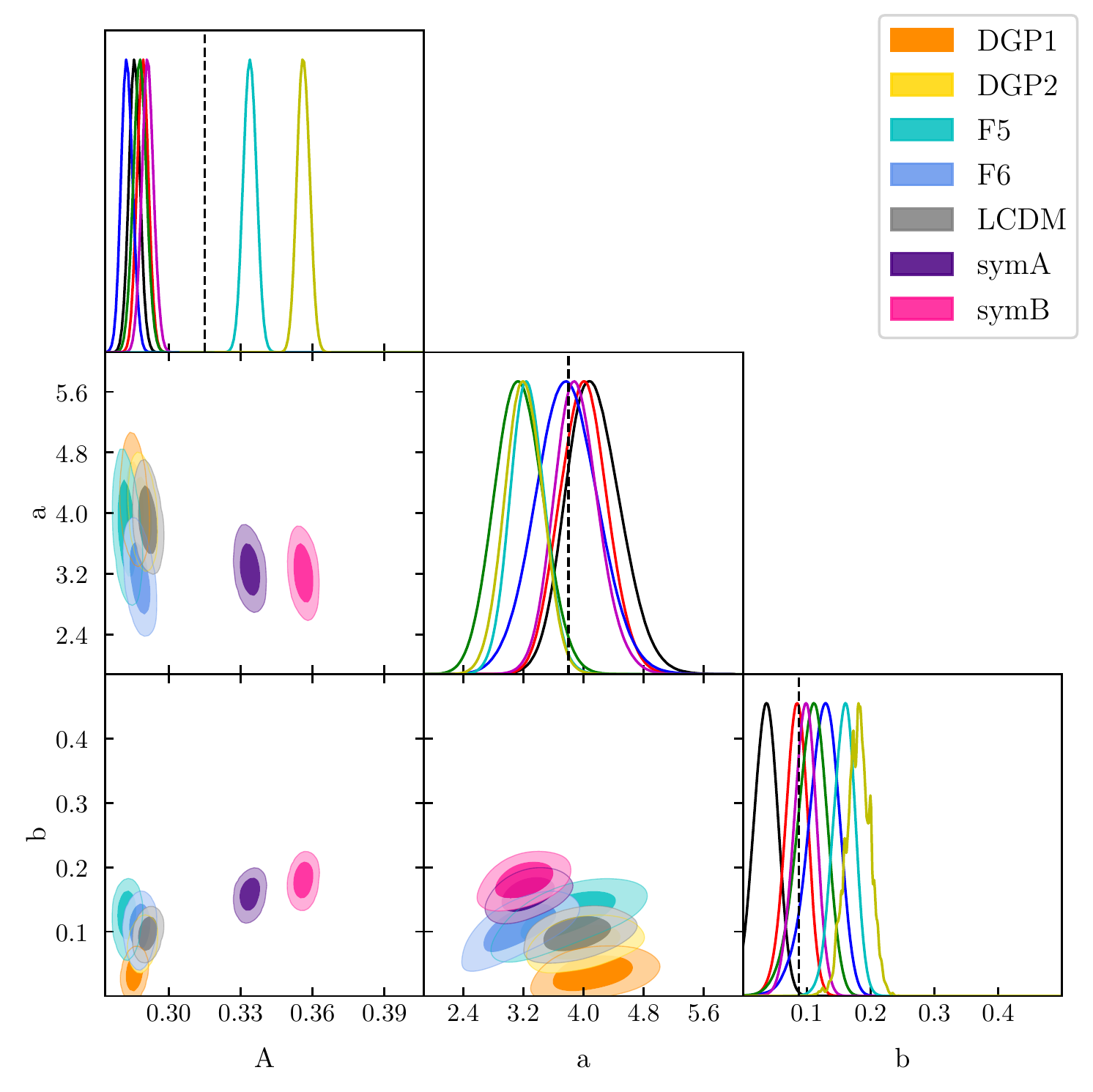}} \\
    \subfloat[MG: RockStar halofinder, 30 bins]{\label{chains_new_mg/Jenkins_DGP1_lenm_30_nlive_500_RS_}
    \includegraphics[keepaspectratio,height=0.45\textheight]{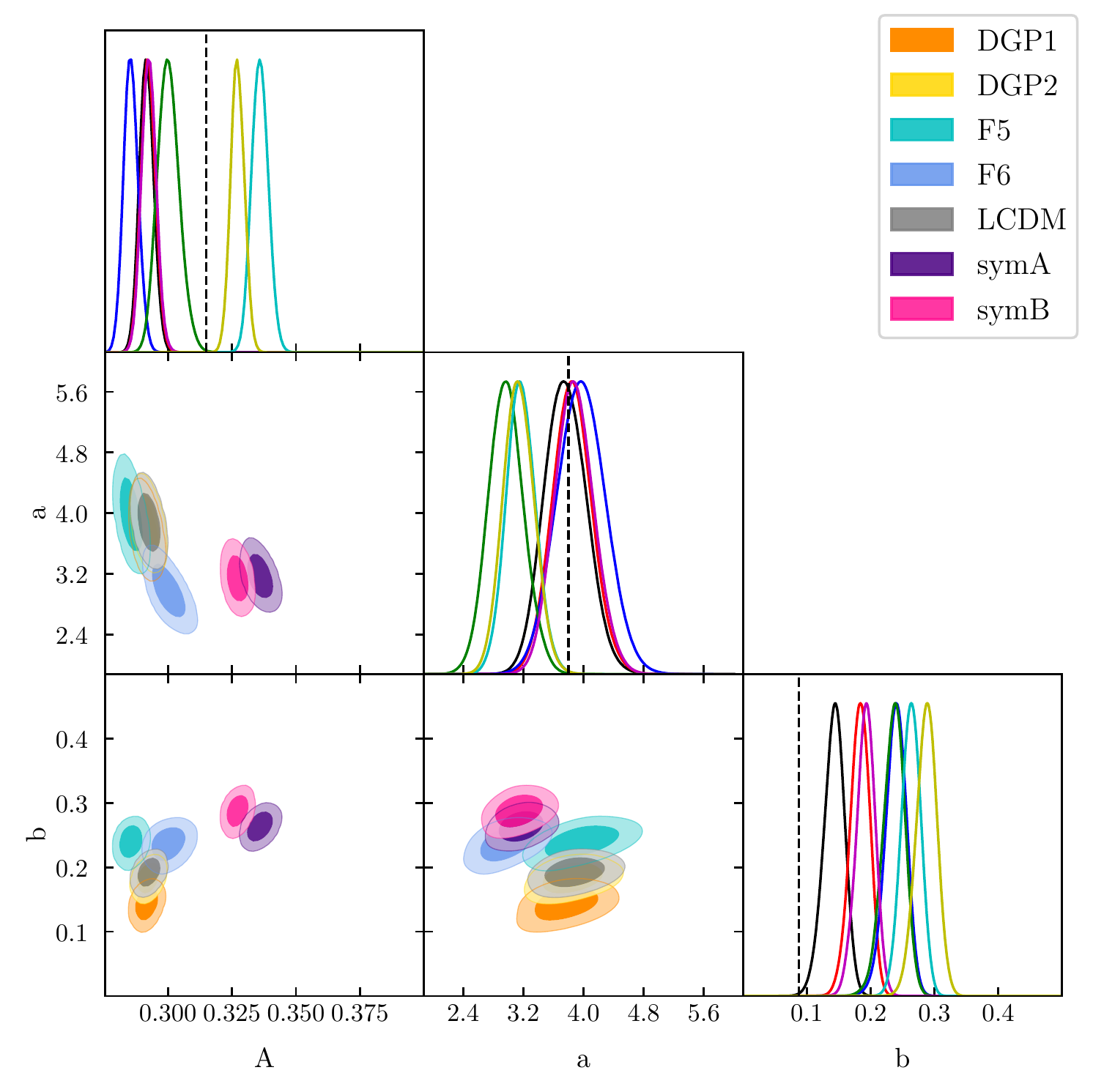}}
  \end{center}
  \caption{Triangle plot showing the 1-$\sigma$ (dark) and 2-$\sigma$ (light) posterior credible regions for the Jenkins HMF assuming the same MG fitting functions as the model in the $N$-body data (coloured regions). The black dashed lines show the values proposed by \cite{2001MNRAS.321..372J}.}
  \label{fig:.plots/chains_new_mg/Jenkins_DGP1_lenm_30_nlive_100}
\end{figure}

\begin{figure}
  \begin{center}
    \subfloat[MG: MatchMaker halofinder, 30 bins]{\label{chains_new_mg/Peacock_DGP1_lenm_30_nlive_500_MM_}\includegraphics[keepaspectratio,height=0.45\textheight]{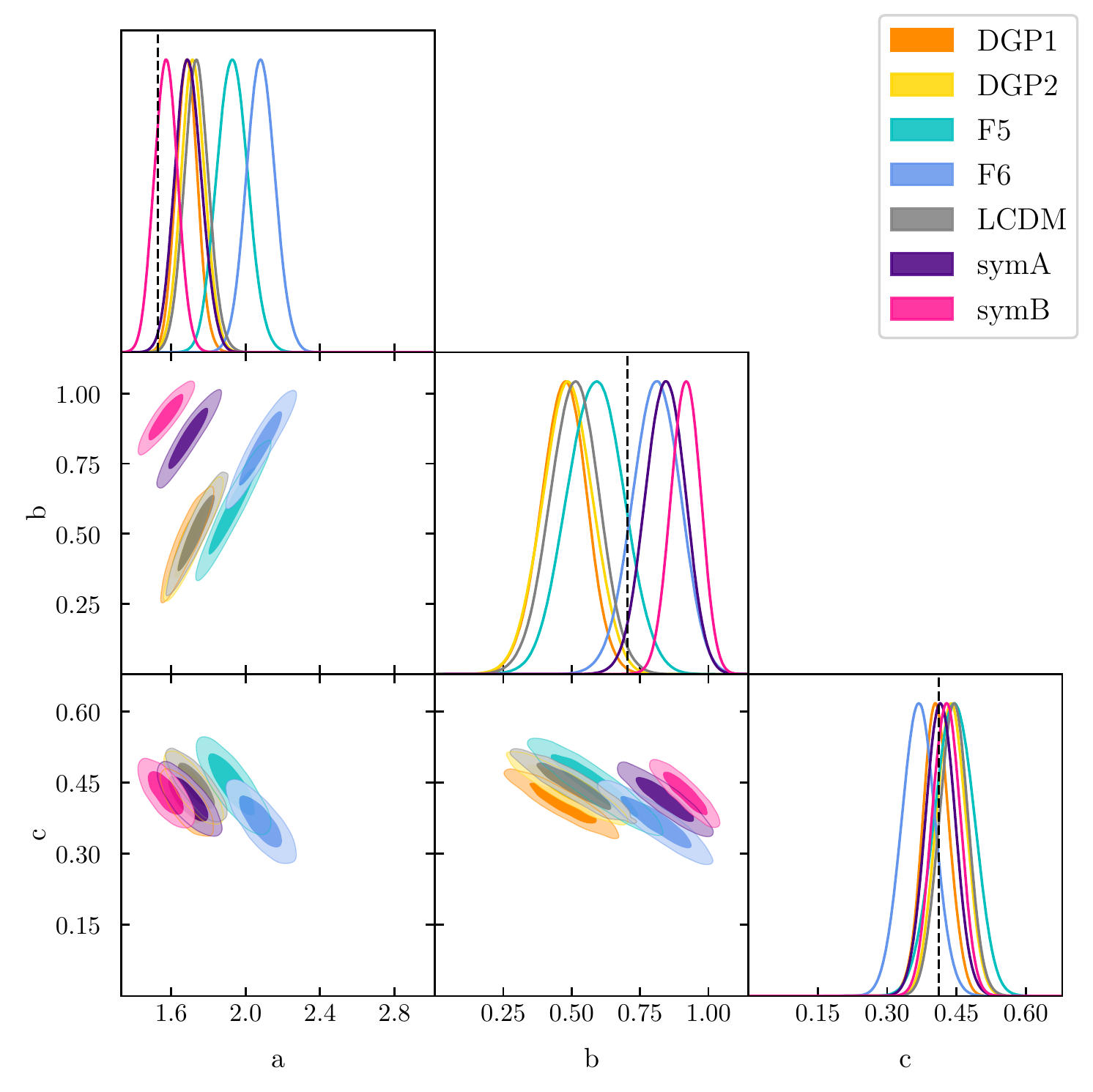}} \\
    \subfloat[MG: RockStar halofinder, 30 bins]{\label{chains_new_mg/Peacock_DGP1_lenm_30_nlive_500_RS_}\includegraphics[keepaspectratio,height=0.45\textheight]{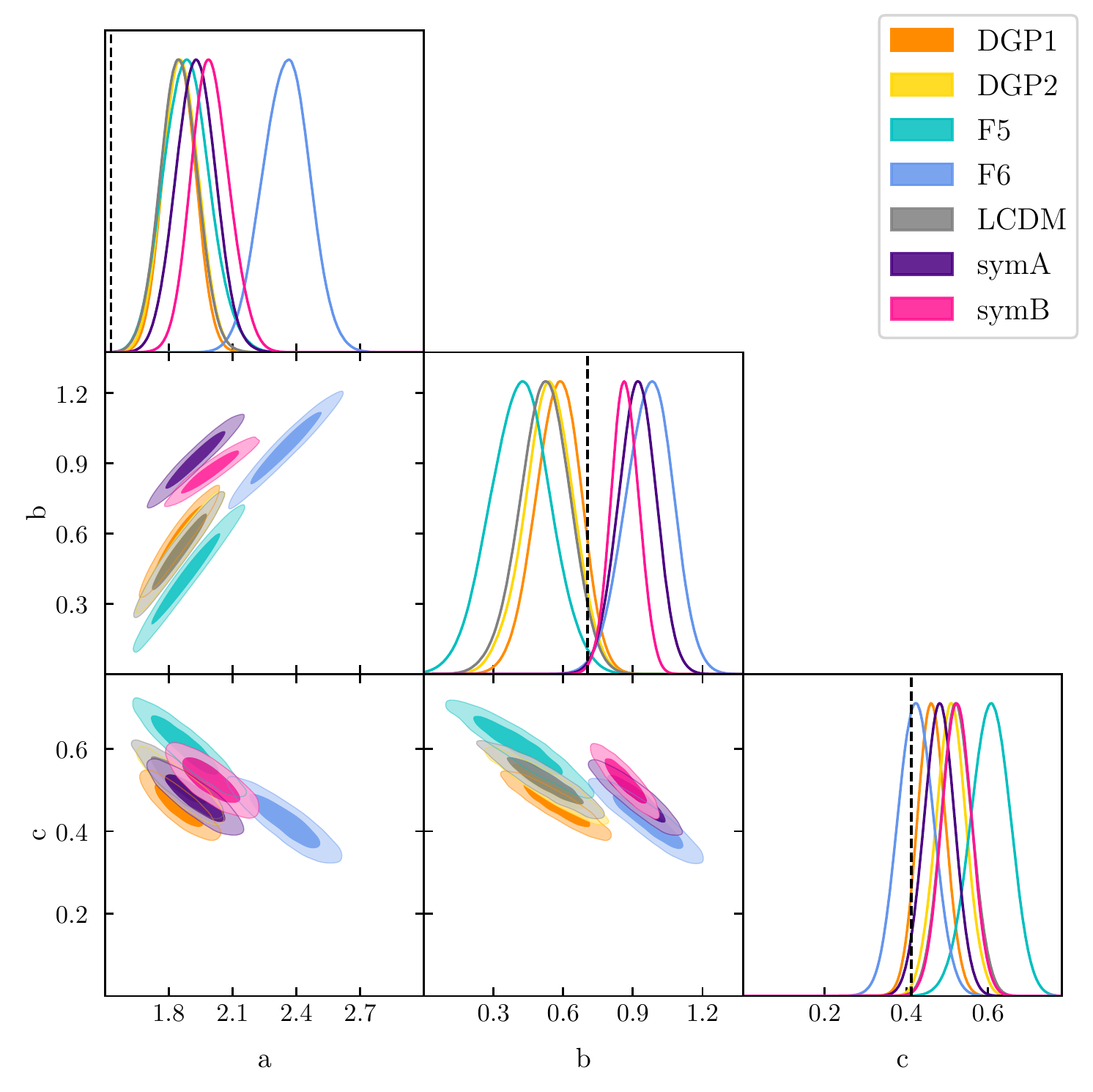}}
  \end{center}
  \caption{Triangle plot showing the 1-$\sigma$ (dark) and 2-$\sigma$ (light) posterior credible regions for the Peacock HMF assuming the same MG fitting functions as the model in the $N$-body data (coloured regions).  The black dashed lines show the values proposed by \cite{Peacock2007}.}
  \label{fig:.plots/chains_new_mg/Peacock_DGP1_lenm_30_nlive_100}
\end{figure}

\begin{figure}
  \begin{center}
    \subfloat[MG: MatchMaker halofinder, 100 bins]{\label{chains_new_mg/SMT-Courtin_DGP1_lenm_30_nlive_500_MM_}\includegraphics[keepaspectratio,height=0.45\textheight]{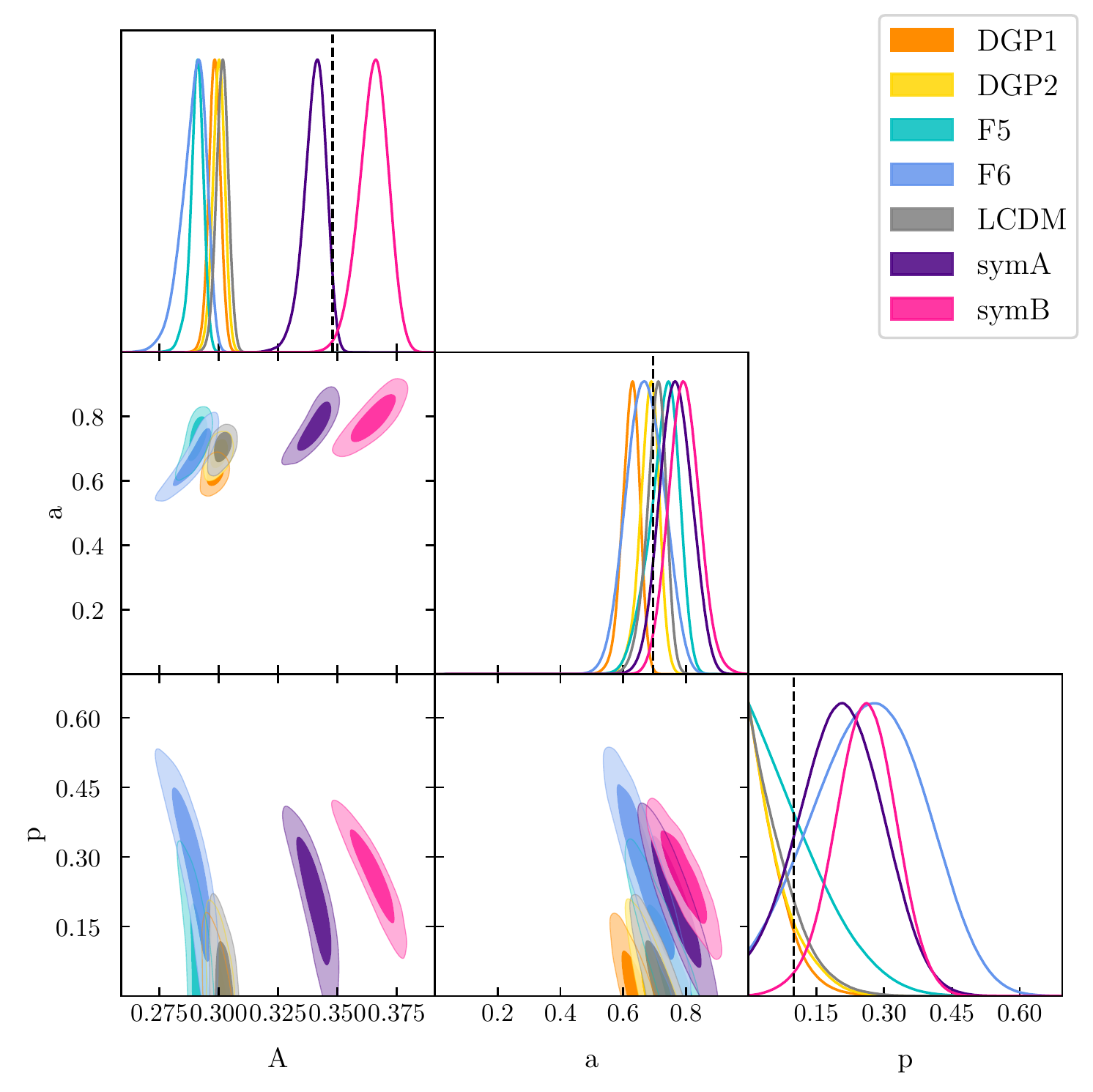}} \\
    \subfloat[MG: RockStar halofinder, 100 bins]{\label{chains_new_mg/SMT-Courtin_DGP1_lenm_30_nlive_500_RS_}\includegraphics[keepaspectratio,height=0.45\textheight]{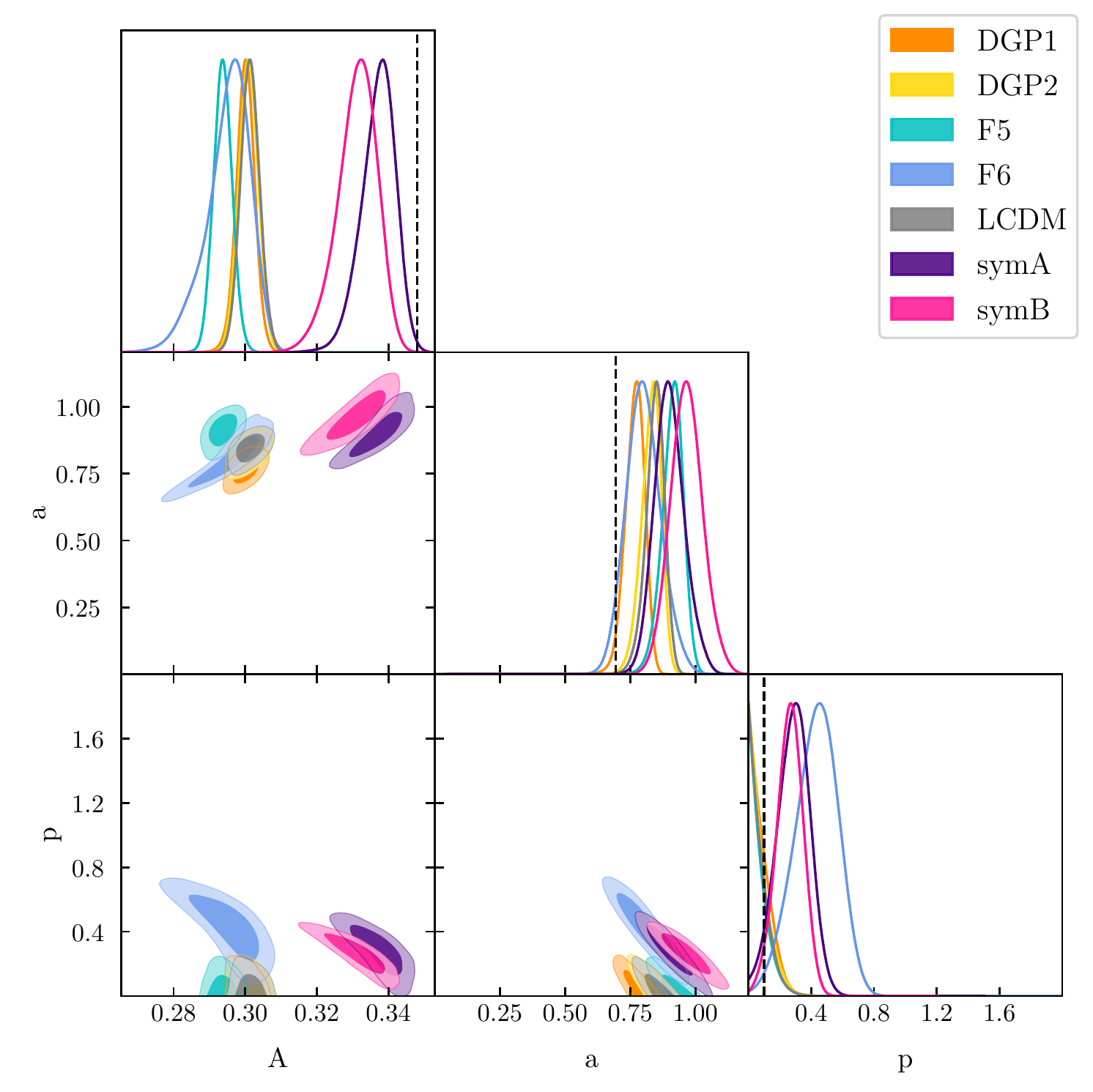}} 
  \end{center}
  \caption{Triangle plot showing the 1-$\sigma$ (dark) and 2-$\sigma$ (light) posterior credible regions for the SMT-Courtin HMF assuming the same MG fitting functions as the model in the $N$-body data (coloured regions).  The black dashed lines show the values proposed by \cite{2011MNRAS.410.1911C}.}
  \label{fig:.plots/chains_new_mg/SMT-Courtin_DGP1_lenm_30_nlive_100}
\end{figure}
  
\begin{figure}
  \begin{center}
    \subfloat[MG: MatchMaker halofinder, 30 bins]{\label{chains_new_mg/Warren-Crocce_DGP1_lenm_30_nlive_500_MM_}\includegraphics[keepaspectratio,height=0.45\textheight]{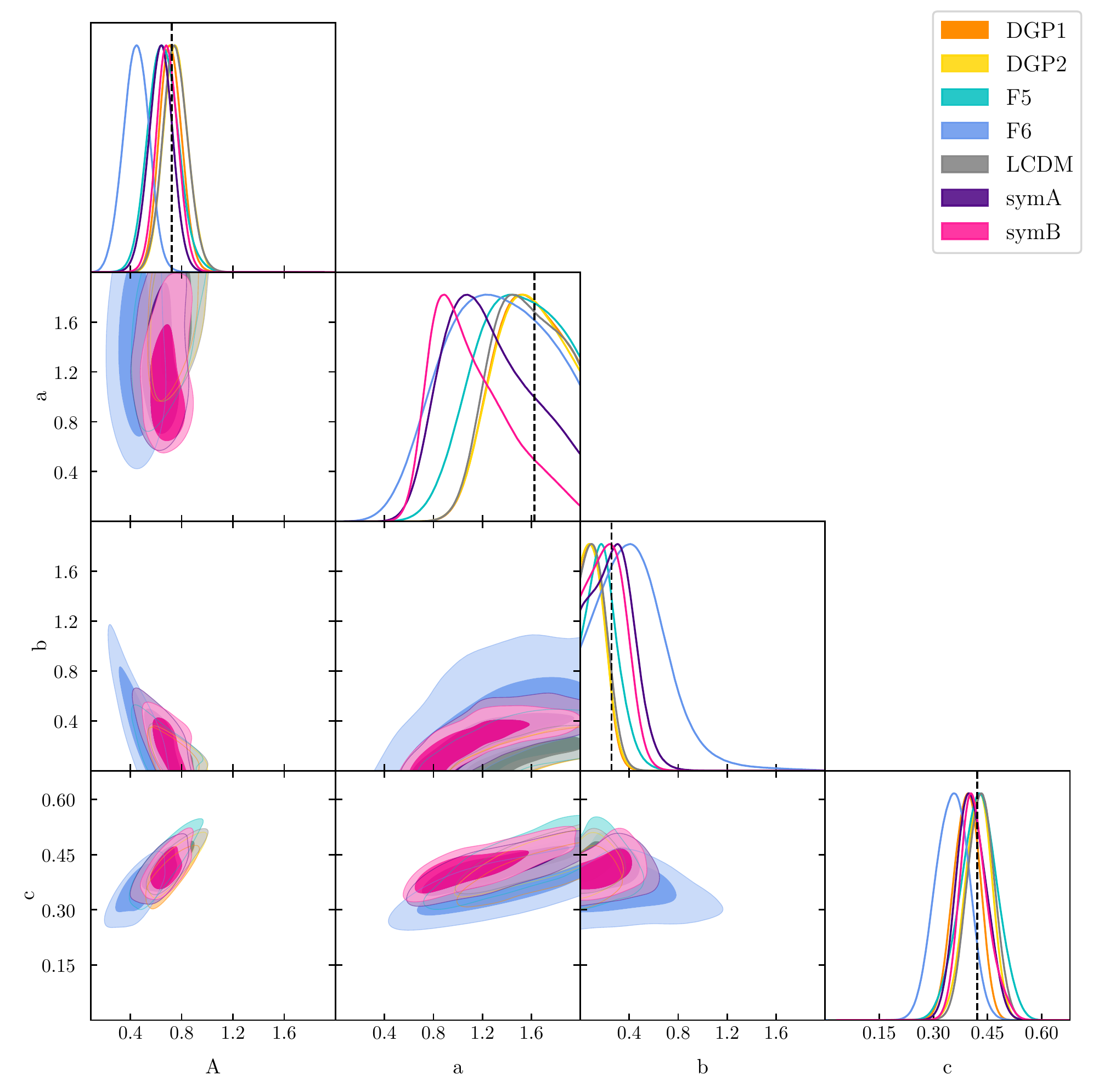}} \\
    \subfloat[MG: RockStar halofinder, 30 bins]{\label{chains_new_mg/Warren-Crocce_DGP1_lenm_30_nlive_500_RS_}\includegraphics[keepaspectratio,height=0.45\textheight]{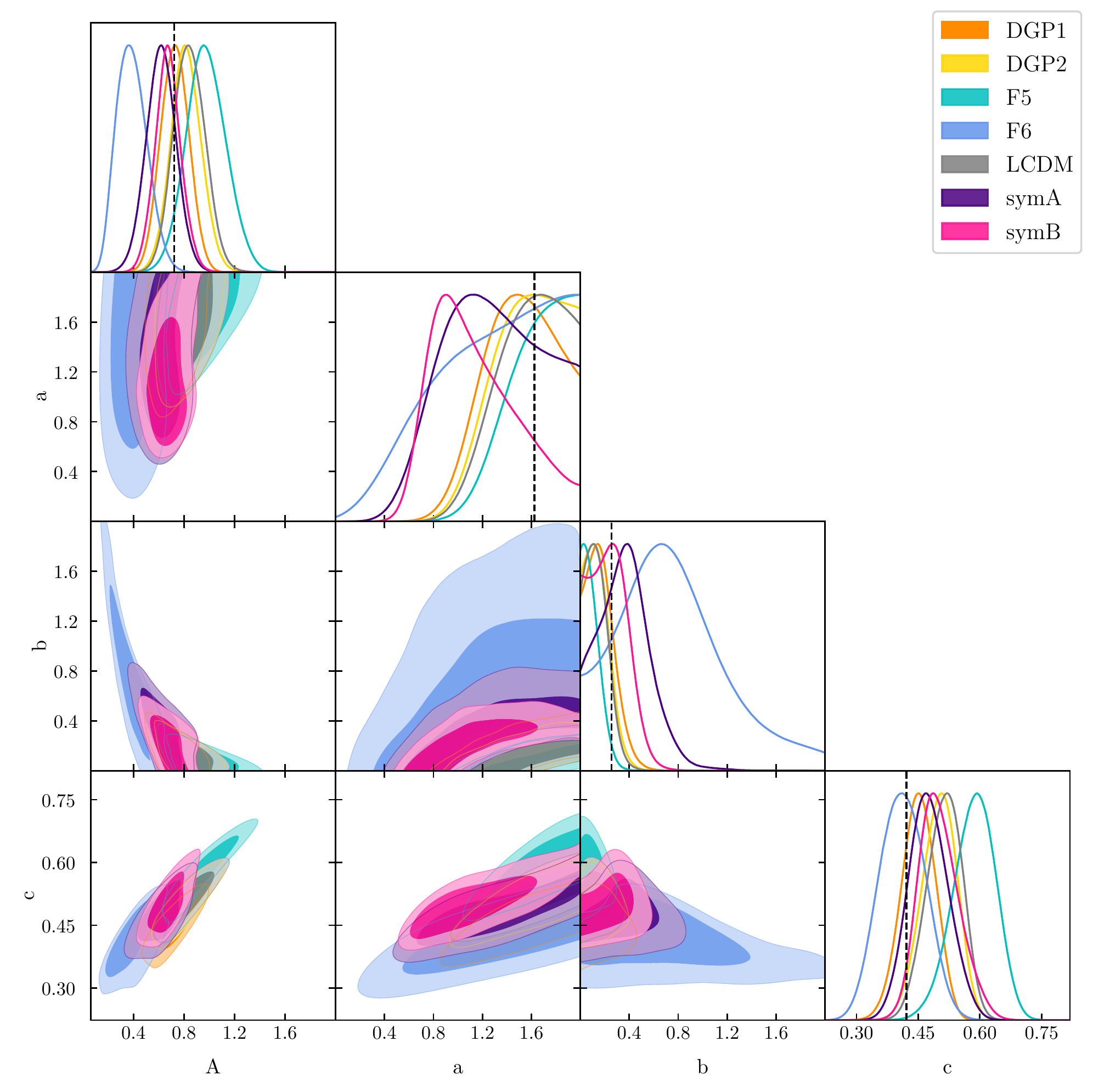}}
  \end{center}
  \caption{Triangle plot showing the 1-$\sigma$ (dark) and 2-$\sigma$ (light) posterior credible regions for the Warren-Crocce HMF assuming the same MG fitting functions as the model in the $N$-body data (coloured regions).  The black dashed lines show the values proposed by \cite{2006ApJ...646..881W}.}
  \label{fig:.plots/chains_new_mg/Warren-Crocce_DGP1_lenm_30_nlive_100}
\end{figure}

\subsection{Recalibrating best-fit parameters} 
\label{sub:recalibrating_best_fit_parameters}
 
Now we explore the opposite question to the previous one, namely: how universal is the halo mass function?  In this subsection, we use the same extended gravity model for both the $N$-body data and the collapse density for the halo mass function.  First we define what we mean by \lq\lq{}universality\rq\rq{} in the context of MG HMFs.  Then we explain why our \lcdm{} results do not recover the best-fit values of the free parameters proposed by the original authors of the fitting functions.  Next we discuss the results for each fitting function in detail.

We use the term \lq\lq{}universal\rq\rq{} in the sense that a given fitting function is insensitive to changes in the gravity theory underlying the spherical collapse of the haloes.  Specifically, provided that we account for the changes in MG by using the correct collapse density in $\nu$, we expect a universal HMF to predict the same best-fit values for the free parameters in the fitting function.  In this way, $\nu$ would account for changes in gravitational models, without requiring recalibration of the fitting function.  This is entirely analogous to the definition of a universal HMF in GR, where the universality refers to the fitting function being insensitive to changes in redshift and cosmological parameters.  One of the drivers behind the proliferation of fitting functions in \cref{tab:hmf_functions} is the quest for such au universal function in \lcdm{} (and earlier, in other CDM cosmologies).  This would vindicate the current practice to derive and calibrate HMF functions in \lcdm{} $N$-body simulations alone---with the advantage that they could then be reliably applied to any scalar theory of MG with the same cosmological parameters.
  
Should we expect the LCDM results from our calculations to equal those from the original values of their respective papers?  We do not expect to recover these results for a variety of reasons:
 
\begin{itemize}
\item The value of $\delta_{c}$ changes with $\Omega_{m0}$ and $H_{0}$
\item The value of $\sigma$ depends upon the power spectrum and value of $\sigma_{8}$
\item Differing halo finders and settings
\item The presence (or absence) of corrections to the $N$-body halo data
\end{itemize}
Let us examine each of these in turn.
 
\cref{tab:cosmologies} summarises the cosmological parameters used for each of the papers which provided "best-fit" parameters for particular fitting functions.   The dependence of $\delta_{c}^{\Lambda}$ on $\Omega_{m0}$ is given in parametric form by \cite{1995MNRAS.274L..73E} as $\delta_{c} = 1.68/\Omega_{m0}^{0.28}$.  Our value is $\delta_{c}^{\Lambda} \approx 1.675$.   This contrasts with most of the papers (bar \cite{2011MNRAS.410.1911C}).  Some \cite{Sheth:2001dp,Peacock2007,2010MNRAS.403.1353C} use the SCDM value $\delta_{c} \approx 1.686$ (following \PS{} \cite{1974ApJ...187..425P}) regardless of the actual cosmology of their simulations.  Others \cite{2001MNRAS.321..372J,2008ApJ...688..709T,1203.3216,2013MNRAS.433.1230W,2006ApJ...646..881W} absorb the value of $\delta_{c}$ into their free parameters because they use $\sigma$ as the independent variable rather than $\nu$, so their choice---which we need to convert back to $\nu$---is ambiguous.  In all cases we have assumed the exact SCDM collapse value $\delta_{c} = 3 \cdot (12 \pi)^{2/3} / 20$ when converting from $\sigma$ to $\nu$.  Only \cite{2011MNRAS.410.1911C} vary $\delta_{c}$ according to the solution of the spherical collapse equation for a variety of cosmologies.   
Thus we have a different numerator for $\nu$.  Moreover, it is unlikely that our $P(k)$ is precisely equal to any of those used in the papers.

Even if it did match, the normalisation $\sigma_{8}$ differs.  The variance $\sigma$ which forms the denominator of $\nu$ therefore also differs from previous publications.  Since the independent variable $\nu$ (or more precisely, our mapping $\ln M \rightarrow \nu$ from counts to \fcd{}) differs in this paper (and indeed in all of the others), our free parameters must change to compensate.  Even if we had exactly the same numbers of counts and the same bins, and our best-fit HMF had the same $n(M)$ values as a preceding paper, we would see a change in $f(\nu)$.  This is a small contribution to the movement of the maximum-likelihood peaks from the published values.
 
The choice of halo finder further abstracts the problem.  All of the published values except Peacock and Watson (both of which are FoF-only) arise from a compromise between the best-fit values for multiple halo finders.  For example, Jenkins \cite{2001MNRAS.321..372J} uses both FoF and SO halo finders to find a fitting function which has a residual of within 20\% compared to their $N$-body HMFs simulated for a range of cosmologies (not just \lcdm{}).  Even when restricting their data to FoF-only and SO-only (whereby they obtain different best-fit values), the FoF linking length is changed between cosmologies ($b=0.2$ for $\tau$CDM, $b=0.164$ for \lcdm{}), so we cannot disentangle the cosmological effects from the reduction of the $N$-body \lq\lq{}data\rq\rq{}.  As we can see by comparing the results for \texttt{MatchMaker} and \texttt{RockStar} in \cref{fig:.plots/chains_new_mg/Jenkins_DGP1_lenm_30_nlive_100,fig:.plots/chains_new_mg/Peacock_DGP1_lenm_30_nlive_100,fig:.plots/chains_new_mg/SMT-Courtin_DGP1_lenm_30_nlive_100,fig:.plots/chains_new_mg/Warren-Crocce_DGP1_lenm_30_nlive_100}, the same distribution of mass within the $N$-body simulations produces a different HMF according to the halo finder used.  \cite{2011MNRAS.415.2293K} discuss this in great detail for \lcdm{} and we have no reason to disagree with their findings.
 
Differing authors also treat the distribution $n(M)$ obtained from their halo finder(s) in a variety of ways.  As we briefly covered in \cref{sec:algorithms}, the mass cutoffs are controlled by the simulation box size and the mass resolution.  \cref{tab:hmf_functions} summarises the mass range used by each paper.  This shifts the sampling along a subsection of the actual HMF, changing the influence of each parameter in the likelihood function.  An extreme example of this is the Reed 2003 fit, whose parameter $c$ is completely unconstrained by masses $M \leq 10^{15} M_{\odot}$ \cite{2003MNRAS.346..565R}.  Sometimes the individual halo masses are systematically \lq\lq{}corrected\rq\rq{}, \eg{} \cite{2017MNRAS.467.3454B} or (more relevantly here) the Warren correction: a halo of $N$ particles was updated to $N(1-N^{-0.6})$ to account for perceived flaws in the halo finders.  These factors all affect the final \fcd{} of the data, against which is compared the \fcd{} of the model (converted to counts per bin) which contains the values of the free parameters at a chosen point in the parameter volume.
 
Given these factors, it is not surprising that our LCDM values do not always align with the published values.   In  \cref{fig:.plots/chains_new_mg/Jenkins_DGP1_lenm_30_nlive_100,fig:.plots/chains_new_mg/Peacock_DGP1_lenm_30_nlive_100,fig:.plots/chains_new_mg/SMT-Courtin_DGP1_lenm_30_nlive_100,fig:.plots/chains_new_mg/Warren-Crocce_DGP1_lenm_30_nlive_100} the grey lines show our \lcdm{} posteriors, while the black dashed lines show the values from one of the relevant papers respectively.  (We do not show all of the published values for every fitting function because this would lead to up to five lines on some plots, which would be confusing.)  Of particular interest is the variety of best-fit values for the SMT fitting function.  Our changes in all HMFs are the same order of magnitude as the changes other authors have found when altering the cosmological parameters, halo finders and data reduction techniques used to derive the HMF.    Since we are utilising a single cosmology and separating the effects of our two halo finders, it is not surprising that our values do not reproduce the existing ones.
 
Now we extend our discussion to our MG results.  \cref{fig:.plots/chains_new_mg/Warren-Crocce_DGP1_lenm_30_nlive_100,fig:.plots/chains_new_mg/Peacock_DGP1_lenm_30_nlive_100,fig:.plots/chains_new_mg/SMT-Courtin_DGP1_lenm_30_nlive_100,fig:.plots/chains_new_mg/Jenkins_DGP1_lenm_30_nlive_100} show the 1,2-$\sigma$ credible regions for the free parameters in each of the fitting functions which passed our tests in \cref{app:calibration_of_multinest}.  As in the previous subsection, the main diagonal shows the 1-$d$ posteriors marginalised over all other parameters, while the off-diagonal plots show correlations between pairs of parameters via the 2-$d$ credible regions at $1,2$-$\sigma$.  The black dashed lines show the published values for each HMF, \ie{} those proposed by their original authors.  (Where it is not visible, the value is outside the range of the plot provided by \texttt{GetDist}.)  We examined the behaviour of both halo finders, showing the \texttt{MatchMaker} in top subplots and \texttt{RockStar} in the bottom subplots.

Some HMFs have precisely the same \lq\lq{}good behaviour\rq\rq{} (or lack thereof) in both cases.  The Jenkins HMF has the same good behaviour in \cref{fig:.plots/chains_new_mg/Jenkins_DGP1_lenm_30_nlive_100} (full MG $\nu$) as we already saw in \cref{fig:.plots/chains_new_mg/Jenkins_DGP1_lenm_30_nlive_100} (\lcdm{} $\nu$).  Similarly, the Peacock HMF is well-behaved in both \cref{fig:.plots/chains_new_mg/Peacock_DGP1_lenm_30_nlive_100} (full MG $\nu$) and \cref{fig:.plots/chains_new_lcdm/Peacock_DGP1_lenm_30_nlive_100} (\lcdm{} $\nu$).    It is notable that the Peacock HMF---an empirically-derived fit to the Watson-FoF HMF accurate to 1\% in \lcdm{}---does behave well when generalised, whereas the original Watson fit does not.  (In fact, we discarded the Tinker-Angulo-Watson fitting function in calibration in \cref{app:calibration_of_multinest}.)  This suggests that the behaviour is not caused by over-simplifying the gravitational collapse, but by the underlying form of the fitting function itself.  Despite the fact that we have used the full MG modifications to $\nu \left( S \given \denv, \Senv \right)$, the behaviour of these HMFs mirrors that of the previous subsection, in which we assumed $\nu = \delta_{c}^{\Lambda} / S$.  In order to vindicate our more complex hypothesis as the correct gravitational model underlying the $N$-body data, we would need to calculate the evidence factors for our hypotheses.

In some cases, accounting for the mass dependence of the critical density does improve the behaviour of the HMF.  While SMT-Courtin has the same good behaviour for $A$ (left column in triangle plot) and $a$ (centre column) in \cref{fig:.plots/chains_new_mg/SMT-Courtin_DGP1_lenm_30_nlive_100} (full MG $\nu$) and \cref{fig:.plots/chains_new_mg/SMT-Courtin_DGP1_lenm_30_nlive_100}, the behaviour of $p$ improves.  Whereas in our \lcdm{}-$\nu$ results, all of the MAP values for $p$ were zero (bottom-right of triangle plot), the posteriors peak away from zero for SymA, SymB and F6 in our MG-$\nu$ results, regardless of halo finder.  Warren-Crocce shows some improvement as well.  Again, we see good behaviour for $A$ (left column in triangle plot) and $c$ (right column) in \cref{fig:.plots/chains_new_mg/SMT-Courtin_DGP1_lenm_30_nlive_100} (full MG $\nu$) and \cref{fig:.plots/chains_new_lcdm/SMT-Courtin_DGP1_lenm_30_nlive_100} (\lcdm{} $\nu$).  The imrpovement in the remaining parameters $a$ and $b$ depends upon the MG theory.  Instead of the posteriors peaking at $b \approx 0$ in \cref{fig:.plots/chains_new_lcdm/Warren-Crocce_DGP1_lenm_30_nlive_100}, the ones in \cref{fig:.plots/chains_new_mg/Warren-Crocce_DGP1_lenm_30_nlive_100} are well-defined for F5,6 models using the \texttt{MatchMaker} halo finder (top plots); moreover \texttt{RockStar} behaves (relatively) well for everything apart from SymB in the bottom plot.  Thus we see that generalising $\nu$ from its \lcdm{} value (proportional to $\sigma^{-1}$) to MG does improve the overall behaviour of the HMFs, when the barrier density is not a constant.
 
We can use the HMFs to examine the universality of $\nu$: specifically, whether accounting for the excursion set behaviour of the MG models is sufficient to render our fitting functions independent of MG.  While it is evident from \cref{fig:.plots/chains_new_mg/Jenkins_DGP1_lenm_30_nlive_100,fig:.plots/chains_new_mg/Peacock_DGP1_lenm_30_nlive_100,fig:.plots/chains_new_mg/SMT-Courtin_DGP1_lenm_30_nlive_100} that no single value works for every MG model, we do find clustering between families.  The Jenkins posteriors (\cref{fig:.plots/chains_new_mg/Jenkins_DGP1_lenm_30_nlive_100}) show that the credible regions for the two symmetron models are quite distinct from the other five models, all of which have overlapping credible regions. 
The Peacock function shows a higher degree of universality than Jenkins: particularly in \cref{chains_new_mg/Peacock_DGP1_lenm_30_nlive_500_MM_}, but slightly less in \cref{chains_new_mg/Peacock_DGP1_lenm_30_nlive_500_RS_}.  In particular $c$ is practically universal, but there is a spread of overlapping values for the other two parameters. 
SMT-Courtin (\cref{fig:.plots/chains_new_mg/SMT-Courtin_DGP1_lenm_30_nlive_100}) has very similar behaviour to the Jenkins HMF, with the two Symmetron models distinct from each other as well as the clustering--but here it is four of the other five models, with F6 closer to the Symmetron results.
Unlike the preceding fitting functions, Warren-Crocce (\cref{fig:.plots/chains_new_mg/Warren-Crocce_DGP1_lenm_30_nlive_100}) has posteriors which largely overlap, perhaps with the exception of F6, and more clustering in \texttt{MatchMaker} than \texttt{RockStar}.  This fitting function has no credible regions which are isolated from one other.  Thus, the fitting functions display a range of behaviours, but most show that the best-fit parameter values for multiple MG models do overlap.
 
Where the symmetron models are distinguishable from the others, the differences are driven by the \lq\lq{}normalisation\footnote{While $A$ originally played this role in the SMT-Courtin function \cite{Sheth:2001dp}, we no longer require the cumulative mass fraction to tend to unity.} factor:\rq\rq{} the symmetron models underpredict $n(M)$, requiring a systematic upwards shift by increasing the multiplicative factor $A$.  These models have a drifting barrier $\delta_{c}(S)$ which includes mass dependence.  However, this dependence via the collapse density ODE is not accounting for all of the actual behaviour of the haloes in non-linear collapse.  Considering that we use a simple model of a collapsing spherical top-hat, we may be oversimplifying the effect of the symmetron fifth-force.
 
There is also the issue of the F5 and F6 models exhibiting greater spread (albeit still with overlapping credible regions) than the LCDM and DGP results.  This is particularly visible in Warren-Crocce.  Recall that here we fully account for the excursion-set barrier density using the Volterra solution.  However, we use this to scale the unconditional HMF, so we approximate the integral over the environmental dependence of the HMF via the value at the peak of the environment distribution (which happens to be $\denv = 0$).  Under these circumstances, it is remarkable that our approximation does produce such a universal result.
 
The two DGP models cluster strongly with \lcdm{} in all the HMFs.  This is possibly because all of these models use a flat barrier, so there is no additional mass- or environment-dependence to be included in $\nu$, so no additional excursion set behaviour which needs to be approximated by the change in the independent variable. 
 
While we do not find a strong degree of overall universality, we can see that the different screening mechanisms do behave similarly.  This is independent of the choice of halo finder, so this clustering is not caused by systematic effects or scatter in the $N$-body data.
 
This illustrates the caution which must be employed when using fitting function originally calibrated in \lcdm{} in the context of MG.  Although the generalisation of the fitting functions from \lcdm{} to MG does not produce a completely universal fitting function, there is a degree of universality in the clustering of the credible regions for different screening mechanisms.  This is because the effects of the fifth-force are largely encapsulated by the modified Poisson equation which appears in the ODE for gravitational collapse.  The resulting $\delta_{c}$ clearly does not contain all of the non-linear collapse information (otherwise we \textit{would} have a universal HMF) but it does incorporate enough into $\nu$ that the resulting fitting function depends only on the type of screening, rather than the values of the fifth-force parameters.

 
 
\section{Conclusions} 
\label{sec:conclusions}

This section reiterates the salient points of this paper.  We outline the method we have used, before describing avenues for generalisation and other possibilities for further work.  We conclude by summarising the key results of this paper.

\subsection{Summary} 
\label{sub:summary}

In this paper, we explored the use of the halo mass function in screened MG theories.  We selected a range of theories which have different screening mechanisms (\cref{sub:dgp_gravity,sub:fr_gravity,sub:symmetron_gravity}) and derived their additional contribution to the Poisson equation.  We summarised a variety of HMFs and described the nature of their universality in GR and how to transform this into the equivalent in MG (\cref{sec:the_halo_mass_function}).  The $N$-body simulations from which we extracted halo catalogues to compare to our empirical fits are described in \cref{sub:n_body_simulations_haloes}.  Similarly, the Bayesian methodology for estimating maximum-likelihood parameters and the relative likelihood of the different models is outlined in \cref{sub:bayesian_inference}.  The key steps of our method are:
\begin{enumerate}
\item Conversion of the GR HMF from $\sigma$ to $\nu$ (if necessary).
\item Calculation of the effective fifth-force $F_{\textrm{eff}}$ to insert into the spherical collapse ODE.
\item Calculations of the collapse density $\delta_{c}(S,\denv,\Senv)$ to incorporate into $\nu$.
\item Use of an appropriate excursion-set technique to account for the barrier density $\delta_{c}$.
\item MCMC estimation of the best-fit free parameter values and their credible regions.
\item Output of the corresponding best-fit HMF.
\end{enumerate}
 
Our main results (\cref{sec:results_and_discussion}) are as follows.  

We found a broad spectrum of possible methods---some newly-proposed in this paper---by means of which we can incorporate MG into fitting functions originally designed for \lcdm{} alone.  Of the various techniques for extending the fitting functions to chameleon MG, we have found that some are more suited to certain applications (\eg{} the cosmic web approach) or halo finders (\eg{} the two density-marginalised methods) than others.  This demonstrates the additional complexity which environment dependence produces in chameleon screening compared to symmetron- and Vainshtein-screened theories.  We cannot neglect this and simply substitute the unconditional HMF if we wish to produce a useful empirical function to use in lieu of deriving one from $N$-body simulations.
 
We found that the effects of MG can be interpreted as a change in best-fit parameters in the \lcdm{} HMF for all of the fitting functions.  Alternatively, the relation can be inverted to judge the universality of the HMF, \ie{} its independence on the underlying theory of gravity.  Although we found no completely universal HMF, the parameter values did cluster according to the type of screening mechanism, with Jenkins, Peacock and SMT-Courtin being the least universal and Warren-Crocce the most.  The former group required very different best-fit parameters for the two Symmetron models, whereas in the latter all of the models had overlapping credible regions.  The results suggest that a single, best-fit HMF might be used for each type of screening, independent of the parameters in the MG model.  This demonstrates that the additional complexity of the gravitational collapse in screened MG theories cannot always be accounted for using the techniques developed in GR.  However, it is unnecessary to develop new fitting functions and calibrate them on a case-by-case basis. 
 
We have demonstrated that it is possible to generalise some of the halo mass functions in common use in GR to incorporate MG theories with a variety of screening mechanisms.  However, the calibration of these fitting functions has a number of caveats which are not encountered in the \lcdm{} framework for which they were initially developed.  Nonetheless, it is remarkable that our method can incorporate much of the non-linear collapse behaviour of screened MG in a simple and efficient mechanism.  This is in direct contrast to the difficulties encountered in performing $N$-body simulations in screened MG.  Thus we have provided an excursion-set-motivated alternative in MG to the need to replicate the time-consuming development of accurate halo mass functions which took place (and is ongoing) in GR.
 
 
\subsection{Further work} 
\label{sub:further_work}
 
The method presented in this paper for calculation of the MG halo mass function using the fitting functions derived from \lcdm{} has many avenues for generalisation.  Most straightforward of these is the application to other fitting functions as they become available, provided that these functions can be expressed in terms of the \lq\lq{}universal parameter\rq\rq{} $\nu$ rather than the variance $\sigma$. 
 
The universality of the halo mass function can be further extended to higher redshifts.  The collapse ODE (derived in \cite{2012MNRAS.421.1431L}) has a new stopping condition that $y_{h}(z_{c}) = 0$, but the same bijection scheme can be applied to calculate the collapse density $\delta_{c}(z_{c})$.  The variance $\sigma(z)$ is obtained from the present-day value via the growth factor $D(z)$.  However, to a good approximation, these modifications cancel, leaving $\nu$ independent of $z$ \cite{Zentner:2006vw}.  This generalisation is particularly relevant given the ongoing discussion on the $z$-independence of $f(\nu)$ in \lcdm{}.  It would be particularly interesting to determine the influence of the fifth-force on the evolution of the HMF.
 
The calibration techniques are applicable to any MG theory which satisfies the following:
\begin{itemize}
\item Existence of a modified Poisson equation to approximate the modifications to gravity
\item Well-posedness of the corresponding spherical-collapse ODE
\item Background expansion similar to \lcdm{} (so that the \lcdm{} growth factor can be used and the halo environment treated as \lcdm{} in the collapse ODE)
\end{itemize}
Galileon MG is an example of a screened theory for which this technique may be used.  However, it may also be applied to MG theories which do not involve screening, but have some other method of being observationally-viable.  It would be interesting to investigate whether the results we have found are unique to screening models, or whether they extend to non-screened theories.
  
The cosmology-dependence of these results can also be explored.  This would be a daunting task, requiring $N$-body simulations for a grid of cosmological parameters, especially given the additional complexity of incorporating a fifth-force into the simulations.  Nonetheless, this would permit comparison with the investigation of the cosmological-dependence of the HMF in GR (\eg{} \cite{2011MNRAS.410.1911C}).  Moreover, if using changes in the best-fit GR parameters for a given fitting function to suggest a deviation from \lcdm{}, it would highlight the potential degeneracy between a change in MG and a change in the GR cosmological parameters.  This is important if we are to use the HMF as a probe of MG in future surveys.
 
The many avenues for generalisation illustrate that the same attention to detail can be applied to the HMF in both \lcdm{} and MG.  Having illustrated a number of caveats---the choice of fitting function, likelihood and the dependence of the results on both halo finder and bin width---we nonetheless show that three common HMFs can be used and calibrated in both GR and MG.  Without applying the same calibration techniques in both theories, we are not making a like-for-like comparison when analysing the behaviour of the HMF, especially when constructing theoretical HMFs to compare to observations.
  
 
\section*{Acknowledgements}
We would like to thank Hans Winther for useful feedback on the many drafts and Pedro Ferreira for useful comments and discussions. 

\appendix

\section{Screened gravity theories} 
\label{app:screened_gravity_theories}

\subsection{$\fr$ gravity} 
\label{sub:fr_gravity}

An $f(\R)$ theory can be defined in the Jordan frame via the action:
\begin{equation} \label{eq:fr_jordan_action}
S_J = \int d^4 x\sqrt{-g}\, \frac{1}{2} \left[ \R + f(\R) \right] + {\mathcal{L}}_{\rm m} \left[ \Phi_i,g_{\mu\nu} \right] 
\end{equation}
where we have (temporarily) chosen units such that $8\pi G = 1$, the function $f(\R)$ is a general function of the Ricci scalar $\R$ and $\Phi_i$ denotes all matter fields. 

The $f(\R)$ modification in the Jordan frame translates to a scalar-tensor theory of gravity in the Einstein frame where the scalar field $\phi$ is coupled to matter.  The fact that we have this conformal transformation is the essential ingredient behind the mapping between $\fr$ and chameleon-screened theories \cite{KhouryWeltman}. 

The explicit $f(\R)$ model which we apply here is the Hu-Sawicki model of \cite{0705.1158}. This is a well studied model known to exhibit chameleon screening \cite{2010arXiv1011.5909K}.  The value of $\fr[]$, the value of $\fr$ in the cosmological background evaluated at $z=0$:
\begin{equation}
|f_{\R 0}| = n_{f(\R)} \frac{|c_1|}{c_2^2}\left(\frac{\Omega_m}{3(\Omega_m + 4\Omega_\Lambda)}\right)^{1+n_{f(\R)}}
\end{equation}
 determines the magnitude of variations from \lcdm{}. Throughout this paper we set $n_{f(\R)} = 1$ and will only consider $\fr[5]$ and $\fr[6]$.

To see how screening works, let us consider a top-hat over-density of radius $R_{\textrm{TH}}$ and mass $M_{\rm TH}$.  As shown in \cite{2008PhRvD..78j4021B} the enhancement of the gravitational force on a test-mass of mass $m$ outside the top-hat is approximately given by
\begin{subequations} \label{eq:feff_fr}
\begin{align}
    F_{\textrm{eff}}(a, R_{\rm TH},\rho_{\rm TH},\rho_{\rm env}) &= \frac{1}{3} \left[ 3 \left( \frac{\Delta R}{R_{\textrm{TH}}} \right)
    - 3 \left( \frac{\Delta R}{R_{\textrm{TH}}} \right)^{\!2}
    + \left( \frac{\Delta R}{R_{\textrm{TH}}} \right)^{\!3} \right] \\
    \frac{\Delta R}{R_{\textrm{TH}}} &=\min\left\{\frac{3|f_{\R}^{\rm TH} - f_{\R}^{\rm env}|}{2\Phi_N} ,1\right\}
\end{align}
\end{subequations}
and $f_{\R}^{\rm TH} = f_{\R}(\rho_{\rm TH})$ and $f_{\R}^{\rm env} = f_{\R}(\rho_{\rm env})$ are the scalar field values inside and outside the body respectively.  When the over-density is massive or is located in a very dense environment then $\frac{\Delta R}{R_{\textrm{TH}}} \ll 1$ and the fifth-force is screened. In contrast, when the over-density is not massive, then $\frac{\Delta R}{R_{\textrm{TH}}} \approx 1$ and the force is $4/3$ the value of the Newtonian prediction.  This completes our discussion of $\fr$.


\subsection{Symmetron gravity} 
\label{sub:symmetron_gravity}

The symmetron mechanism adds to \lcdm{} a scalar field with an artificially-imposed $\mathbb{Z}_{2}$ symmetry and a coupling to matter \cite{2014PhRvD..89b3523T,2010PhRvL.104w1301H,2005PhRvD..72d3535P}.  The breaking of this symmetry occurs when the environmental density drops below a critical value, which causes the matter-scalar coupling to become non-zero \cite{2014PhRvD..89b3523T}.

The free parameters in our model are \cite{2015PhRvD..91l3507W}:
\begin{enumerate}
  \item The range of the field at which $\rho = 0$ in $\mathrm{Mpc}/h$ : $\lambda_{0} = \frac{1}{\mu\sqrt{2}} $ 
  \item A dimensionless coupling constant $\beta_0 = \phi_0 \frac{M_{Pl}}{M_0} = \frac{\mu}{\sqrt{\lambda}} \frac{M_{Pl}}{M_0} $
  \item The scale factor at which the background density takes the value required for symmetry breaking in the cosmological background $ a_{SSB} = \frac{\rho_0}{\mu^{2} M_{0}^{2}} $
\end{enumerate}
We set $\lambda_{0} = 1$, $\beta_0 = 1$ and $ a_{SSB} = 0.33,\, 0.5$ in our simulations.

The effective gravitational potential can be expressed similarly to that of $\fr$.  The scalar field value in the cosmological background and inside the halo are:
\begin{subequations} \label{eq:symm_feff}
\begin{align}  
  F_{\mathrm{env}} &= \sqrt{1 - \left( \frac{\assb}{a} \right)^3} \\
  F_h &= \sqrt{1 - (1 + \delta) \assb^3)} \\
\intertext{which leads to a thin-shell factor}
  \frac{\Delta R}{R_{\mathrm{TH}}} &= \frac{\Omega_{m0}}{\assb^{3}} \frac{\lambda_{0}^{2}}{\Phi_{N}} 
  \frac{\abs{F_h - F_{\mathrm{env}}}}{F_{\mathrm{env}}} \\
\intertext{and as per $\fr$ the effective factor is:}
  \Feff &= 2 \left( \beta_{0}F_{\mathrm{env}} \right) ^{2} \mathrm{min} \left\{ 3 \frac{\Delta R}{R_{\mathrm{TH}}}, 1 \right\} 
\end{align}
\end{subequations}

By analogy with the chameleon potential in \cref{sub:fr_gravity}, the modification falls into an unscreened regime where $\phi_{\mathrm{in}} \approx \phi_0 $ and the scalar field cannot relax to the value at the effective minimum of the potential $V(\phi)$; and a screened regime where $\phi_{\mathrm{in}}$ is suppressed exponentially compared to $\phi_0$.  Thus, we see that despite the different mechanism by which screening occurs, the result is very similar for both the chameleon screening in $\fr$ gravity and symmetron gravity.

\subsection{DGP gravity} 
\label{sub:dgp_gravity}

Dvali-Gabadadze-Porrati (DGP) gravity  is a braneworld model, in which the usual $\left( 3+1 \right)$ foliated hypersurfaces are a brane embedded in a(n otherwise empty) higher-dimensional spacetime known as the bulk.  While the Standard Model interactions are limited to the brane, gravitational interactions extend into the bulk.  The DGP model employs Vainshtein screening to remain observationally viable.

Dvali-Gabadadze-Porrati gravity embeds the FLRW manifold into a 5D Minkowski manifold, where the free parameter is the crossover scale:
\begin{equation} \label{eq:r_c}
  r_{c} \equiv \frac{\scalar(5\mathrm{D}){\kappa}}{ 2 \scalar(4\mathrm{D}){\kappa}} 
  \quad\text{where } \scalar(4\mathrm{D}){\kappa} = \left( 8 \pi G_{N} \right) \text{ as usual} \qedhere
\end{equation}
On scales $r \gg r_{c}$, the 5D effects are unscreened, whereas on scales $r \ll r_{c}$, the brane is unaffected by the presence of the bulk and the dynamics are screened \cite{0264-9381-28-16-164011}.  

In the weak-field, quasi-static limit, the equation of motion for the scalar field and the equations for the potentials can be combined to obtain a modified Poission equation with:
\begin{equation} \label{eq:dgp_feff}
  \Feff{} (r,a) = \frac{2}{3 \beta(a)} \frac{\sqrt{1 + x^{-3}} - 1}{x^{-3}}
  \quad\text{where } x(r,a) \equiv \frac{r}{R_{\ast}}
\end{equation}
where $R_{\ast}$ is the Vainshtein radius for the mass $M(r)$ enclosed inside the radius $r$
\begin{subequations} \label{eq:dgp_eqns}
\begin{align}  
  R_{\ast}(r,a) &= \left( r_{c}^{2} \frac{16 c^{2} G_{N} \delta M(r)}{9\beta^{2}(a)} \right)^{\frac{1}{3}} \\
\intertext{where we define}
  \beta(\phi) &\equiv 1 + 2 r_{c} H(a) \left( 1 + \frac{\dot{H}(a)}{3H^{2}(a)} \right)
\end{align}
\end{subequations}
The scalar field is unscreeened outside the Vainshtein radius and screened within it.  This completes our discussion of DGP gravity.


\section{Technical aspects of spherical collapse in MG} 
\label{app:technical_aspects_of_spherical_collapse_in_mg}

This section covers two main points which we mentioned in passing in the main text: the probability density function for the environment overdensity and the use of the \lcdm{} power spectrum even in MG.

\subsection{Choice of environment density function} 
\label{sec:choice_of_environment_density_function}

\begin{figure}
  \begin{center}
    \includegraphics[keepaspectratio,width=0.75\textwidth]{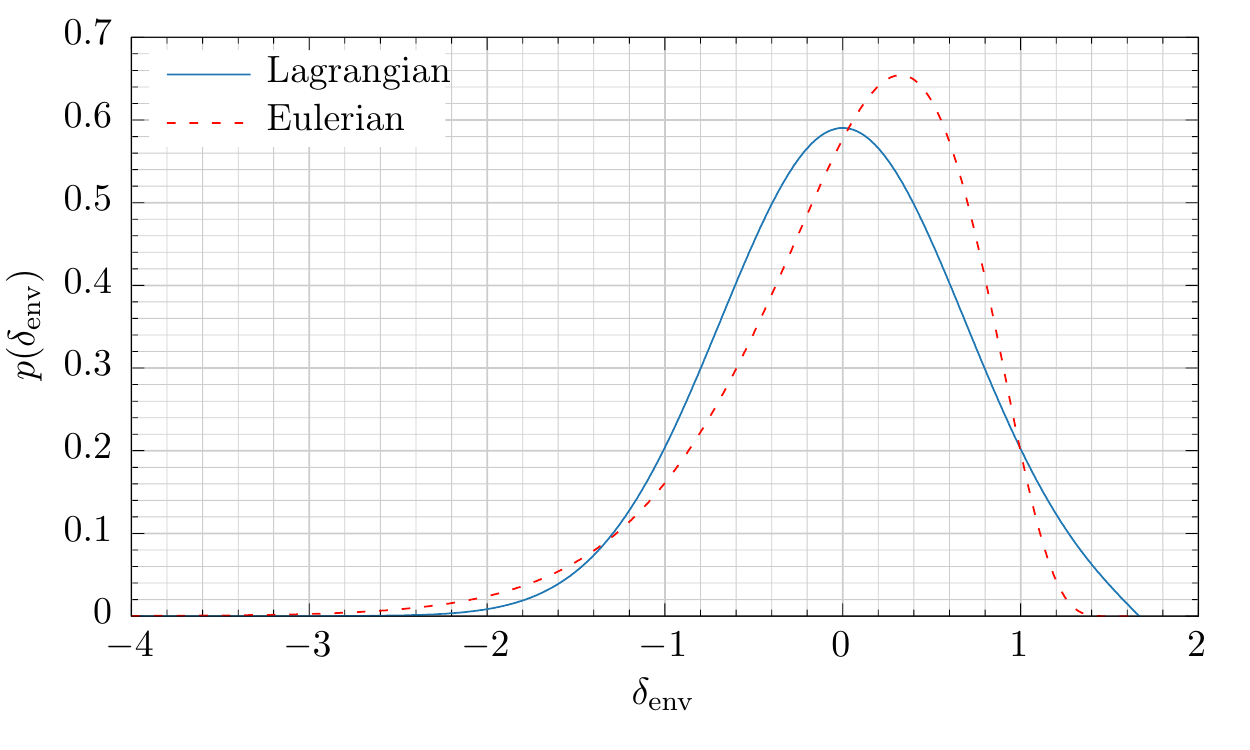}
  \end{center}
  \caption{The \pdf{}s for Lagrangian and Eulerian environment density according to \cref{eq:pdf_denv_L,eq:pdf_denv_E} respectively.  This assumes a smoothing scale of $10\; \mathrm{Mpc}/h$ for both environments.}
  \label{fig:Lag_vs_Eul_denv}
\end{figure}

We briefly describe our two choices for the distribution of the environment overdensity---namely the Lagrangian and Eulerian distributions---and justify avoiding a distribution derived from $N$-body simulations.

Three options are frequently used for the environment density in the literature:
\begin{enumerate}
  \item \label{item:denv_L} Lagrangian distribution from excursion set theory
  \item \label{item:denv_E} Eulerian equivalent to \cref{item:denv_L}
  \item \label{item:denv_N} Numerical distribution from $N$-body simulations
\end{enumerate}

Recall that we assume that the environment behaviour is on sufficiently large scales that it is well-approximated by \lcdm{}.  Then \cite{1991ApJ...379..440B} have shown that the environment distribution can be obtained using the same excursion set method by which we have derived the PS mass function.  The probability that the trajectory starting from $(0,0)$ survives to up-cross the environment density $\denv$ at resolution $\Senv$ is \cite{1991ApJ...379..440B}:
\begin{equation} \label{eq:pdf_denv_L}
  p \left( \denv  \right)
  = \frac{ \mathcal{H} \left( \delta_{c}^{\Lambda} - \denv \right) }{\sqrt{2 \pi \Senv}}
  \left[ \exp \left( -\frac{\denv^{2}}{2\Senv} \right) - \exp \left( -\frac{\left( \delta_{c}^{\Lambda} - \denv \right)^{2}}{2\Senv} \right) \right]
\end{equation}
where $\delta_{c}^{\Lambda} \approx 1.676$ is the barrier density for \lcdm{} and $\mathcal{H}$ is the Heaviside step function.

\cref{item:denv_E} can be obtained from \cref{eq:pdf_denv_L} by treating the non-linear Eulerian over-density $\Delta$ as the barrier function in the excursion set method:
\begin{subequations} \label{eq:Eulerian_pdf}
\begin{align}
  \Delta(S) &= \delta_{c} \left[ 1 - \left( \frac{R_{\mathrm{env}}}{8 \; \mathrm{Mpc}/h} \right)^{3/\delta_{c}} \left( \frac{S}{\sigma_{8}^{2}} \right)^{1/\omega} \right]  
\intertext{which produces the distribution\footnotemark{} for the Eulerian linear density contrast}
  q \left( \denv \right) &=
  \frac{\beta^{\omega/2}}{\sqrt{2\pi}} \left[ 1 + \left( \omega - 1 \right) \frac{\denv}{ \delta_{c}^{\Lambda} }  \right]
  \left[ 1 - \frac{\denv}{ \delta_{c}^{\Lambda} } \right]^{-(\omega/2+1)}
  \exp{ \left[ - \frac{\denv^{2}}{2} \left( \frac{\beta}{1 - \frac{\denv}{ \delta_{c}^{\Lambda} }} \right)^{\!\!\omega\;} \right] } \label{eq:pdf_denv_E}
\end{align}
\begin{equation*}
  \text{where}\qquad
  \omega = -\delta_{c} \left. \frac{\dx \ln S}{\dx \ln M} \right|_{\mathrm{env}}
  \qquad\text{and}\qquad
  \beta = \left( \frac{R_{\mathrm{env}}}{8 \; \mathrm{Mpc}/h} \right)^{3/\delta_{c}} \left( \frac{1}{\sigma_{8}} \right)^{2/\omega}  
\end{equation*}
\end{subequations}
\footnotetext{We correct an error in \cite{Lombriser:2013wta} Equation (49) and \cite{2012MNRAS.425..730L} Equation (15) by restoring the $\denv^{2}$ term in the exponent.  Without it, the distribution does not normalise to unity.  Furthermore \cite{Lombriser:2013wta} Equation (49) has an error in the definition of $\beta$, which is correct in \cite{2012MNRAS.425..730L} Equation (15).}
The values of $\sigma_{8} = 0.8$ and $n_{s} = 0.966,\; 1.0$ for the Eulerian \pdf{} are set by the simulations (\cref{sub:n_body_simulations_haloes}).

The corresponding Lagrangian and Eulerian \pdf{}s for $R_{\mathrm{env}} = 10\; \mathrm{Mpc}/h$ are shown in \cref{fig:Lag_vs_Eul_denv}.  The Lagrangian PDF is quasi-symmetric: whereas the PDF asmyptotes smoothly to zero for $\denv \leq -\delta_{c}^{\Lambda}$, it is artificially set to zero for values $\denv \geq \delta_{c}^{\Lambda}$ because these environments would already have formed a halo.  The Eulerian PDF is asymmetric, with a greater proportion of over-densities (\ie{} where $\denv \geq 0$) than voids.  However, it does not asymptote to zero for negative densities as rapidly as the Lagrangian PDF.   We analyse the effect of the choice of PDF on the HMF in \cref{sub:including_denv_barrier_in_mg}.

Whereas both of these are analytic methods, perhaps the most accurate solution is simply to extract the distribution from the $N$-body simulations.  This would isolate the effects of the MG excursion set prediction from those of assuming a theoretical approximation to the $\denv$ distribution.  (Implicit in the use of this distribution is the assumption that the environment in the box is an unbiased sub-sampling of the cosmological environment density.)  However, each environment \pdf{} would differ in each simulation, so it would be difficult to disentangle the effects of averaging over a different $p(\denv)$ from the effects of the gravity model on the conditional HMF.  Thus we discard this option.

\subsection{Justification for using the \lcdm{} power spectrum} 
\label{sec:justification_for_using_the_lcdm_power_spectrum}

\begin{figure}
    \begin{center}
        \includegraphics[keepaspectratio,width=0.75\textwidth]{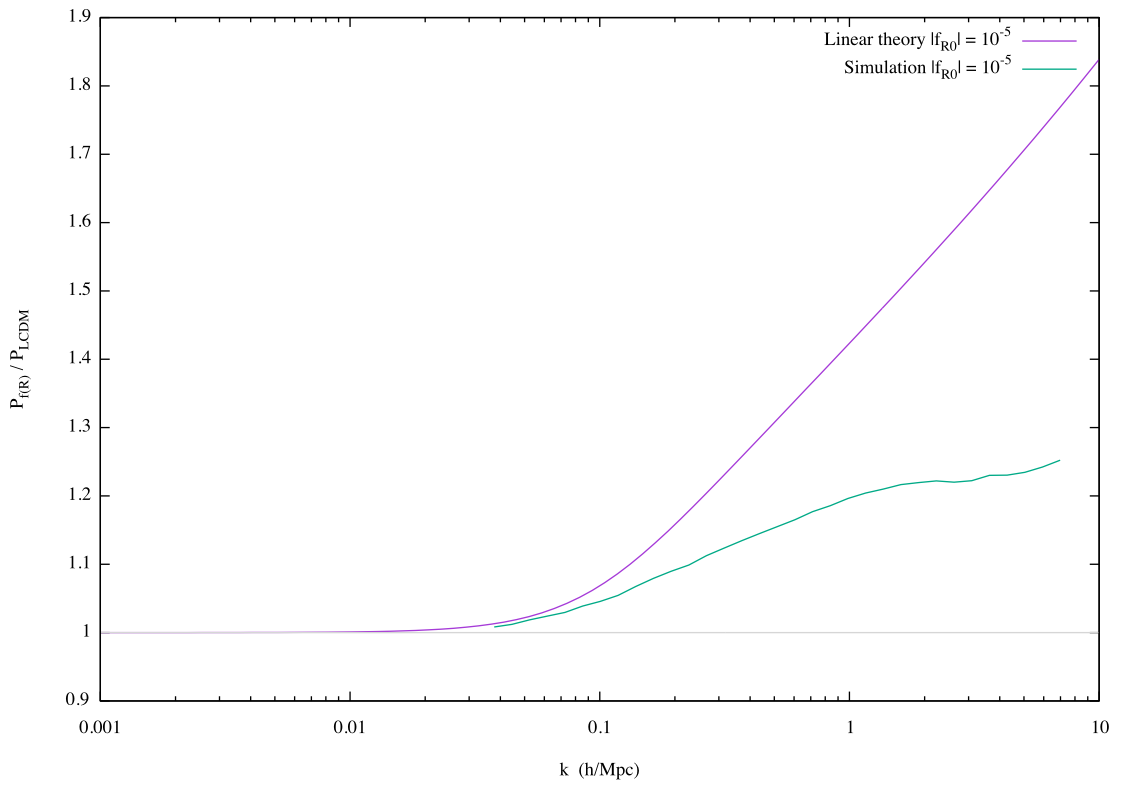}
    \end{center}
    \caption{The ratio of the $f(R)$ power spectrum to that of \lcdm{} for the linear (purple) and non-linear (green) case.  The effect of screening is clearly visible at increasingly large $k$: the non-linear result is suppressed, while the linear result increases rapidly.}
    \label{fig:pofk_f5}
\end{figure}

We have chosen to use the linear $P(k)$ computed in the $\Lambda$CDM model in \cref{eq:resolution} when computing the $f(\R)$ gravity predictions.  For a more comprehensive argument see our earlier paper \cite{2017JCAP...03..012V}.  Briefly, the reason for this choice is that linear theory massively overestimates the clustering in modified gravity theories such as $f(\R)$. \cref{fig:pofk_f5} shows that the ratio of the linear power-spectrum in $\fr[5]$ to the linear power-spectrum in $\Lambda$CDM exhibits a much larger deviation than the ratio of the corresponding non-linear power spectra obtained from simulations.  An alternative would be to consider a compromise approach where we use the linear $\Lambda$CDM power-spectrum corrected with a boost-factor $P_{f(\R)}(k) /P_{\Lambda\rm CDM}(k) $ computed from simulations, however this would require explicit simulations which goes against the appeal of using the excursion set approach---which is to extract observables without having to perform expensive numerical calculations.


\subsection{The scale of the environment} 
\label{sub:the_scale_of_the_environment}

\begin{figure}
  \centering
  \includegraphics[width=0.75\textwidth]{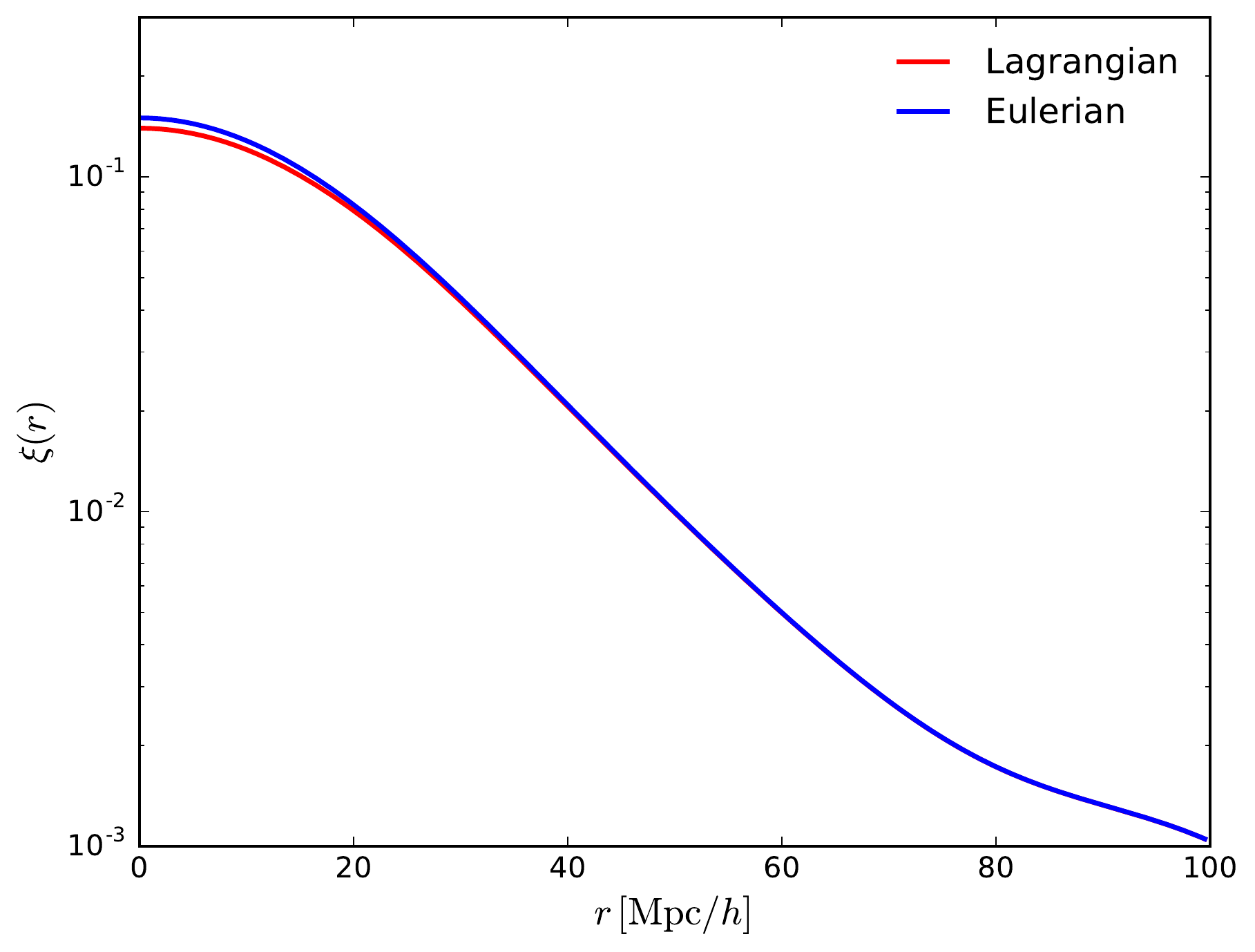}
  \caption{Two-point correlation function of the Lagrangian (red) and Eulerian (blue) matter density fields, the latter approximated by a log-normal transformation of the former. The results are shown for a Gaussian smoothing scale $R_{\rm env}=10\,{\rm Mpc}/h$. The relative difference between both curves is smaller than 7\% on all scales.}  \label{fig:evsl}
\end{figure}

In a strict sense, when relating the excursion-set predictions to simulated or real data, the scale of the environment $S_{\rm env}$ should correspond to its Lagrangian scale (\ie{} in the initial conditions) instead of its Eulerian size.  As pointed out by \cite{2012MNRAS.426.3260L,2012MNRAS.425..730L} these effects could be relevant for modified-gravity theories.  However, for the environment scales studied in this work ($R_{\rm env}\geq10\,{\rm Mpc}/h$), we expect these two quantities to be very similar.  Let us approximate the effects of the Eulerian evolution on the Lagrangian density field via a log-normal transformation.  Then \cref{fig:evsl} shows the negligible difference between the correlation functions of our Eulerian and evolved-Lagrangian density fields for $R=10\,{\rm Mpc}/h$.  Thus we can safely ignore the true Eulerian scale of the environment. 



\section{Calibration of Multinest} 
\label{app:calibration_of_multinest}
 
\begin{figure}
  \begin{center}
    \subfloat[lenm = 10]{\label{Peacock_fake_data_poisson_10}\includegraphics[keepaspectratio,width=0.49\textwidth]{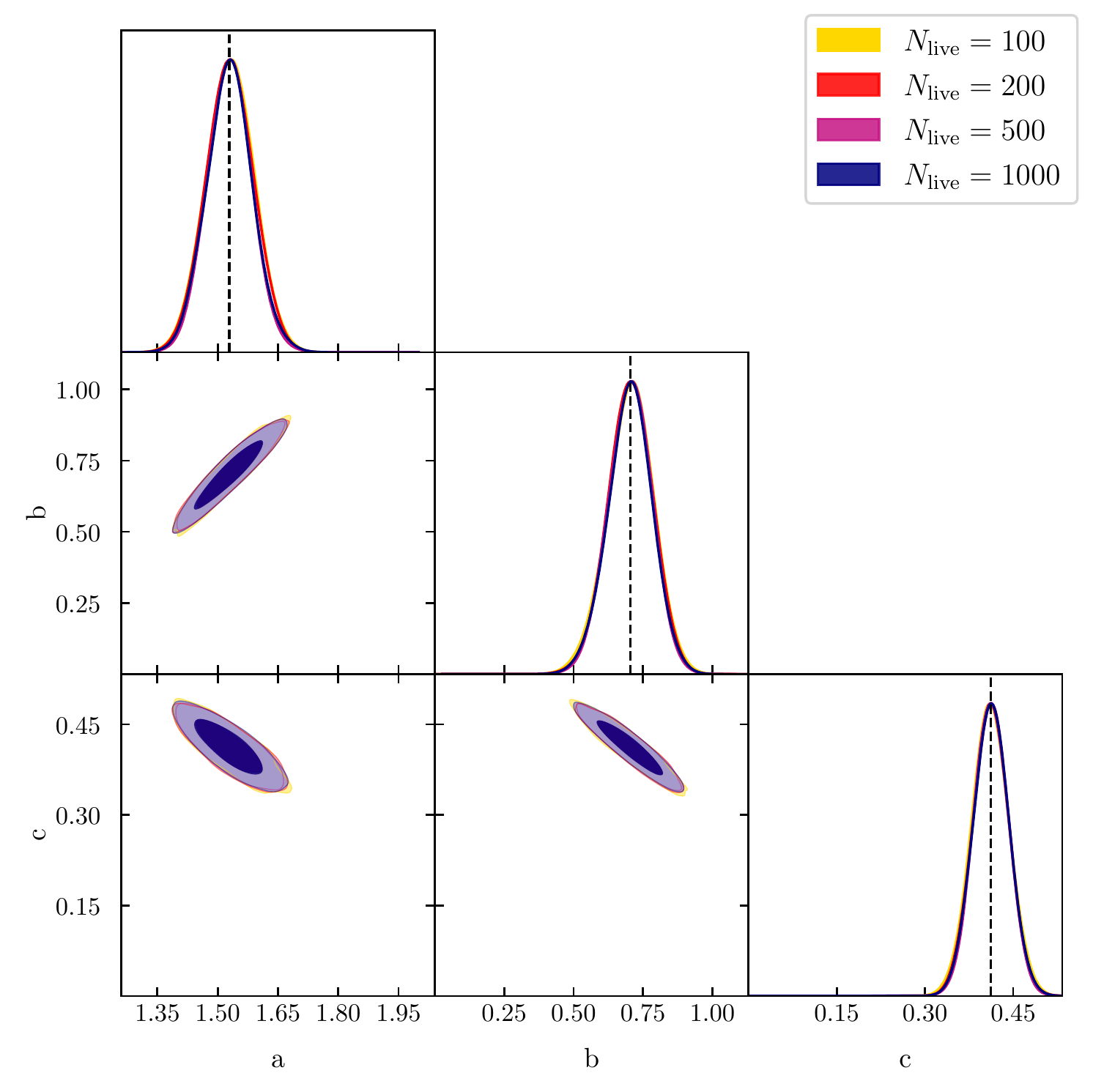}} 
    \subfloat[lenm = 30]{\label{Peacock_fake_data_poisson_30}\includegraphics[keepaspectratio,width=0.49\textwidth]{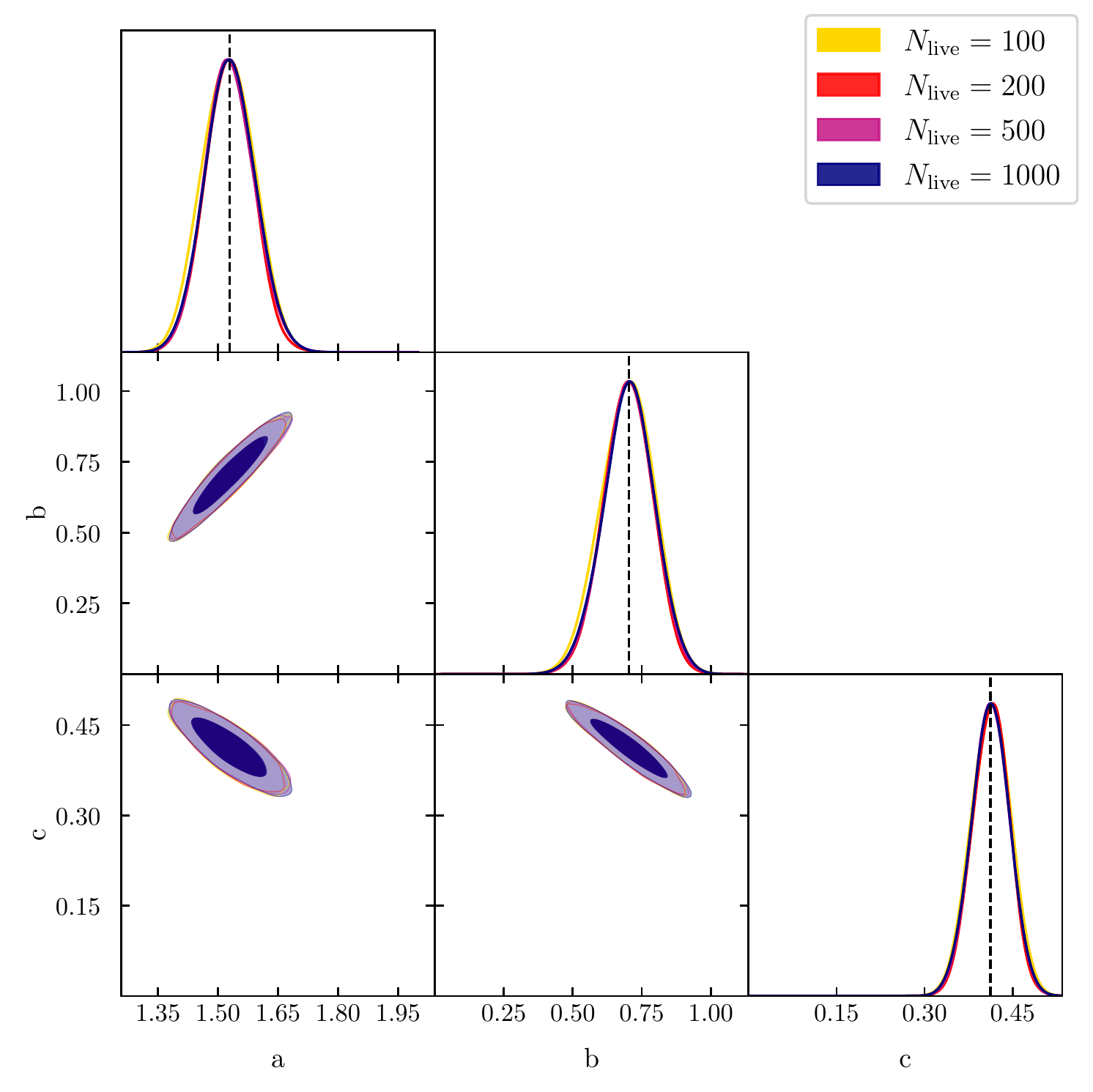}} \\
    \subfloat[lenm = 100]{\label{Peacock_fake_data_poisson_100}\includegraphics[keepaspectratio,width=0.49\textwidth]{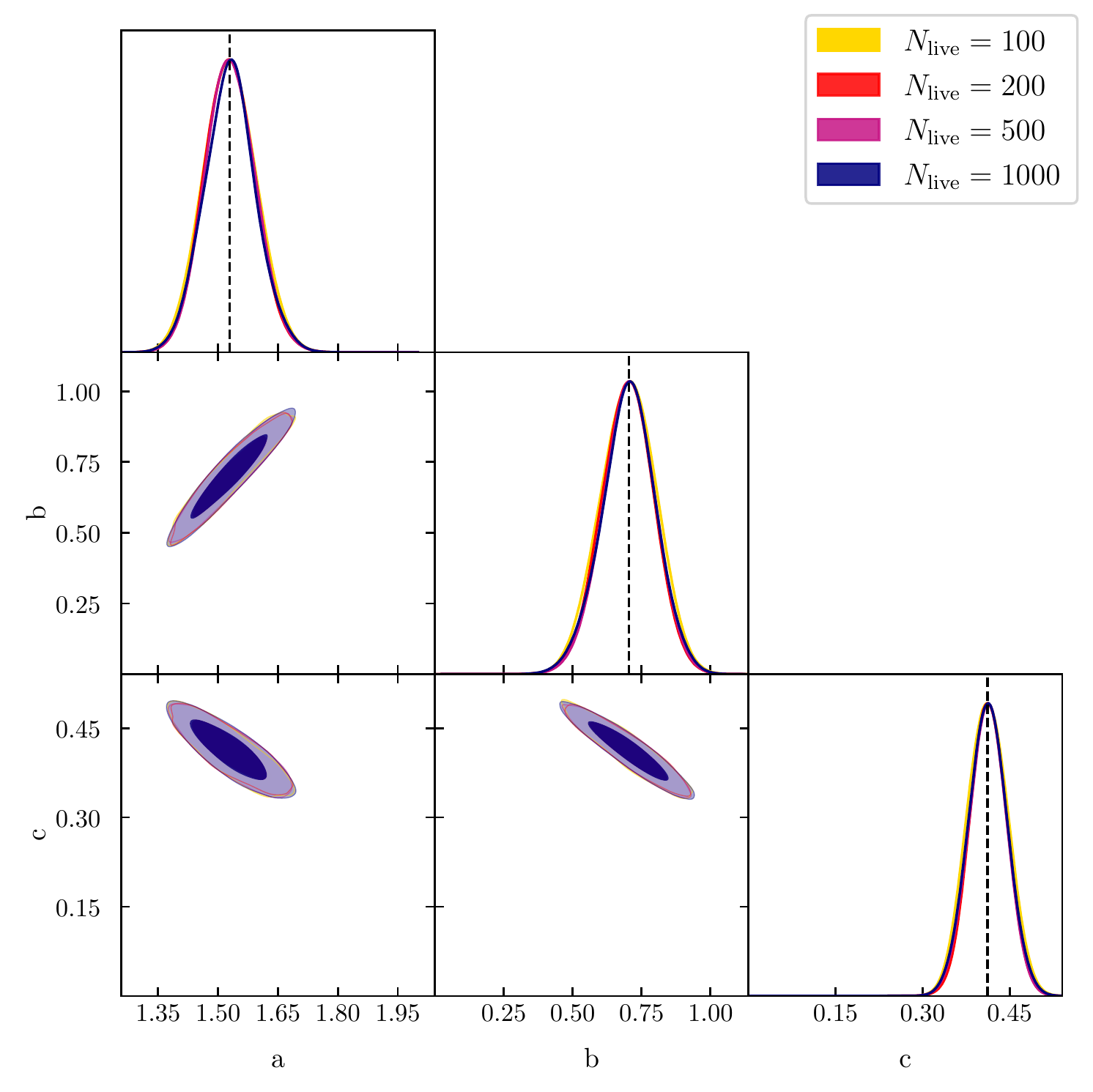}}
  \end{center}
  \caption{The effect of changing bin width on the Peacock HMF.  The triangle plot is the same as \cref{fig:Peacock_fake_data}.  The subfigure captions show the number of bins used in calculating the input HMF and discretising the model HMFs proposed by the nested sampling routine.}
  \label{fig:Peacock_fake_data_change_lenm}
\end{figure}

We used the \texttt{MultiNest} nested sampling algorithm of \cite{2009MNRAS.398.1601F} to carry out the integrals in the previous subsections.  This section describes the calibration process we used to tune the algorithm.
 
The simulated data in this section was made by generating an HMF with the \lq\lq{}default\rq\rq{} parameter values for each fitting function, i.e. the values in the original publication.  The Poisson error was simulated by scattering each data point by a value in $[-\sqrt{N} \, , \sqrt{N}]$, where $N$ is the number of haloes per bin.    We used 10, 30 and 100 data points, equivalent to binning our halo data into equal-$\ln M$ bins.  (Recall that the number of bins suggested by the optimisation in \cref{sub:data_errors} was $\sim 30$.)  The \texttt{MultiNest} parameters were selected until both the Poisson and Gaussian likelihoods returned credible regions which included the original parameter.  
 
The key settings are tabulated in \cref{tab:multinest_calibration}.  

\begin{longtable}{*{3}{l}}
\toprule 
  Description &
  Values tested &
  Final value \\
\midrule
\endhead 
\bottomrule \\[-.1in]
\caption{\label{tab:multinest_calibration} Calibration parameters used to tune \texttt{MultiNest}.  The same values were used for both likelihood functions.}
\endfoot 
Importance Nested Sampling & True, False & True \\ 
Identify multi-modal posteriors & True, False & False \\ 
Number of live points & 100 200 500 1000 & 500 \\ 
Tolerance factor & 1d-4, 5d-4, 5d-3, 5d-2, 5d-1 & 1d-4 \\ 
Sampling efficiency & 3d-1, 5d-1, 8d-1, 1, 2 & 8d-1 \\ 
\end{longtable}
\clearpage  
Having calibrated \texttt{MultiNest}, the algorithm was used to recover (as best as possible) the input parameter values.  The properties which we varied were:
\begin{enumerate}
  \item The likelihood function: Gaussian or Poissonian
  \item The number of live points
  \item The number of bins: 10, 30 or 100 (equivalently, the bin width)
\end{enumerate}
Over the next three subsections we analyse the effect of the bin width, the number of live points and the choice of likelihood function on simulated data.  Finally we eliminate certain fitting functions on the grounds that they have too many free parameters to be constrained by our data.

\begin{figure}
  \begin{center}
      \includegraphics[keepaspectratio,width=0.75\textwidth]{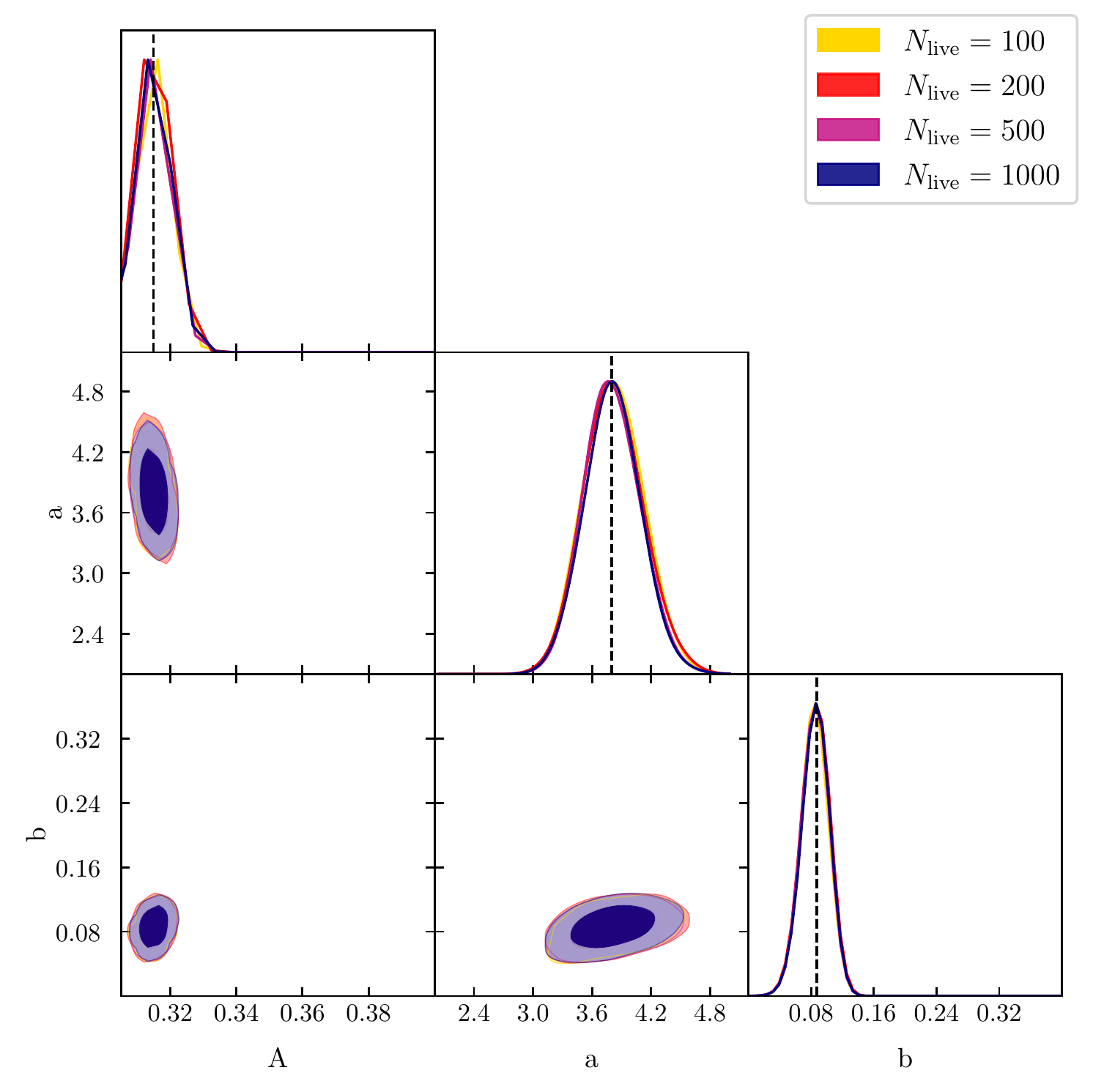}
  \end{center}
  \caption{Posteriors for the Jenkins HMF using different numbers of live points (coloured).  The main diagonal shows the 1-d posteriors marginalised over all other parameters, while the off-diagonal plots show correlations between pairs of parameters via the 2-d $1,2\sigma$ credible regions. The black dashed lines show the input values, while the coloured lines show the PDF of the values recovered by nested sampling, with 30 bins, assuming a Poissonian likelihood.}
  \label{fig:Jenkins_fake_data}
\end{figure}
  
\begin{figure}
  \begin{center}
    \includegraphics[keepaspectratio,width=0.75\textwidth]{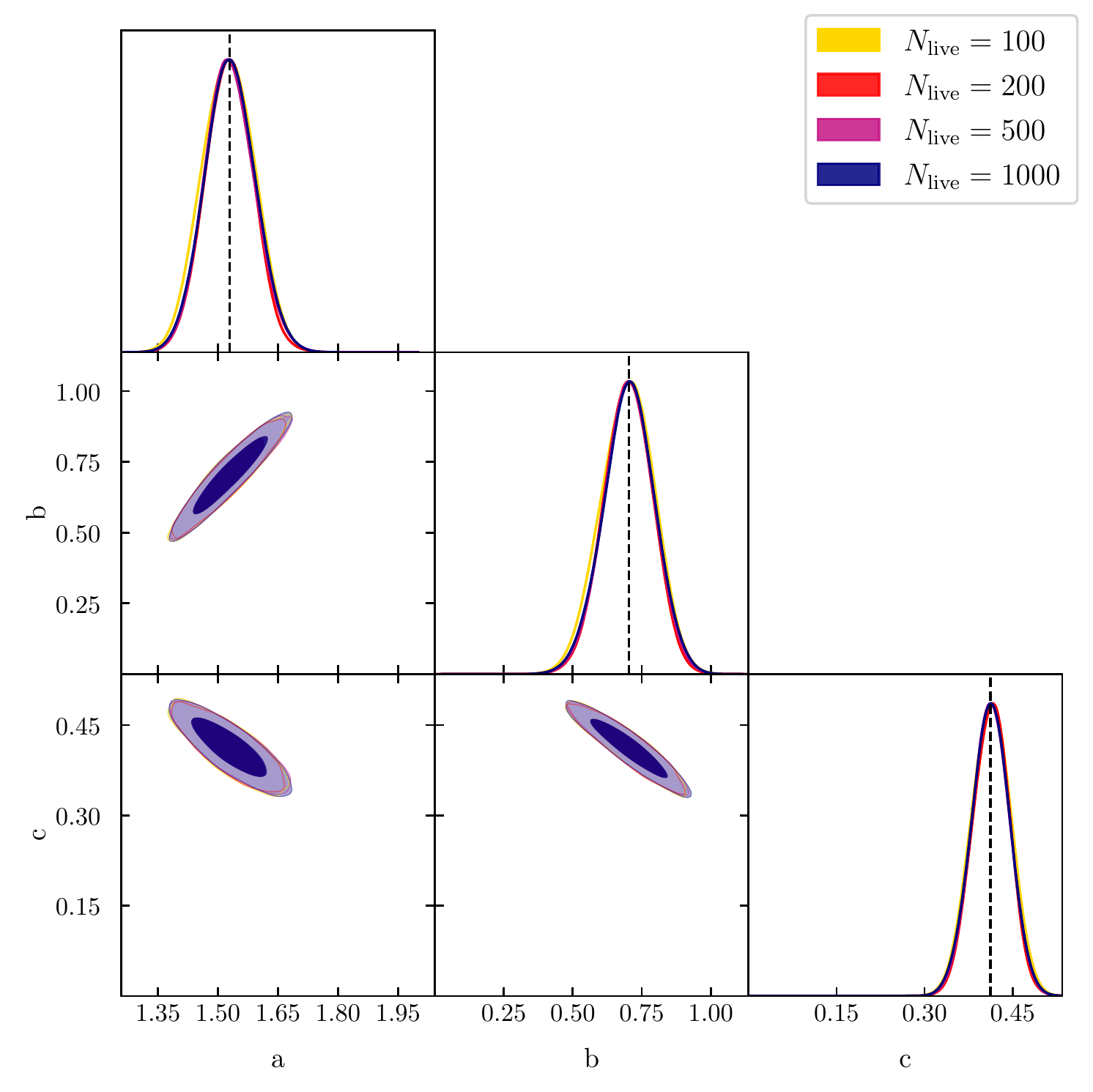}
   \end{center}
  \caption{Posteriors for the Peacock HMF using different numbers of live points (coloured).  The main diagonal shows the 1-d posteriors marginalised over all other parameters, while the off-diagonal plots show correlations between pairs of parameters via the 2-d $1,2\sigma$ credible regions. The black dashed lines show the input values, while the coloured lines show the PDF of the values recovered by nested sampling, with 30 bins, assuming a Poissonian likelihood.}  
  \label{fig:Peacock_fake_data}
\end{figure}
  
\begin{figure}
  \begin{center}
    \includegraphics[keepaspectratio,width=0.75\textwidth]{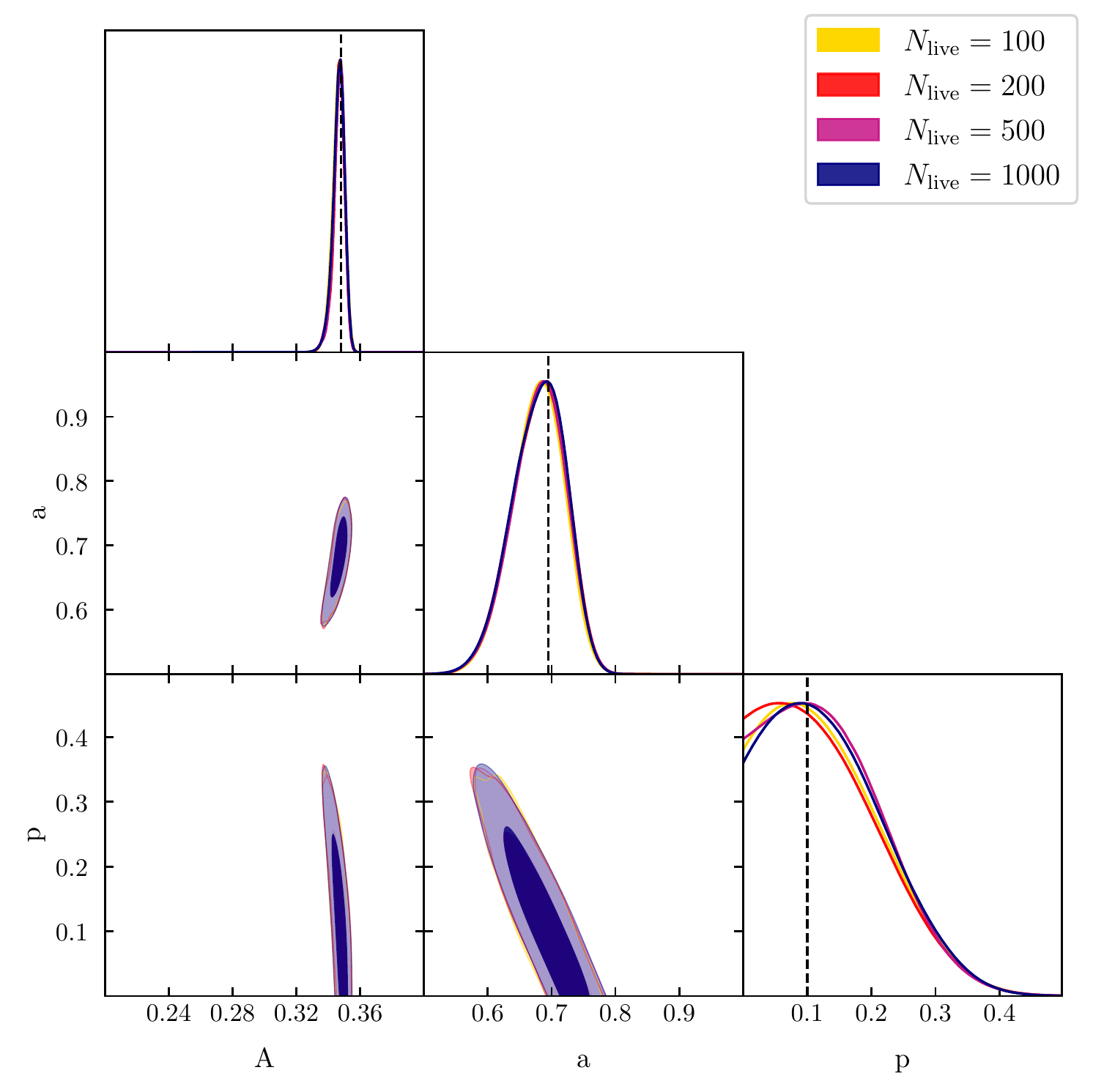}
  \end{center}
  \caption{Posteriors for the SMT-Courtin HMF using different numbers of live points (coloured).  The main diagonal shows the 1-d posteriors marginalised over all other parameters, while the off-diagonal plots show correlations between pairs of parameters via the 2-d $1,2\sigma$ credible regions. The black dashed lines show the input values, while the coloured lines show the PDF of the values recovered by nested sampling, with 30 bins, assuming a Poissonian likelihood.}  \label{fig:SMT-Courtin_fake_data}
\end{figure}
   
\begin{figure}
  \begin{center}
    \includegraphics[keepaspectratio,width=0.75\textwidth]{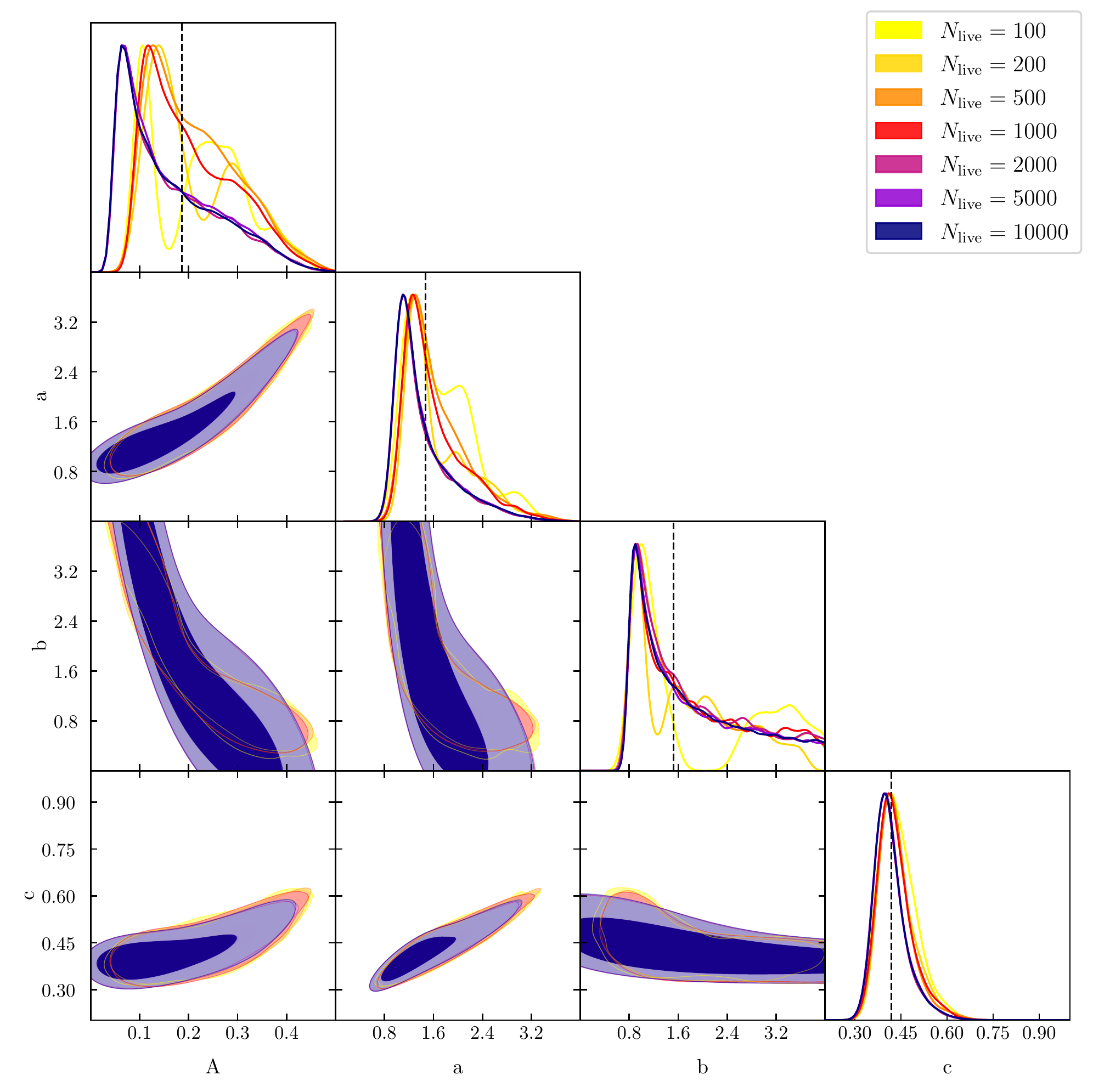}
  \end{center}
  \caption{Posteriors for the Tinker-Angulo-Watson HMF using different numbers of live points (coloured).  The main diagonal shows the 1-d posteriors marginalised over all other parameters, while the off-diagonal plots show correlations between pairs of parameters via the 2-d $1,2\sigma$ credible regions. The black dashed lines show the input values, while the coloured lines show the PDF of the values recovered by nested sampling, with 30 bins, assuming a Poissonian likelihood.}
  \label{fig:Tinker-Angulo-Watson_fake_data}
\end{figure}
 
\begin{figure}
  \begin{center}
    \includegraphics[keepaspectratio,width=0.75\textwidth]{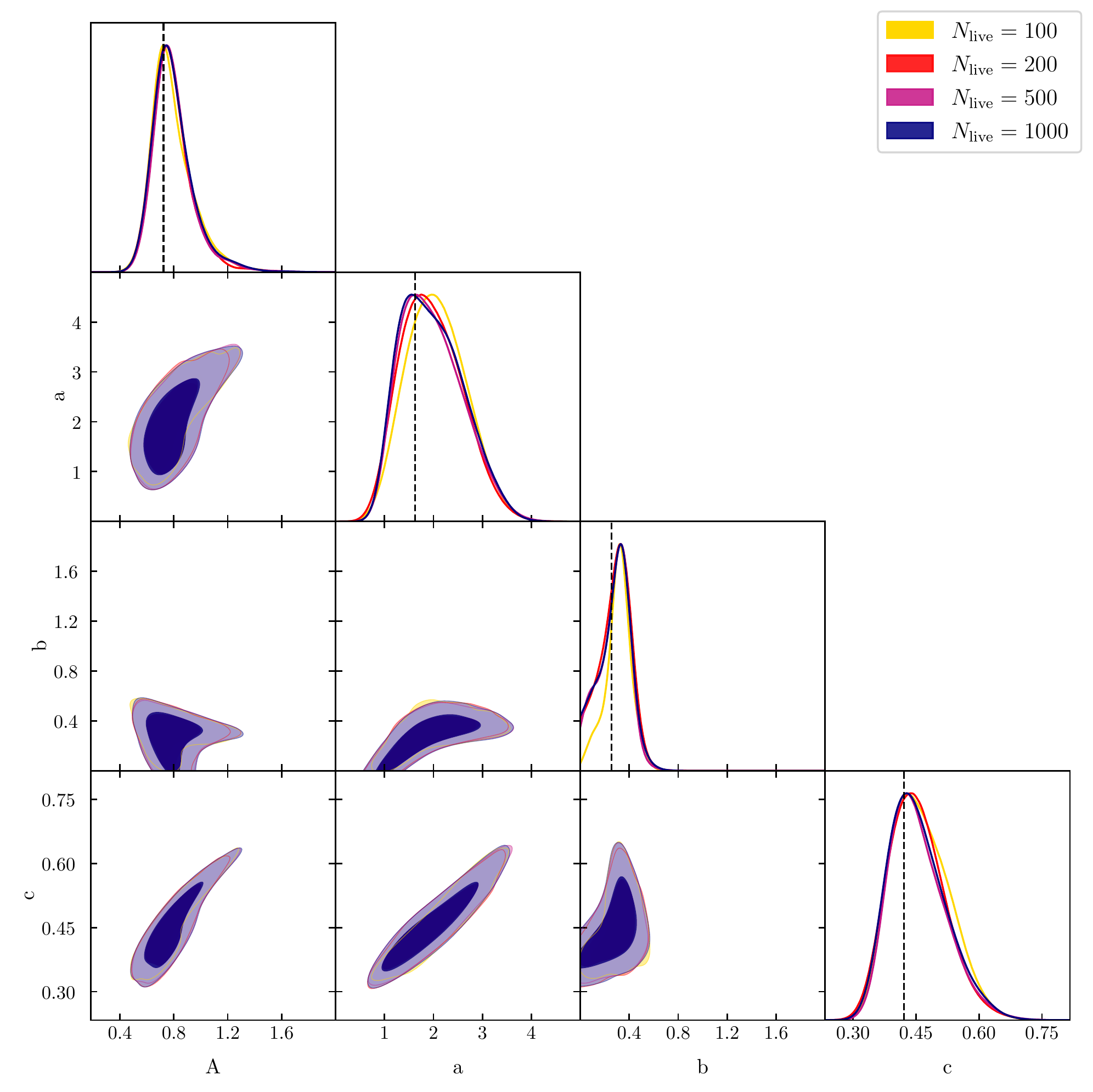}
  \end{center}
  \caption{Posteriors for the Warren-Crocce HMF using different numbers of live points (coloured).  The main diagonal shows the 1-d posteriors marginalised over all other parameters, while the off-diagonal plots show correlations between pairs of parameters via the 2-d $1,2\sigma$ credible regions. The black dashed lines show the input values, while the coloured lines show the PDF of the values recovered by nested sampling, with 30 bins, assuming a Poissonian likelihood.}
  \label{fig:Warren-Crocce_fake_data}
\end{figure}
 
\subsection{Choice of likelihood function} 
\label{sub:choice_of_likelihood_function}

The choice of likelihood function largely reflects one's probabalistic education.  The two schools of thought are typically a \lq\lq{}Gaussian\rq\rq{} likelihood based upon the frequentist chi-squared estimator and a \lq\lq{}Poissonian\rq\rq{} likelihood based upon maximum-entropy results from Bayes theorem.  We choose between:
\begin{enumerate}
\item \label{item:slike_Poisson} The Poisson likelihood (the last term is zero when $n_{i}$ is zero):
\begin{equation} \label{eq:slike_Poisson}
  \ln \mathcal{L} = - \sum_{i} \left( \mu_{i} - n_{i} + n_{i} \ln \frac{n_{i}}{\mu_{i}}  \right)
\end{equation}
\item \label{item:slike_Gsymmetric} The Gaussian likelihood with symmetric errors
\begin{equation} \label{eq:slike_Gsymmetric}
  \ln \mathcal{L} = - \frac{1}{2} \sum_{i} \left[ 2\pi V_{0} + \frac{(\mu_{i} - n_{i})^{2}}{V_{0}}  \right] 
\end{equation}
\item \label{item:slike_Gasymmetric} The Gaussian likelihood with asymmetric errors
\begin{equation} \label{eq:slike_Gasymmetric}
  \ln \mathcal{L} = - \frac{1}{2} \sum_{i} \left[ 2\pi \left( V_{0} + V_{1} (\mu_{i} - n_{i}) \right) + \frac{(\mu_{i} - n_{i})^{2}}{V_{0} + V_{1} (\mu_{i} - n_{i})} \right] 
\end{equation}
\end{enumerate}
where $\mu_{i}(\mathbf{q})$ is the number of counts (\fcd{}) given by the parameter set $\mathbf{q}$ and $n_{i}$ is that given by the data in the $i$-th bin and the choice between counts and \fcd{} is determined by the use of the Poisson and Gaussian likelihood respectively.  When applied to counts, $n$ is always a positive integer, whereas $\mu$ is a positive real number.

Bayesian inference implies that \cref{eq:slike_Poisson} will produce the most accurate result, whereas the standard frequentist approach is equivalent to \cref{eq:slike_Gsymmetric}.  The last approach \cref{eq:slike_Gasymmetric} is no longer equivalent to the frequentist approach because the errors are asymmetric.  

The choice of likelihood function has a common behaviour across all of the HMFs: that the Poissonian likelihood is superior to both the Gaussian ones. 

We do not show posterior figures for the Gaussian data here because we found that the Gaussian likelihood produces broad rather than peaked posteriors---in fact the $1\sigma$ regions occupy most of the parameter space.  Expanding the prior volume did not affect this result, which indicates that the posteriors are prior driven rather than constrained by the data.  The credible regions do not shrink appreciably when the number of live points is dramatically increased (from $10^1$ to $10^4$), so this is a consequence of the likelihood rather than the algorithm used to calculate it. Moreover, this is not a consequence of the scatter in the data, because the Poisson likelihood \textit{does} produce narrow, peaked posteriors with the same data.  We find that these trends are independent of the bin width used.  The only way to utilise the Gaussian likelihood is to replace the Poisson errors with much smaller ones.  For example, \cite{2017MNRAS.467.3454B} run multiple $N$-body simulations and use jack-knife errors which are much smaller.  However, it is impractical to replicate this in the synthesised data.  Therefore we do not apply the Gaussian likelihood to our halo data.  This illustrates the importance of correctly selecting a likelihood function using maximum entropy principles (\ie{} based upon the underlying distribution in the data).


\subsection{Number of bins} 
\label{sub:number_of_bins}

The aim is to verify that the seemingly arbitrary bin choice (which varies in every HMF paper) has minimal effect on the calibration of the free parameters.  We adopted three constant-width bin methods, explained in the next paragraphs.
 
The constant bin width method is that most widely adopted in the literature.  The haloes are typically binned in constant intervals of $\log_{10} M$ \cite{2016JCAP...12..024C}, or $\ln \sigma^{-1}$ \cite{2001MNRAS.321..372J}, rather than $\ln \nu$, so the same $f(\nu)$ will produce a different (discrete) HMF in different MG theories.  We adopt $\ln M$ to ensure a ready comparison between different gravity theories.  A non-arbitrary method for selecting the bin width is to decide upon the maximum Poisson error tolerated and find the bin width ensuring this error in the most massive bin.  (Lower-mass bins will have more haloes, hence less Poisson error.)  However this produces an impractically small number of bins, of order the number of free parameters $(3-6)$ in the fitting functions.  The advice of \cite{2006ApJ...646..881W} is to leave the most massive bin variable and set a constant width for the other bins.  While this method has the advantage of being directed by the simulation data, rather than arbitrarily selected, it produces relatively few bins. 
Instead, we applied $N = 10$ and $N = 100$ bins in order to approximate a choice of too few or too many bins.  The former has difficulty accurately representing the steep gradient at the high-mass end, whereas the latter has such narrow bins that monotonic behaviour of the HMF is not observed, or even punctuated by zero-occupancy bins.  This illustrates two possible dangers of arbitrarily selecting a number of bins.
 
In order for the bin width to be driven by the data, one can use Bayesian analysis to find an optimal number of bins.  There exist a number of ways to find a \lq\lq{}best\rq\rq{} number of bins from a distribution by minimising the $L_{2}$ norm between the data and the underlying (unknown) smooth density function (\eg{} \cite{Freedman1981}).  However, it is possible to minimse the required assumptions using the method of \cite{2006physics...5197K}, which computes the likelihood that the binned data is drawn from the smooth distribution.  The bin width, \ie{} amount of discretisation, which maximises this likelihood value is judged to be optimal.  The downside of this method is the computational expense of the MCMC calculations required (to ensure that the uncertainty in the posterior mean is much less than half, \ie{} we are certain to within a unit of one bin).  However, for a one-dimensional histogram computed using $\sim 50 000$ haloes, we found that this required $\sim 1$ minute per simulation.  The results differ according to the gravity model used and the halo finder (as expected, for each has a different HMF).

 \cref{fig:Peacock_fake_data_change_lenm} shows the effect of our three bin choices on the Peacock HMF.  The number of bins is a compromise between the amount of information provided by the data and the scatter in this information.  Thinner bins produce more data points with which to constrain the parameter values, so we expect smaller credible regions with $100$ than with $30$ bins.  However, the finite number of haloes in the simulation limit the number of samples drawn from the continuous (underlying) distribution (which would be obtained with an infinite number of haloes as the bin width approached zero).  Therefore---particularly at the high-mass end with few haloes---bins may be insufficiently wide to preserve the monotonicity of the HMF, or may be empty altogether.  This does not only widen the posteriors, but may also shift the peaks, if the low-mass end does not provide sufficient constraints where the scatter is minimal.  In our simulated \lq\lq{}data,\rq\rq{} we have maximised the scatter (to within Poisson errors) to produce a pessimistic scenario.  The performance of each HMF is similar, so we only show the Peacock HMF.  We find that this produces little to no effect on the location or width of the $1\sigma$ credible regions.  This behaviour is largely independent of the number of live points as well.  Consequently, we fixed the number of bins when processing the halo data in \cref{sec:results_and_discussion}.

  
\subsection{Number of live points} 
\label{sub:number_of_live_points}

\cref{fig:Peacock_fake_data_change_lenm,fig:Peacock_fake_data,fig:Jenkins_fake_data,fig:Warren-Crocce_fake_data,fig:SMT-Courtin_fake_data,fig:Tinker-Angulo-Watson_fake_data} show---for a given HMF---the posteriors produced by the Poissonian likelihood functions.  The number of live points is indicated by the colour of the credible regions.  The dashed black line shows the value of the parameter at input.

The number of live points is of interest as we need to find a compromise between efficiency and accuracy.  The higher the dimensionality of the parameter space, the more live points are typically required to constrain the posterior.  However, we find that once the number is increased beyond $10^{2}$, there is usually little difference between the $100$, $200$, $500$ and $1000$ posteriors.  (\texttt{MultiNest} advises using fewer than 1000 live points when the Importance Nested Sampling mode is used, due to memory constraints.)  The credible regions all largely overlap, indepedent of the number of live points.  The marginalised posteriors usually overlap as well (\eg{} all parameters in \cref{fig:Jenkins_fake_data,fig:Peacock_fake_data}; $\left\{ A,a \right\}$ in \cref{fig:SMT-Courtin_fake_data}; $\left\{ A,b,c \right\}$ in \cref{fig:Warren-Crocce_fake_data}).  In some cases ($p$ in \cref{fig:SMT-Courtin_fake_data}; $a$ in \cref{fig:Warren-Crocce_fake_data}) the peaks of the posteriors move further from or closer to the original value.  This does not affect the result that in all cases the input parameters lie within the 1-$\sigma$ credible regions.  This illustrates the need to find a minimum number of live points for which the results converge and ensuring that adding more live points does not move the ML and MAP estimates outside the existing credible regions.  This demonstrates that our estimates of the posteriors have already converged to an accurate representation of the likelihood with only $\sim 10^{2}$ live points.
 

\subsection{Number of free parameters} 
\label{sub:number_of_free_parameters}
 
Finally we examine the effect of the number of free parameters.  Again we refer to
\cref{fig:Peacock_fake_data_change_lenm,fig:Peacock_fake_data,fig:Jenkins_fake_data,fig:Warren-Crocce_fake_data,fig:SMT-Courtin_fake_data,fig:Tinker-Angulo-Watson_fake_data} which show---for a given HMF---the posteriors produced by the Poissonian likelihood functions.  The number of live points is indicated by the colour of the credible regions.  The dashed black line shows the value of the parameter at input.

The HMFs with three free parameters---Jenkins (\cref{fig:Jenkins_fake_data}), Peacock (\cref{fig:Peacock_fake_data}) and SMT-Courtin (\cref{fig:SMT-Courtin_fake_data}) ---can all be reasonably constrained by the available data.  Of the two four-parameter fits, Warren-Crocce (\cref{fig:Warren-Crocce_fake_data}) also recovers the required values whereas Tinker-Angulo-Watson (\cref{fig:Tinker-Angulo-Watson_fake_data}) does not produce such an accurate result.  Although \texttt{MultiNest} did find the ML values to within a few percent of the input values, the posteriors created by \texttt{GetDist} showed a broad tail in its credible regions for $\left\{ A,a,b \right\}$, while $c$ was fine.  Although the chains do converge as we increase the number of live points, they converge to the wrong values when examining the peaks in the posterior (as processed by \texttt{GetDist}).  The nested sampling ellipses exclude the ML value at some point during the sampling, from which point onwards the sample can only be drawn from within the likelihood iso-surface created by the ellipses.  (This is hardly surprising given that $A(b\nu)^{a}$ is automatically degenerate.  Such behaviour was also seen in \cite{Buchner2016}).  The Reed 2007 fit has six parameters.  (We omit this figure.)  In this case the Poissonian likelihood cannot identify the original parameter values used to create the data.  Many pairs of parameters in this fitting function are degenerate (\cref{tab:hmf_functions}).  This may be a failure of the data as well: counts must be rounded to an integer value (to correspond with the behaviour in our $N$-body derived data) and, once scattered, could be negative.  Rounding the counts mapped a small portion of the parameter space (used for calculating the \fcd{}) to the same counts.  The two factors which affect this conversion are the box size and the bin width (larger boxes create more haloes; wider bins include more haloes).  The behaviour of the fit agrees with statements in \cite{2003MNRAS.346..565R,2007MNRAS.374....2R} that some of their parameters can only be constrained by masses $M \geq 10^{15} M_{\odot}$, of which we have very few in the halo data.  The nested sampling algorithm is not infallible.  For this reason, we exclude the Reed and Tinker-Angulo-Watson fits from our halo results in \cref{sec:results_and_discussion}.


Having explored the effects of the likelihood function, the number of bins and the number of live points, we have found that the optimal choices are the Poissonian likelihood function with at least $100$ live points and that the number of bins is irrelevant.  Thus we set $30$ bins and $500$ live points for our data proper in \cref{sec:results_and_discussion}.  We also exclude the Tinker-Angulo-Watson and Reed-07 fits on the basis that we have insufficient data to properly constrain their free parameters.

\clearpage
\bibliographystyle{JHEP}
\bibliography{main}
\end{document}